%% file: main.tex
\pdfoutput=1
\documentclass[preprint,12pt,authoryear]{elsarticle}

\input{definitions}

\journal{}

\usepackage{natbib}
\begin{document}
\interfootnotelinepenalty=10000
\begin{frontmatter}

\title{Incorporating physical constraints in a deep probabilistic machine learning framework for coarse-graining dynamical systems}


\author[mymainaddress]{Sebastian Kaltenbach}
\ead{sebastian.kaltenbach@tum.de}

\author[mymainaddress]{Phaedon-Stelios Koutsourelakis\corref{mycorrespondingauthor}}
\cortext[mycorrespondingauthor]{Corresponding author}
\ead{p.s.koutsourelakis@tum.de}

\address[mymainaddress]{Professorship of Continuum Mechanics, Technical University of Munich}



\begin{abstract}
\input{abstract.tex}

\end{abstract}


\end{frontmatter}


\section{Introduction}
\label{sec:intro}

\input{introduction.tex}

\section{Methodology}
\label{sec:meth}
\input{methodology.tex}

\section{Numerical Illustrations}
\label{sec:num}
\input{numerical-illustrations.tex}

\section{Conclusions}
\label{sec:conclusions}
\input{conclusions.tex}
\newpage
\bibliography{refs}

\end{document}

%% file: definitions.tex
\usepackage{amssymb}
\usepackage{graphicx}
\usepackage{latexsym}
\usepackage{amsmath}
\usepackage{bbold}      
\usepackage[algoruled, linesnumbered]{algorithm2e}

\usepackage{url}
\usepackage{caption}
\usepackage{subcaption}
\usepackage{lineno}

\usepackage{bm}
\usepackage{xcolor}
\usepackage{algorithm2e}

\usepackage{enumerate}
\usepackage{sidecap}
\usepackage{wrapfig}

\usepackage{comment}    
\usepackage{makecell}   

\usepackage{tikz}
\usetikzlibrary{bayesnet}
\usetikzlibrary{calc,fit,arrows,arrows.meta, backgrounds}
\pgfdeclarelayer{background}
\pgfdeclarelayer{foreground}
\pgfsetlayers{background,foreground}

\newcommand{\x}{\mathcal{X}}  

\definecolor{newcolor}{rgb}{.8,.349,.1}

\newcommand{\bx}{\bs{x}}

\newcommand{\bxx}{\bs{X}}

\newcommand{\bt}{\bs{\theta}}

\newcommand{\bp}{\bs{\phi}}

\newcommand{\beps}{\bs{\epsilon}}

\newcommand{\be}{\begin{equation}}
\newcommand{\ee}{\end{equation}}
\newcommand{\bc}{\begin{center}}
\newcommand{\ec}{\end{center}}
\newcommand{\bd}{\begin{description}}
\newcommand{\ed}{\end{description}}
\newcommand{\bi}{\begin{itemize}}
\newcommand{\ei}{\end{itemize}}
\newcommand{\pa}{\partial}
\newcommand{\bs}{\boldsymbol}

\def\RR{ \mathbb R}
\newcommand{\expe}{\mathbb{E}}

\newcommand{\refeq}[1]{Equation (\ref{#1})}

\newcommand{\bmat}{\begin{pmatrix}}
\newcommand{\emat}{\end{pmatrix}}
\newcommand{\bsmat}{\left(\begin{smallmatrix}}
\newcommand{\esmat}{\end{smallmatrix}\right)}
\newcommand{\bes}{\begin{equation}\begin{split}}
\newcommand{\ees}{\end{split}\end{equation}}

\newcommand{\review}[1]{\textcolor{black}{#1}}

\definecolor{TUMGruen}{RGB}{162,173,0} 
\definecolor{TUMOrange}{RGB}{227,114,34} 

%% file: abstract.tex
Data-based  discovery of effective, coarse-grained (CG) models of high-dimensional dynamical systems presents a unique challenge in computational physics and particularly in the context of multiscale problems. 
The present paper offers  a data-based, probabilistic perspective that enables the quantification of predictive uncertainties.
One of the outstanding problems has been the introduction of physical constraints in the probabilistic machine learning objectives. The primary utility of such constraints  stems from the undisputed physical laws such as conservation of mass, energy etc. that they represent. Furthermore and apart from leading to physically realistic predictions, they can significantly reduce the requisite amount of training data which for high-dimensional, multiscale systems are expensive to obtain (Small Data regime). 
 We formulate the coarse-graining  process by employing  a probabilistic state-space model and account for the aforementioned equality constraints as virtual observables in the associated densities.  We demonstrate how   deep neural nets in combination with probabilistic inference tools  can  be employed to identify the coarse-grained variables and their evolution model without ever needing to define a fine-to-coarse (restriction) projection and without needing time-derivatives of state variables.

We advocate  a sparse Bayesian learning  perspective which avoids overfitting and  reveals the most salient features in the CG evolution law.
The formulation adopted enables the quantification of a crucial, and often neglected, component in the CG process, i.e. the predictive uncertainty due to information loss.
Furthermore, it is capable of  reconstructing  the evolution of the full, fine-scale system and therefore the observables of interest need not be selected a priori.
 We  demonstrate the efficacy of the proposed framework by applying it to  systems of interacting particles and a series of images  of a nonlinear pendulum. In both cases we identify the underlying coarse dynamics and can generate extrapolative predictions including the forming and propagation of a shock for the particle systems and a stable trajectory in the phase space for the pendulum.

%% file: introduction.tex
High-dimensional, nonlinear dynamical systems are ubiquitous in applied physics and engineering. The computational resources needed for their solution can grow exponentially with the dimension of the state-space as well as with the smallest time-scale that needs to be resolved and which  determines the discretization time-step. Hence the ability to construct reduced, {\em coarse-grained} descriptions and models that are nevertheless predictive of various observables and at time-scales much larger than the inherent ones, is  an important task \citep{givon_extracting_2004}.

One strategy for  learning such coarse-grained (CG) models is based on  data generated by simulations of the fine-grained (FG) system. This can yield  an automated solution especially in cases where domain knowledge is limited or absent. The derivation of CG models from  data is also particularly relevant in domains where  FG models are not available, such as in social sciences or biophysics, but data abound \citep{bialek_biophysics:_2012,alber_integrating_2019}. Data-based methodologies have also been fueled by recent advances in statistical- \citep{ghahramani_probabilistic_2015} or machine-learning \citep{lecun2015deep} which, in large part, have been enabled by large datasets (and the computational means to leverage them). We note nevertheless that coarse-graining tasks based on FG simulation data exhibit some fundamental differences \citep{koutsourelakis2016big}. Firstly, the acquisition of FG simulation data is by definition expensive and the reduction of the required FG simulations is one of the objectives of CG model development. Secondly, in  physical applications, significant  information  about the underlying physical/mathematical structure of the problem, and of the CG model in particular, is available. This information might come in the form of constraints that reflect e.g undisputed physical principles such as conservation laws (e.g. mass, momentum, energy).  Injecting this prior information into the CG models in combination with FG data in an automated fashion represents a significant challenge \citep{marcus_rebooting_2019}, especially in the context of {\em probabilistic} models \citep{stinis2019enforcing}. Such a capability would be instrumental not only in reducing the required amount of FG data, but more importantly, in enabling predictions under {\em extrapolative} settings  as those arising  e.g. when the initial conditions of the FG system are different from the ones  in the training  data.


In this paper, we propose a generative, probabilistic (Bayesian) machine learning framework \citep{koutsourelakis_scalable_2011} which employs FG simulation data augmented by {\em virtual observables} to account for constraints. \review{The latter concept which we elucidate in the sequel, enables the incorporation of domain knowledge in probabilistic models and represents, in our opinion the most novel contribution of this paper. Furthermore and within the Bayesian framework advocated, it allows us to introduce appropriate priors that promote  the discovery of {\em slow-varying}  CG state-variables which is a highly-desirable feature for multiscale systems \citep{kevrekidis_equation-free_2003}.}
In contrast to most existing techniques which consider the problems of CG state variable discovery and CG model construction in two or more steps \citep{schmid_dynamic_2010,williams_datadriven_2015,wu_variational_2017,froyland_computational_2014}, we address both simultaneously \citep{felsberger_physics-constrained_2019}. 
The framework proposed consists of two building blocks: a probabilistic coarse-to-fine map \citep{schoberl2017predictive} and an evolution law for the CG dynamics. The former can be endowed with great flexibility in discovering appropriate CG variables  when combined with deep neural nets  \citep{raissi2017physics,raissi2019physics,yang2019conditional}, \review{which is especially challenging if the number of training data is small\footnote{In the dynamical systems investigated the size of the dataset depends on the length of the FG time-sequences as well as the number of such sequences employed for training.}. We demonstrate nevertheless the efficacy of such an approach when  physical information is incorporated  a-priori into the model.} The CG variables identified are not restricted to  indicator functions of sub-domains of the state-space as in other generative models  \citep{mardt_vampnets_2018,wu_variational_2017,wu2018deep} and which are difficult to learn when the simulation data is limited and has not sufficiently populated all important regions of the state-space.

The second component of the proposed framework pertains to the discovery of the CG evolution law which is learned by employing a large vocabulary of feature functions and sparsity-inducing priors. This leads to interpretable solutions \citep{duncker_learning_2019}, even in the {\em Small Data} regime that avoid overfitting and reveal salient characteristics of the CG system \citep{grigo2019physics}. The premise of sparsity \citep{pantazis_unified_2019} has been employed in the past for the discovery of the CG dynamics as e.g. in the  SINDy method \citep{brunton2016discovering,kaiser2018sparse, champion2019data}. This however requires the availability of time-derivatives of the CG variables   and does not directly lead to a posterior on the model parameters that can reflect inferential uncertainties. Nonparametric models for the CG dynamics have also been proposed \citep{ohkubo_nonparametric_2011} but have been restricted to low dimensions. The learned CG dynamics are in general nonlinear in contrast to efforts based on transfer operators \citep{klus_data-driven_2018} and particularly the Koopman operator \citep{koopman_hamiltonian_1931,mezic2005spectral,brunton_koopman_2016}. While the associated  theory guarantees the existence of a linear operator,  this is possible in the infinite dimensional space of observables, it does not specify how many should be used to obtain a good approximation, and more importantly, how one can predict future FG states given predictions on the evolution of those observables i.e. the reconstruction step.\\

The latter constitutes the main difference of the proposed  model with non-generative ones based  e.g. on information-theoretic concepts \citep{katsoulakis2013information,harmandaris_path-space_2016,katsoulakis2019data} or on the Mori-Zwanzig (MZ) formalism \citep{mori1965transport,zwanzig1973nonlinear,chorin2007problem}. Apart from the difficulties in approximating the right-hand-side of the MZ-prescribed CG dynamics, and particularly the memory term \citep{lei2016data,zhu_estimation_2018}, this can only guarantee correct predictions of the CG variables' evolution. If  observables  not depending on CG variables are of interest, then a reconstruction operator would need to be added. In contrast, in the proposed model this reconstruction operator is represented by the probabilistic coarse-to-fine map which is simultaneously learned from the data and can quantify  predictive uncertainties associated with the information loss that unavoidably takes place in any CG process as well as due to the  fact that finite (and preferably, small) data has been used for training. 
The enabling computational technology for training the proposed model is based on probabilistic inference. In order to resolve the intractable  posterior on latent variables and model parameters  in our Bayesian framework, we make use of Stochastic Variational Inference \citep{hoffman2013stochastic} as MCMC is cumbersome in high dimensions. We operate on the discretized time domain  \citep{archambeau_approximate_2011} and  demonstrate how amortized \citep{krishnan2017structured, fortuin2019multivariate} and non-amortized approximations can be employed.



The remainder of the paper is structured as follows: In Section \ref{sec:meth} we present the general methodological framework with special attention on  the two building  blocks of the state-space model proposed  i.e. the transition law for the CG dynamics  and the incorporation of virtual observables (section \ref{sec:transition}), as well as the  the emission law which provides the link between CG and FG description through a probabilistic {\em coarse-to-fine} map (section \ref{sec:emission}). Computational aspects related to inference  and prediction are discussed in sections  \ref{sec:inference} and \ref{sec:predictions} respectively. 
Section \ref{sec:num} contains illustrative applications  involving  coarse-graining of  high-dimensional systems of interacting particles (section \ref{sec:particles}) as well as learning the dynamics of a  nonlinear pendulum (section \ref{sec:pendulum}) from a sequence of images. 
We conclude in section \ref{sec:conclusions} which also contains a discussion on possible extensions. 


%% file: methodology.tex
In general, we use the subscript $f$ or lower-case letters  to denote variables  associated with the (high-dimensional) fine-grained(FG)/full-order  model and the subscript $c$ or upper-case letters for quantities of the (lower-dimensional) coarse-grained(CG)/reduced-order description. We also use a circumflex  \textbf{$~\hat{}~$}   to denote observed/known variables. 
We begin with the presentation of the FG and the CG model and subsequently explain the essential ingredients of the proposed formulation.

\subsection{The FG and CG models}
We consider a, generally high-dimensional, FG system with state variables $\bx$ of dimension $d_f$ ($d_f>>1$) such that   $\bx \in \mathcal{X}_f \subset  \mathbf{R}^{d_f}$. 
The dynamics of the FG system are dictated by system of deterministic or stochastic ODEs i.e.,\\
\begin{equation}
\dot{\bx_t} = \bs{f}(\bx_t,t), \quad t>0    
\label{eq:fg}
\end{equation}
 The initial condition  $\bx_0$ might be deterministic or  drawn from a specified distribution. In the following we do not make explicit use of the FG dynamics  but rely purely on FG  data  i.e. time sequences simulated from \refeq{eq:fg} with a time-step, say $\delta t$. That is, our observables consists of $n$ data sequences over $T+1$ FG time-steps $\delta t$ i.e., 
 \be
 \mathcal{D}_{T,n}=\{ \hat{\bx}_{ 0: T \delta t}^{(1:n)} \} 
 \label{eq:fgdata}
\ee
We denote the (unknown) CG state variables by $\bxx$  and assume   $\bxx \in \mathcal{X}_c \subset  \mathbf{R}^{d_c} $, where $d_c$ is the dimension of the CG  system. 
We presuppose Markovian dynamics\footnote{As discussed in section \ref{sec:num}, this assumption can be relaxed.} for the CG system of the form:
\begin{equation}
\dot{\bxx_t} = \bs{F}(\bxx_t,t) 
\label{eq:cg}
\end{equation}
which we discretize using a linear multistep method and a CG time step $\Delta t$: 
\be
\bs{R}_l(\bxx)=\sum_{k=0}^K \left( \alpha_k \bxx_{(l-k)\Delta t} + \Delta t \beta_k \bs{F}(\bxx_{(l-k) \Delta t}) \right) =0, \qquad l=K,K+1, \ldots 
\label{eq:cgdiscr}
\ee
 where $\alpha_k,\beta_k$ are the parameters of the discretization scheme and $\bs{R}_l$ the corresponding residual at time step $l$ \citep{butcher2016numerical}. We note that depending on the values of the parameters $K, \alpha_k,\beta_k$, several of the well-known, explicit/implicit, numerical time-integration schemes can be recovered. 
In this work, our goal is two-fold:
\bi
\item[a)] to identify the CG state-variables $\bxx$ and their relation with the FG description $\bx$,
\item[b)] to identify the right-hand side of \refeq{eq:cg},
\ei
in view of enabling predictions of the FG system over longer time horizons. 
Traditionally, the aforementioned tasks are {\em not} considered simultaneously. Usually  the CG state variables are specified a priori using domain-knowledge (physical insight) or based on the observables of interest \citep{harmandaris_path-space_2016}. In other  efforts,  linear or non-linear dimensionality reduction  procedures are first employed  in order to identify such a lower-dimensional set of collective variables $\bxx$ (e.g. \citep{coifman_diffusion_2008}).
In both of these cases, $\bxx$ are defined using a {\em fine-to-coarse}, projection map e.g. $\bxx= \Pi(\bx)$ where $\Pi: \mathcal{X}_f \subset \RR^{d_f} \to \mathcal{X}_c \subset \RR^{d_c}$. Irrespective of whether this map is prescribed from the physics or learned from data, it is  generally a many-to-one function that does not have an inverse i.e. if the CG states $\bxx$ are known one cannot readily reconstruct $\bx$ \citep{trashorras_mesoscale_2010}. 

We note that that this has nothing to do with the quality of the CG evolution law (problem b) above). Even if the Mori-Zwanzig (MZ) formalism were employed,  which in principle provides an exact,  closed system of evolution equations for any observable of the FG states and therefore for $\bxx=  \Pi(\bx)$, even if all the terms in the right-hand side were available, one would simply be able to predict the future evolution of $\bxx$ but not $\bx$. This might be sufficient for a lot of problems of practical interest where  the CG variables (or observables thereof) are of sole interest. Our  goal however is a bit more ambitious, i.e. we seek to find a $\bxx$ that would allow us to reconstruct as accurately as possible the whole FG vector $\bx$ into the future. As with any coarse-graining process, we recognize that this would unavoidably imply some information loss which in turn will give rise to predictive uncertainty \citep{katsoulakis_information_2006}. In this work, we advocate a probabilistic framework that quantifies this uncertainty.      

With regards to problem b) above, we note that its solution hinges  upon the CG variables $\bxx$ employed (problem a)). Irrespective of the breadth of the model forms considered (i.e. functions $\bs{F}$ in \refeq{eq:cg}), the evolution of some $\bxx$ might fall outside this realm. For example, it is known from MZ theory that memory terms can become significant for certain observables. It is  well-known that such memory terms can be substituted or approximated by additional variables \citep{kondrashov_data-driven_2015} which would  in turn  imply  an augmented CG description $\bxx$ in \refeq{eq:cg} that contains these auxiliary internal state variables \citep{coleman_thermodynamics_1967}.   

\begin{figure}
\def\svgwidth{\linewidth}
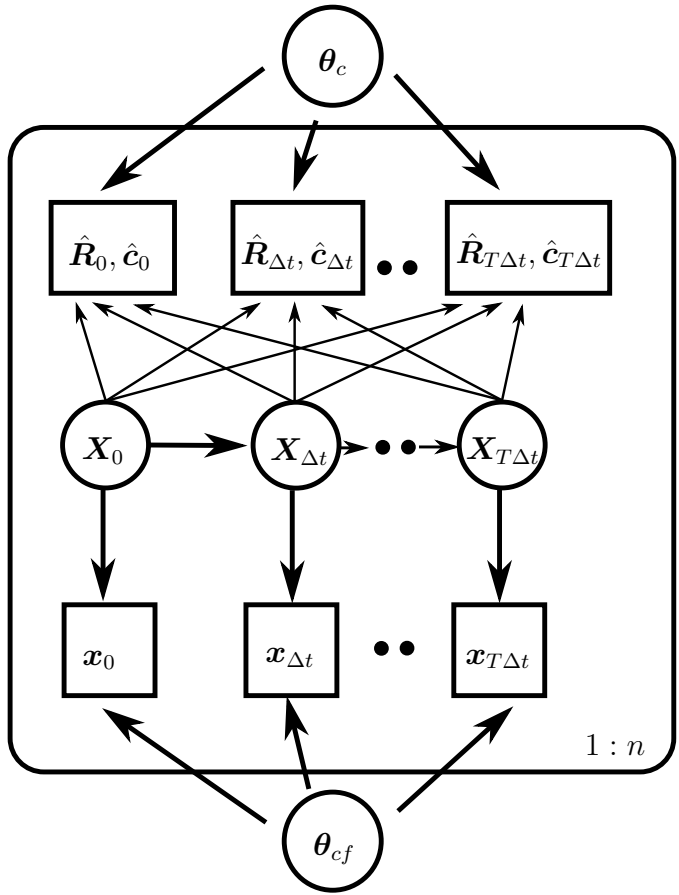
\vspace{-2cm}
\caption{Proposed probabilistic graphical model. The CG variables $\bxx$ are latent and are inferred together with the parameters $\bt_c$ and $\bt_{cf}$. Apart from the  the FG states $\bx$, the observables are augmented by {\em virtual observables} $\hat{\bs{R}},\hat{\bs{c}}$ (see section \ref{sec:transition}). \review{These virtual observables can depend on all CG variables but more often this dependence is restricted to only a few of them.}} 
\label{fig:pgm}
\end{figure}

We address problems a) and b) in the coarse-graining process {\em simultaneously} by employing a probabilistic state-space model. This consists of two densities i.e. 
\bi
\item the transition law which dictates the evolution of the CG variables $\bxx$ (section \ref{sec:transition}). Special attention is paid to the definition of {\em virtual observables} with which the CG states and their dynamics can be injected with physical information. 
\item the emission law which provides the link between CG and FG description through a probabilistic {\em coarse-to-fine} map  (section \ref{sec:emission}, \citep{felsberger_physics-constrained_2019}).
\ei
We emphasize that in our formulation, the CG state-variables $\bxx$ are implicitly defined as latent generators of the FG description $\bx$. As discussed in detail in the sequel, this enables a straightforward, {\em  probabilistic} reconstruction of $\bx$ when $\bxx $ is known. The inverse map (analogous to $\Pi$ above) arises naturally through probabilistic inference as explained in section \ref{sec:inference}.  
An overview of the essential elements of the proposed model can be seen  in the probabilistic  graphical model of Figure \ref{fig:pgm}.

\subsection{Transition Law: CG dynamics and virtual observables}
\label{sec:transition}
Typical  state-space models \citep{cappe_inference_2005,ghahramani_unsupervised_2004,durstewitz_state_2017,krishnan2017structured} postulate Markovian, stochastic dynamics for the hidden variables $\bxx$, in the form of a diffusion process,  
 which are subsequently discretized {\em explicitly} using e.g. a  Euler-Maruyama scheme with time step $\Delta t$.
This gives rise to a, generally Gaussian, conditional density $p(\bxx_{(l+1)\Delta t} |\bxx_{l \Delta t})$  which can be stacked over multiple time-instants in order to formulate a generalized prior on the CG-space. 

When the CG state-variables $\bxx$ are given (in part or in whole) physical meaning (e.g. as thermodynamic state variables), then some of the equations for their evolution  are prescribed by associated physical principles e.g. conservation of mass, momentum, energy. These can be reflected in the residuals $\bs{R}_l$ of the governing equations as in \refeq{eq:cgdiscr}
 or alternatively as equality constraints of the form:
 \be
 \bs{c}_l(\bxx_{l \Delta t}) = \bs{0}, \quad l=0,1,\ldots
 \label{eq:constraintl}
 \ee
 which must hold at each time-step. The function $\bs{c}_l: \mathcal{X}_c \subset \RR^{d_c} \to \RR^{M_c}$  enforces these known constraints at each time-step (see specific examples in  section \ref{sec:num}) and the only requirement we will impose is that of differentiability of $\bs{c}_l$ (see section \ref{sec:inference}).
 In order to account for the aforementioned constraints in the transition law of the CG state variables, we employ the novel (to the best of our knowledge) concept of {\em virtual observables}.
 In  particular for each of the residuals $\bs{R}_l$ in \refeq{eq:cgdiscr}, we define a new variable/vector $\hat{\bs{R}}_l$ which relates to  $\bs{R}_l$ as follows:
 \be
 \hat{\bs{R}}_l = \bs{R}_l (\bxx) +\sigma_R \beps_R, \qquad \beps_R \sim \mathcal{N}(\bs{0}, \bs{I})
 \label{eq:voresl}
 \ee
 We further assume that $\hat{\bs{R}}_l$ have been {\em virtually observed} and $\hat{\bs{R}}_l=0$ leading to an augmented version of the  data in \refeq{eq:fgdata}, by  a set of virtual observations and therefore virtual likelihoods of the type:
 \be
 p(\hat{\bs{R}}_l=\bs{0}~ | ~\bxx, \sigma_R) = \mathcal{N}(\bs{0} ~| ~\bs{R}_l (\bxx), \sigma^2_R \bs{I})
 \label{eq:vlikeres}
 \ee
 The ``noise" parameter $\sigma_R$ determines the intensity of the enforcement of the virtual observations and is analogous to the tolerance parameter with which residuals are enforced in a deterministic solution of the dynamics. Similarly, for constraints of the form  of \refeq{eq:constraintl}, additional variables and virtual observables of the type:
 \be
 \bs{0}=\hat{\bs{c}}_l=\bs{c}_l(\bxx_{l \Delta t})+\sigma_c \beps_c, \quad \beps_c \sim \mathcal{N}(\bs{0}, \bs{I})
 \label{eq:vocon}
 \ee
 can be defined which would lead to an augmented (virtual) likelihood with terms of the type:
 \be
 p\left(\hat{\bs{c}}_l=\bs{0} ~| \bxx_{l \Delta t}, \sigma_c \right) = \mathcal{N}\left( \bs{0} ~| ~\bs{c}_l (\bxx_{l\Delta t}), \sigma^2_c \bs{I} \right)
 \label{eq:vlikecon}
 \ee
 where the role of $\sigma_c^2$ is analogous to $\sigma_R^2$ above.
 
Since the goal is to identify the right-hand side of the evolution laws in \refeq{eq:cg}, we denote by $\bt_c$ the parameters appearing in  $\bs{F}$ i.e. $\bs{F}(\bxx_t, t; ~\bt_c)$. Accordingly, the virtual observations in \refeq{eq:voresl} or \refeq{eq:vocon} would depend on $\bt_c$. We defer until section \ref{sec:num} a detailed discussion on the form, the parametrization  as well as the prior specifications  in the Bayesian setting adopted. The latter plays an important role as with sparsity-inducing priors we can  avoid overfitting and obtain a parsimonious and physically-interpretable solution for $\bs{F}$. 
 \review{We finally remark that physical information taking the form of equalities can also be available for the FG states $\bx$. While this can be incorporated using appropriate virtual observables as above, the inference framework would exhibit significant differences (in brief,  FG states would need to be inferred as well) and in order to avoid confusion we do not discuss such cases here. }
 
\subsection{Emission law: Coarse-to-Fine map}
\label{sec:emission}

We make use of a probabilistic {\em generative} model in the definition of the CG state-variables through a {\em coarse-to-fine} map \citep{felsberger_physics-constrained_2019} as opposed to traditional,  many-to-one maps from the FG description to the CG one.
We denote the associated (conditional) density by:
\be
p_{cf}(\bx_t |~\bxx_t; ~\bt_{cf})
\label{eq:pcf}
\ee
where $\bt_{cf}$ denote the (unknown) parameters that will be learned from the data.
The  form of $p_{cf}$ can be adapted to the particulars of the problem and can be endowed with various levels of domain knowledge. In section \ref{sec:num}, we provide various examples, from particle-systems where $p_{cf}$ is fully determined by the physics, to a more abstract case  where  deep neural networks are employed in order to learn the full $p_{cf}$. 
We note finally that a (probabilistic) fine-to-coarse map can still be learned in the current setting, and would correspond to the {\em posterior} of $\bxx_t$ given $\bx_t$. We discuss this as well as all aspects pertaining to inference and learning in the next section.

\subsection{Inference and Learning}
\label{sec:inference}

We start this section by summarizing the main elements of the model presented (i.e. data, latent variables and parameters - see also Table \ref{tab:variables}) and subsequently describe a fully Bayesian inference  scheme based on Stochastic Variational Inference (SVI, \citep{hoffman2013stochastic}) tools. 

\begin{table}
    \begin{tabular}{l|c|l}
       &  $\hat{\bx}_{ 0: T \Delta t}^{(1:n)} $ & FG simulation Data  \\
  Observables $\mathcal{D}$      & $\hat{\bs{R}}_{0:T}^{(1:n)}$ & Virtual Observables corresponding to CG model residuals\\
         & $\hat{\bs{c}}_{0:T}^{(1:n)}$ & Virtual Observables corresponding to CG constraints \\
         \hline
 Latent variables &                 $\bxx_{ 0: T \Delta t}^{(1:n)}$ & CG state variable\\
 \hline
  Model parameters $\bt$  &      $\bt_{cf}$ & parameters in the coarse-to-fine mapping\\
    &     $\bt_c$ & parameters in the CG evolution law\\
    \end{tabular} 
    \caption{Data, latent variables and model parameters}
    \label{tab:variables}
    
\end{table}

We  adopt an enlarged definition of {\em data} which we cumulatively denote by $\mathcal{D}$ and which encompasses:
\bi
\item FG simulation data as in \refeq{eq:fgdata} consisting of $n$ sequences of the FG state-variables. As the likelihood model implied by the $p_{cf}$ in \refeq{eq:pcf} involves only the observables at each {\em coarse} time-step we denote those by $\{\hat{\bx}_{ 0: T \Delta t}^{(1:n)} \}$. We assume that the number of observations in each sequence is the same although this is not necessary. In fact, the length  of each time-sequence and the number of time-sequences needed could be the subject of an active learning scheme. This would be particularly important in cases where very expensive, high-dimensional FG simulators are employed.  \review{The generative, proposed formulation can account for any type of (in)direct or  (in)complete/partial,  experimental or computational observations relating to FG states which we omit here for simplicity of the presentation. We nevertheless illustrate this capability of the model in the example of  section \ref{sec:pendulum}}.

\item Virtual observables relating to the CG states $\bxx$  at each time-step $l$ consisting of residuals $\hat{\bs{R}}_l^{(1:n)}$  as in \refeq{eq:voresl} and/or constraints $\hat{\bs{c}}_l^{(1:n)}$ as in \refeq{eq:vocon} (the superscript pertains to the time sequence $i=1,\ldots,n$). Assuming they  pertain to all time-steps, we denote them by $\left \{ \hat{\bs{R}}_{0:T}^{(1:n)}, \hat{\bs{c}}_{0:T}^{(1:n)} \right\}$. 
\ei

The latent (unobserved) variables of the model are represented by the CG state-variables $\left\{\bxx_{ 0: T \Delta t}^{(1:n)} \right\} $ which relate to the FG data through the $p_{cf}$ (in \refeq{eq:pcf}) and to the virtual observables through \refeq{eq:vlikeres} or \refeq{eq:vlikecon}.

Finally, the (unknown) parameters of the model which we denote cumulatively by $\bt$ consist of\footnote{If any of the parameters in this list are prescribed, then they are omitted from $\bt$.}:
\bi
\item $\bt_c$ which parametrize the right-hand-side of the CG evolution law (see end of section \ref{sec:transition}),
\item  $\bt_{cf}$ which parametrize the probabilistic coarse-to-fine map (\refeq{eq:pcf}),
\item $\sigma_R, \sigma_c$ involved in the enforcement of virtual observables in \refeq{eq:voresl} and \refeq{eq:vocon} respectively, and,
\item {\em hyperparameters} associated with the priors employed on the latent variables or the previous parameters.
\ei

Following  a fully-Bayesian formulation, we can express the posterior of the unknowns (i.e. latent variables and parameters) as follows:
\be
p( \bxx_{ 0: T \Delta t}^{(1:n)}, ~\bt~ |~ \mathcal{D}) = \cfrac{ p(\mathcal{D}~ | ~\bxx_{ 0: T \Delta t}^{(1:n)}, \bt) ~p(\bxx_{ 0: T \Delta t}^{(1:n)}, \bt) }{p(\mathcal{D}) }
\label{eq:genbayes}
\ee
where $p(\bxx_{ 0: T \Delta t}^{(1:n)}, \bt)$ denotes the prior on the latent variables and parameters. 

We discuss first the likelihood term $p(\mathcal{D} | \bxx_{ 0: T \Delta t}^{(1:n)}, \bt)$ which can be decomposed into the product of three (conditionally) independent terms, one for each data-type, i.e.:
\be
p(\mathcal{D} ~|~ \bxx_{ 0: T \Delta t}^{(1:n)}, \bt) = p(\hat{\bx}_{ 0: T \Delta t}^{(1:n)} ~|~  \bxx_{ 0: T \Delta t}^{(1:n)}, \bt) ~p( \hat{\bs{R}}_{0:T}^{(1:n)}~  | ~ \bxx_{ 0: T \Delta t}^{(1:n)}, \bt)~ p(\hat{\bs{c}}_{0:T}^{(1:n)} ~|~  \bxx_{ 0: T \Delta t}^{(1:n)}, \bt)
\label{eq:likedecomp}
\ee
We further note that (from \refeq{eq:pcf}):
\be
p(\hat{\bx}_{ 0: T \Delta t}^{(1:n)}~ | ~ \bxx_{ 0: T \Delta t}^{(1:n)}, \bt)  = \prod_{i=1}^n \prod_{l=0}^T p_{cf}( \bx^{(i)}_{l~\Delta t} ~|~  \bxx^{(i)}_{l~\Delta t}, \bt_{cf})
\label{eq:fgdatalikelihood}
\ee
and (from \refeq{eq:vlikeres}):
\be
\begin{array}{ll}
p( \hat{\bs{R}}_{0:T}^{(1:n)}  |  \bxx_{ 0: T \Delta t}^{(1:n)}, \bt)  &  = \prod_{i=1}^n \prod_{l=0}^T \mathcal{N}\left( \bs{0} | \bs{R}_l( \bxx^{(i)}), \sigma_R^2 \bs{I} \right) \\
& \propto \prod_{i=1}^n \prod_{l=0}^T  \frac{1}{\sigma_R^{dim(\bs{R})} } \exp \left\{ -\frac{1}{2 \sigma_R^2} \left| \bs{R}_l( \bxx^{(i)})  \right|^2 \right\}
\end{array}
\ee
and (from \refeq{eq:vlikecon}):
\be
\begin{array}{ll}
p( \hat{\bs{c}}_{0:T}^{(1:n)}  |  \bxx_{ 0: T \Delta t}^{(1:n)}, \bt)  &  = \prod_{i=1}^n \prod_{l=0}^T \mathcal{N}(\bs{0} | \bs{c}_l( \bxx^{(i)}_{l~\Delta t}), \sigma_c^2 \bs{I}) \\
& \propto \prod_{i=1}^n \prod_{l=0}^T  \frac{1}{\sigma_c^{dim(\bs{c})} } \exp \left\{ -\frac{1}{2 \sigma_c^2} \left| \bs{c}_l( \bxx^{(i)}_{l~\Delta t})  \right|^2 \right\}
\end{array}
\ee
While the complexity of the expressions involved imply a non-analytic solution for the posterior, we emphasize that the terms above encode actual and virtual observables (constraints) and they are differentiable, a property that is crucial for carrying out Variational Inference.

Before presenting the inference procedure, we mention an interesting possibility for encoding prior information for the latent CG  states $\bxx_{ 0: T \Delta t}^{(1:n)}$ through the prior term $p(\bxx_{ 0: T \Delta t}^{(1:n)})$.  A desirable property of the CG state-variables is that of {\em slowness} i.e. that they should capture features of the system that evolve over  (much) larger time-scales \citep{kevrekidis_equation-free_2003}. The discovery of such features has  been the goal of several statistical analysis procedures (e.g. Slow Feature Analysis \citep{wiskott_slow_2002}) as well as in  physics/chemistry literature (see a recent review in \citep{klus_data-driven_2018}). In this work we promote the discovery of such slow features by appropriate prior selection, and in particular by penalizing the jumps between two successive time-instants, i.e.:
\be
\begin{array}{ll}
 p( \bxx_{ 0: T \Delta t}^{(1:n)}) &  = \prod_{i=1}^n p_{c,0}(\bxx_{0}^{(i)}) \prod_{l=0}^{T-1} p(\bxx_{(l+1)~\Delta t}^{(i)} | \bxx_{l~\Delta t}^{(i)}, \sigma_X^2 \bs{I}) \\
 & = \prod_{i=1}^n p_{c,0}(\bxx_{0}^{(i)}) \prod_{l=0}^{T-1} \mathcal{N}( \bxx_{(l+1)~\Delta t}^{(i)} | \bxx_{l~\Delta t}^{(i)}, \sigma_X^2 \bs{I})\\
 & \propto \prod_{i=1}^n p_{c,0}(\bxx_{0}^{(i)}) \prod_{l=0}^{T-1} \frac{1}{\sigma_X^{d_c}} \exp \left\{ -\frac{1}{\sigma_X^2} \left| \bxx_{(l+1)~\Delta t}^{(i)} -\bxx_{l~\Delta t}^{(i)}\right|^2 \right\}
\end{array}
\label{eq:cgprior}
\ee
where $p_{c,0}$ is a prior density for the initial CG state. 
We observe that the strength of the penalty is inversely proportional to the hyperparameter $\sigma_X^2$ and in the limit $\sigma_X^2 \to 0$ it implies a constant time history of $\bxx_t$. As the appropriate value for $ \sigma_X^2$ depends on the problem, we include this in the parameter vector $\bt$ that is inferred/learned from the data. 

Given the intractability of the actual posterior, we advocate  in this work Variational Inference. This operates on a parameterized family of densities, say $q_{\bs{\phi}}( \bxx_{ 0: T \Delta t}^{(1:n)}, ~\bt)$ and  attempts to find the one (i.e. the value of $\bs{\phi}$) that most closely approximates the posterior by minimizing their Kullback-Leibler divergence. It can be readily shown \citep{bishop_pattern_2006}, that the optimal $q_{\bs{\phi}}$,   maximizes the Evidence Lower Bound (ELBO) $\mathcal{F}(q_{\bp}( \bxx_{ 0: T \Delta t}^{(1:n)}, ~\bt))$ below:
\be
\begin{array}{ll}
 \log p(\mathcal{D}) & =\log  \int p( \mathcal{D}, ~\bxx_{ 0: T \Delta t}^{(1:n)}, ~\bt ) ~d\bxx_{ 0: T \Delta t}^{(1:n)} ~d\bt \\
 & = \log  \int \cfrac{ p( \mathcal{D} | ~\bxx_{ 0: T \Delta t}^{(1:n)}, ~\bt ) p( \bxx_{ 0: T \Delta t}^{(1:n)}, ~\bt )}{ q_{\bp}( \bxx_{ 0: T \Delta t}^{(1:n)}, ~\bt)} q_{\bp}( \bxx_{ 0: T \Delta t}^{(1:n)}, ~\bt) ~d\bxx_{ 0: T \Delta t}^{(1:n)} ~d\bt \\
 & \ge \int \log \cfrac{p( \mathcal{D} | ~\bxx_{ 0: T \Delta t}^{(1:n)}, ~\bt ) p( \bxx_{ 0: T \Delta t}^{(1:n)}, ~\bt )}{ q_{\bp}( \bxx_{ 0: T \Delta t}^{(1:n)}, ~\bt)} q_{\bp}( \bxx_{ 0: T \Delta t}^{(1:n)}, ~\bt) ~d\bxx_{ 0: T \Delta t}^{(1:n)} ~d\bt \\
 & = \mathcal{F}(q_{\bp}( \bxx_{ 0: T \Delta t}^{(1:n)}, ~\bt))
\end{array}
\label{eq:elboqphi}
\ee

In the examples analyzed we decompose the approximate posterior as:
\be
\begin{array}{ll}
q_{\bp}( \bxx_{ 0: T \Delta t}^{(1:n)}, ~\bt) & = q_{\bp}( \bxx_{ 0: T \Delta t}^{(1:n)}) ~q_{\bp}(\bt)  \\
& = \left[ \prod_{i=0}^{n} q_{\bp}( \bxx_{ 0: T \Delta t}^{(i)}) \right]~~q_{\bp}(\bt)
\end{array}
\label{eq:qfactor}
\ee
where the first line is the so-called mean-field approximation and the second is a direct consequence of the (conditional) independence of the time sequences in the likelihood.
We note that evaluations of the ELBO $\mathcal{F}$ involve expectations with respect to $q_{\bp}$ i.e.:
\be
\begin{array}{ll}
\mathcal{F}\left( q_{\bp}( \bxx_{ 0: T \Delta t}^{(1:n)}, ~\bt) \right) & =\expe_{q_{\bp}} \left[ \log
p( \mathcal{D} | ~\bxx_{ 0: T \Delta t}^{(1:n)}, ~\bt )\right] +  \expe_{q_{\bp}} \left[ \log \cfrac{ p( \bxx_{ 0: T \Delta t}^{(1:n)}, ~\bt )}{  q_{\bp}( \bxx_{ 0: T \Delta t}^{(1:n)}, ~\bt ) }\right] \\ 
\end{array}
\label{eq:ELBOgeneral}
\ee
and in order to maximize it (with respect to $\bp$), gradients of those are needed. Given the intractability of these expectations and their derivatives, we make use of Monte Carlo estimates in combination with stochastic gradient ascent for the $\bp$-updates. In order to reduce the Monte Carlo error in these estimates, we make use of the reparametrization trick \citep{kingma_auto-encoding_2014}, for which the differentiability of the residuals/constraints is necessary. We specify the particulars of the  algorithm more precisely in the numerical illustration section (see e.g. Algorithm \ref{alg:particles} or \ref{alg:pend}).  

We note that maximum likelihood or maximum-a-posteriori (MAP) point estimates for any of the parameters involved can be obtained as a special case of the aforementioned scheme by employing a $q_{\bp}$ that is equal to a Dirac-delta function.
Furthermore, amortized versions of the approximate posterior $q_{\bp}$ i.e. forms that explicitly account on the dependence on the data values, can be employed in part or in whole. These have the capability of being able to transfer information across data points and are necessary in the realm of Big Data. We note though that we operate in the {\em Small Data} regime, i.e. the number of time sequences $n$ (and time-steps $T$) is not particularly large.  %
Hybrid versions between amortized and non-amortized posteriors could also be employed \citep{kim_semi-amortized_2018}.

We note finally that while the ELBO $\mathcal{F}$ is used  purely as the objective function for the determination of the approximate posterior, its role can be quite significant in model validation and refinement. In particular since $\mathcal{F}$ approximates the model evidence (denominator of \refeq{eq:genbayes}), once evaluated, it can be used to comparatively assess different models. These could have different CG states $\bxx$ (in type and/or number) or  different parametrizations $\bt$. In this regard, the ELBO $\mathcal{F}$ could serve as the primary driver for the adaptive refinement of the CG model \citep{grigo2019bayesian} in order to  better explain the observables and  lead to superior predictions.

\subsection{Prediction}
\label{sec:predictions}
An essential feature of the proposed modeling framework is the ability to produce {\em probabilistic } predictive estimates. These encompass the information-loss due to the coarse-graining process as well as the epistemic uncertainty arising from  finite (and small) datasets. 
We distinguish between two settings:
\bi
\item[a)] the {\em "interpolative"} i.e. predictions into the future of a  sequence $i$ observed up to time-step $T$ i.e. $\hat{\bx}^{(i)}_{0:T\Delta t}$ which was used in the training phase - see section \ref{sec:num}, or
\item[b)] the {\em "extrapolative"} i.e. predictions  for a completely new initial condition $\hat{\bx}_0$ - see section \ref{sec:num}.
\ei

We note that any predictions should account for the domain knowledge incorporated in the training through the residuals $\bs{R}_l$ or constraints $\bs{c}_l$. Formally that is, one should enlarge the posterior density defined in \refeq{eq:genbayes}, in order to account for the residuals and/or constraints at future time-steps. This would in turn imply, that future (FG or CG) states should be inferred from such an augmented  posterior i.e. prediction would imply an enlarged inference process. 
In the examples presented we adopt a simpler procedure that retains  the essential features (i.e. probabilistic nature) but is more computationally expedient. 
In particular, for case a) above and if $q_{\bp}(\bxx_{T\Delta t}^{(i)})$ is the (marginal) posterior of the last,  hidden CG state and $q(\bt)$ the posterior of the model parameters, then we (see also Agorithm \ref{alg:preda}):
\bi
\item sample from $q(\bxx_{T\Delta t}^{(i)}), q(\bt)$
\item for each sample, we propagate the CG dynamics dynamics of \refeq{eq:cg} (e.g. by solving the corresponding residual Equations (\ref{eq:cgdiscr})) in order to obtain $\bxx_{(T+1)\Delta t}^{(i)}, \bxx_{(T+2)\Delta t}^{(i)}, \ldots$, and,
\item we sample $\bx_{(T+1)\Delta t}^{(i)}$ from $p_{cf}(\bx_{(T+1)\Delta t}^{(i)} | \bxx_{(T+1)\Delta t}^{(i)}, \bt_{cf})$,  $\bx_{(T+2)\Delta t}^{(i)}$ from $\;\;\;$ $p_{cf}(\bx_{(T+2)\Delta t}^{(i)} | \bxx_{(T+2)\Delta t}^{(i)}, \bt_{cf})$ etc.
\ei
We note that this procedure does not necessarily ensure enforcement of the constraints by future CG states. Nevertheless it gives rise to samples of the full FG state evolution from which any observable of interest as well as the predictive uncertainty can be computed.

\begin{algorithm}[!ht]
\SetAlgoLined
\KwResult{Sample of $\bx_{ (T+P)\Delta t}^{(i)}$}
\KwData{$q_{\bp}(\bxx_{T\Delta t}), q_{\bp}(\bt)$}
 Sample from $q_{\bp}(\bxx_{T\Delta t}^{(i)})$ and $q_{\bp}(\bt)$\;
 \While{Time-step $(T+P)\Delta t$ of interest not reached}{
  Apply the CG evolution law as described in \refeq{eq:cgdiscr}\;
 }
 Sample from $p_{cf}(\bx_{(T+P)\Delta t}~|~ \bxx_{(T+P)\Delta t}, \bt)$
 \caption{ Prediction - Algorithm for {\em interoplative} setting  }
 \label{alg:preda}
\end{algorithm}

For the {\em extrapolative}  setting above, i.e. for a new FG initial condition $\hat{\bx}_0$, the evolution equations of the CG states as well as the emission density $p_{cf}$ can be employed as long as the initial state $\bs{X}_0$ is specified or better yet inferred. For that purpose,  the posterior $p(\bs{X}_0 |\hat{\bx}_0)$  of $\bs{X}_0$ needs to be determined which according to Bayes rule will be proportional to:
\be
p(\bs{X}_0~ |~\hat{\bx}_0) \propto p_{cf}( \hat{\bx}_0 ~|~ \bs{X}_0, \bt_{cf}) ~p_{c,0}(\bs{X}_0)
\label{eq:bayinf}
\ee
where $p_{c,0}(\bs{X}_0)$ is the initial state's prior (see also \refeq{eq:cgprior}). For each sample   of $\bt_{cf}$ from the (approximate) posterior $q_{\bp}(\bt_{cf})$, samples of $\bxx_0$ must be drawn from $p(\bs{X}_0 |\hat{\bx}_0)$ and subsequently propagated as in the 3 steps above in order to obtain predictive samples of the full FG state vector (see Algorithm \ref{alg:predb}). 

\begin{algorithm}[!t]
\SetAlgoLined
\KwResult{Sample of $\bx_{P\Delta t}$ }
\KwData{$p_{\bp}(\hat{\bx}_{0}), q_{\bp}(\bt)$}
 Apply Bayesian Inference as described in \refeq{eq:bayinf} to infer $p(\bs{X}_0 |\hat{\bx}_0)$\;
 Sample from $p(\bs{X}_0 |\hat{\bx}_0)$ and $q(\bt)$\;
 \While{Time-step $P\Delta t$ of interest not reached}{
  Apply the CG evolution law as described in \refeq{eq:cgdiscr}\;
 }
 Sample from $p_{cf}(\bx_{P\Delta t}| \bxx_{P\Delta t}, \bt)$
 \caption{ Prediction - Algorithm for {\em extrapolative} setting }
 \label{alg:predb}
\end{algorithm}
%
%
%

\review{
\subsection{Computational considerations}
}
\review{
We note that in multiscale dynamical systems  of physical interest, the computational cost stems primarily from the simulation of the FG system due to its generally very high-dimensional state-vector $\bx$ and very small time-step $\delta t$. Hence, one of the main objectives of this work is to enable the learning of the CG dynamics  with the fewest possible and shortest possible FG time-sequences.}

\review{We note that once such FG simulation (or experimental) data have been obtained, neither the training phase of the CG model (section \ref{sec:inference}) nor the prediction phase (section \ref{sec:predictions}) require any additional FG simulations. The cost of training depends on the dimension of the CG states $\bxx$ as well as the number of parameters $\bt_c$ (for the CG dynamics), $\bt_{cf}$ (for the coarse-to-fine map) and $\bp$ (for the approximate posterior).}

\review{
We emphasize that this is a one-time, offline cost i.e. once the CG model has been trained, it can be used to produce probabilistic predictive estimates of the whole FG state-vector into the future without any further recourse to the FG model. One needs only to simulate in such case the CG dynamics which due to the lower-dimensional state-vector  $\bxx$ and the much larger CG time-step $\Delta t$ are much less cumbersome than the FG system.
}

\review{Finally, if more FG data (e.g. longer or new sequences) become available at a later stage, the SVI algorithm can be re-initialized from the previous values and incorporate the new likelihood terms. If a modest amount of data is introduced, one would expect small (or even no changes for faraway states) changes and therefore rapid convergence.
Naturally the introduction of observables at new time instants would introduce additional latent variables for the corresponding CG states.
}


%% file: Figures/drawing.pdf_tex
\begingroup%
  \makeatletter%
  \providecommand\color[2][]{%
    \errmessage{(Inkscape) Color is used for the text in Inkscape, but the package 'color.sty' is not loaded}%
    \renewcommand\color[2][]{}%
  }%
  \providecommand\transparent[1]{%
    \errmessage{(Inkscape) Transparency is used (non-zero) for the text in Inkscape, but the package 'transparent.sty' is not loaded}%
    \renewcommand\transparent[1]{}%
  }%
  \providecommand\rotatebox[2]{#2}%
  \newcommand*\fsize{\dimexpr\f@size pt\relax}%
  \newcommand*\lineheight[1]{\fontsize{\fsize}{#1\fsize}\selectfont}%
  \ifx\svgwidth\undefined%
    \setlength{\unitlength}{595.27559055bp}%
    \ifx\svgscale\undefined%
      \relax%
    \else%
      \setlength{\unitlength}{\unitlength * \real{\svgscale}}%
    \fi%
  \else%
    \setlength{\unitlength}{\svgwidth}%
  \fi%
  \global\let\svgwidth\undefined%
  \global\let\svgscale\undefined%
  \makeatother%
  \begin{picture}(1,1.41428571)%
    \lineheight{1}%
    \setlength\tabcolsep{0pt}%
    \put(0,0){\includegraphics[width=\unitlength,page=1]{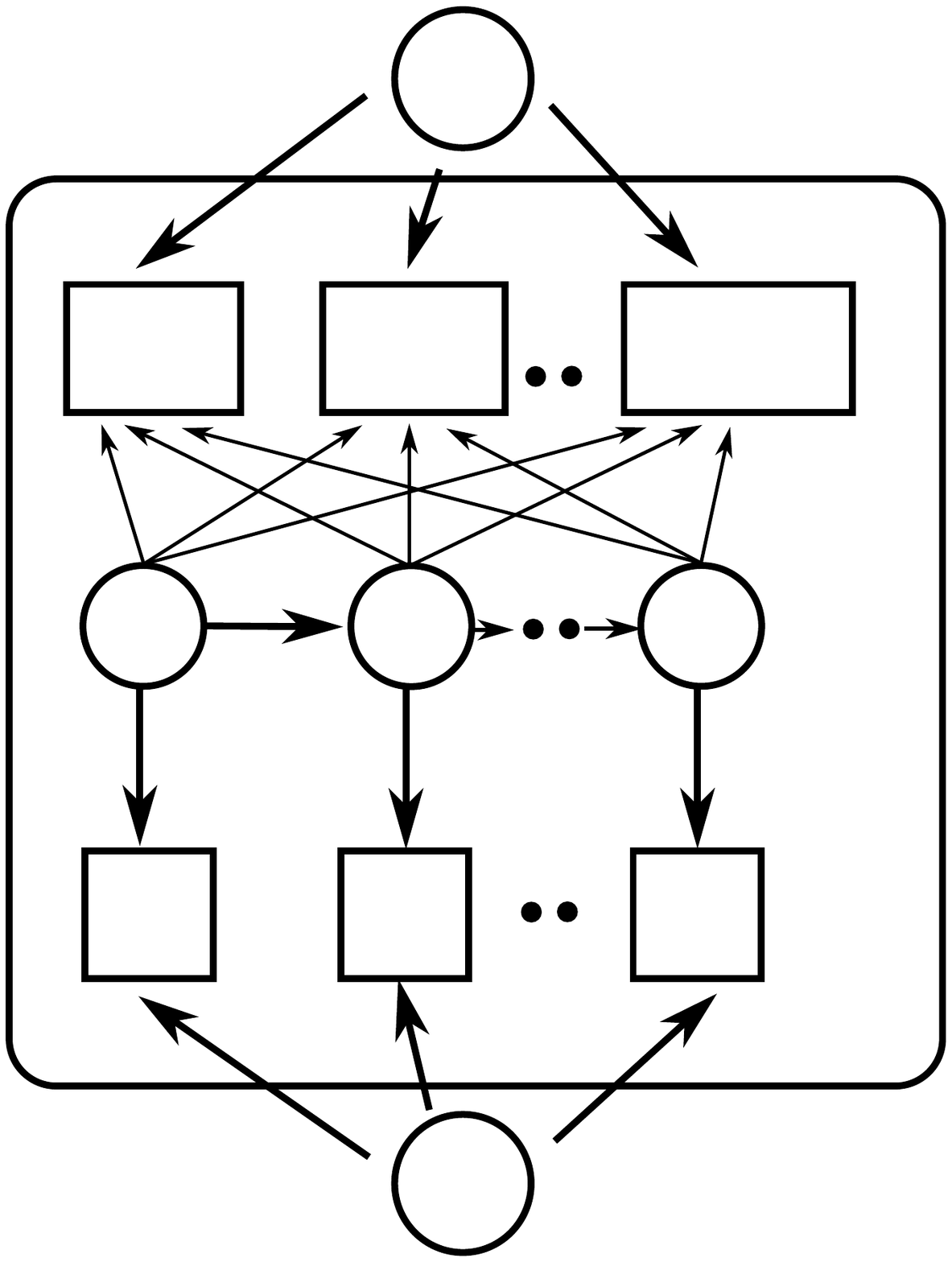}}%
    \put(0.39660512,1.22798849){\color[rgb]{0,0,0}\makebox(0,0)[lt]{\begin{minipage}{0.08280451\unitlength}\raggedright \end{minipage}}}%
    \put(0.6408217,0.84284347){\color[rgb]{0,0,0}\makebox(0,0)[lt]{\begin{minipage}{0.05647923\unitlength}\raggedright \end{minipage}}}%
    \put(1.46681664,-0.68245225){\color[rgb]{0,0,0}\makebox(0,0)[lt]{\begin{minipage}{0.35995498\unitlength}\raggedright \end{minipage}}}%
    \put(0.47742949,0.68612061){\color[rgb]{0,0,0}\makebox(0,0)[lt]{\lineheight{1.25}\smash{\begin{tabular}[t]{l}$\bx_{T\Delta t}$\end{tabular}}}}%
    \put(0.28361617,0.68786301){\color[rgb]{0,0,0}\makebox(0,0)[lt]{\lineheight{1.25}\smash{\begin{tabular}[t]{l}$\bx_{\Delta t}$\end{tabular}}}}%
    \put(0.10615503,0.68531639){\color[rgb]{0,0,0}\makebox(0,0)[lt]{\lineheight{1.25}\smash{\begin{tabular}[t]{l}$\bx_{0}$\end{tabular}}}}%
    \put(0.47769754,0.88462514){\color[rgb]{0,0,0}\makebox(0,0)[lt]{\lineheight{1.25}\smash{\begin{tabular}[t]{l}$\bxx_{T\Delta t}$\end{tabular}}}}%
    \put(0.28817333,0.88368686){\color[rgb]{0,0,0}\makebox(0,0)[lt]{\lineheight{1.25}\smash{\begin{tabular}[t]{l}$\bxx_{\Delta t}$\end{tabular}}}}%
    \put(0.10548486,0.88475916){\color[rgb]{0,0,0}\makebox(0,0)[lt]{\lineheight{1.25}\smash{\begin{tabular}[t]{l}$\bxx_{0}$\end{tabular}}}}%
    \put(0.32985796,0.50034935){\color[rgb]{0,0,0}\makebox(0,0)[lt]{\lineheight{1.25}\smash{\begin{tabular}[t]{l}$\bt_{cf}$\end{tabular}}}}%
    \put(0.33468319,1.26179711){\color[rgb]{0,0,0}\makebox(0,0)[lt]{\lineheight{1.25}\smash{\begin{tabular}[t]{l}$\bt_c$\end{tabular}}}}%
    \put(0.59336904,0.59779213){\color[rgb]{0,0,0}\makebox(0,0)[lt]{\lineheight{1.25}\smash{\begin{tabular}[t]{l}$1:n$\end{tabular}}}}%
    \put(0.09208144,1.07294304){\color[rgb]{0,0,0}\makebox(0,0)[lt]{\lineheight{1.25}\smash{\begin{tabular}[t]{l}$\hat{\bs{R}}_{0},\hat{\bs{c}}_{0}$\end{tabular}}}}%
    \put(0.25922199,1.07388128){\color[rgb]{0,0,0}\makebox(0,0)[lt]{\lineheight{1.25}\smash{\begin{tabular}[t]{l}$\hat{\bs{R}}_{\Delta t},\hat{\bs{c}}_{\Delta t}$\end{tabular}}}}%
    \put(0.46804715,1.0754897){\color[rgb]{0,0,0}\makebox(0,0)[lt]{\lineheight{1.25}\smash{\begin{tabular}[t]{l}$\hat{\bs{R}}_{T \Delta t},\hat{\bs{c}}_{T \Delta t}$\end{tabular}}}}%
  \end{picture}%
\endgroup%

%% file: numerical-illustrations.tex
We demonstrate the capabilities of the proposed framework in  discovering predictive,  coarse-grained evolution laws as well as effective coarse-grained descriptions, on three examples. Two of those involve very high-dimensional systems of stochastically interacting particles (section \ref{sec:particles}, \citep{felsberger_physics-constrained_2019})  and the third,  a nonlinear pendulum, the dynamics of which we attempt to identify simply from sequences of images (section \ref{sec:pendulum}, \citep{champion2019data}). 
In the sequel, we specify the elements of the proposed model that were presented generically in the previous sections and concretize parametrizations and their meaning. 
The goals of the numerical illustrations are:
\bi 
\item to assess the predictive performance of the model  under ``interpolative'' and ``extrapolative" conditions (see section \ref{sec:predictions}). By ``interpolative" we mean the ability to predict the evolution of an FG states-sequence  when  data from this sequence has been used for training. By ``extrapolative", we mean the ability to predict the full FG state evolution from new initial conditions  that were {\em not} used in training.
\item to examine the effect of the number  $n$ and length $T$ of the   data sequences and assess the model's ability to learn the correct structure with small $n, T$ and partial observations.
\item to examine the enforcement of the residuals/constraints (e.g. conservation of mass) in the inferred and  predicted states. 
\item to examine the ability of the model to identify {\em sparse}, interpretable solutions for the CG dynamics.
\item to assess the magnitude and time evolution of the predictive uncertainty estimates.
\item to assess the ability of the model to learn effective CG state variables and accurate coarse-to-fine maps.

\ei

Some of the simulation results  as well as the corresponding  code will be made available at the following github repository\footnote{\url{https://github.com/SebastianKaltenbach/PhysicalConstraints_ProbabilisticCG.git}} upon publication.

\subsection{Particle systems}
\label{sec:particles}
\input{particles.tex}

\subsubsection{Advection-Diffusion system}
\label{sec:ad}
\input{ad.tex}

\subsubsection{Burgers' system}
\label{sec:burgers}
\input{burgers.tex}

\newpage
\subsection{Nonlinear Pendulum}
\label{sec:pendulum}
\input{pendulum.tex}

%% file: particles.tex


\subsubsection{ FG model}
The FG model consists of  $d_f$ identical particles which can move in the bounded one-dimensional  domain $[-1,~1]$ (under periodic boundary conditions). The FG variables $\bx_t$  consist  therefore of the coordinates of the particles at each time instant $t$ and  the dimension of the system $d_f$ is equal to the number of particles. We consider two types of stochastic dynamics that correspond to an advection-diffusion-type (section \ref{sec:ad}) and an inviscid-Burgers-type behavior (section \ref{sec:burgers}). The particulars of the microscopic dynamics are described in the corresponding sections. In the following, we discuss common aspects of both problems that pertain to the CG description, the CG evolution law and the inference procedures.  \\

\subsubsection{CG variables and  coarse-to-fine mapping}
\label{eq:parcg}

For the  CG representation, we employ the normalized particle density $\rho(s,t), ~s\in [-1,~1]$ \citep{li_deciding_2007}  which we discretize in $d_c$ bins. The state vector  $\bs{X}_t=\{ X_{t,j} \}_{j=1}^{d_c}$ contains the  particle density values in each of the bins $j$, i.e. $\sum_{j=1}^{d_c} X_{t,j}=1$  and $X_{t,j} \ge 0 ~\forall t,j$. \review{We emphasize that CG and FG variables are of a different nature (i.e. proportion of particles in each bin vs. coordinates of particles) and, more importantly for practical purposes,  of very different dimension.}

The nature of the CG variables $\bxx_t$ suggests   a  {\em multinomial} for the coarse-to-fine density $p_{cf}$ (section \ref{sec:emission}) i.e.:\\
\begin{equation}
p_{cf}(\bs{x}_t | \bs{X}_t)=  \frac{ d_f !}{m_1(\bx_t)!~m_2(\bx_t)! \ldots m_{d_c}(\bx_t)!} \prod_{j=1}^{d_c}X_{t,j}^{m_j(\bx_t)} 
 , \quad \textrm{} 
 \label{eq:Coarse_to_fine_particle}
\end{equation}
where $m_j(\bx_t)$ is the number of particles in bin $j$. The underlying assumption is that, given the CG state $\bxx_t$, the coordinates of the particles $\bx_t$ are {\em conditionally} independent. This does {\em not} imply that they move independently nor that they cannot exhibit coherent behavior \citep{felsberger_physics-constrained_2019}.
The practical consequence of \refeq{eq:Coarse_to_fine_particle} is that no parameters need to be learned for $p_{cf}$  (in contrast to section  \ref{sec:pendulum}).

\subsubsection{The CG evolution law and the virtual observables}
With regards to the evolution law of the CG states (\refeq{eq:cg}), we postulate a right-hand side $\bs{F}(\bxx_t; \bt_c)=\left\{ F_j(\bxx_t; \bt_c)\right\}_{j=1}^{d_c}$ of the form:

\begin{equation}
 \begin{array}{ll}
  F_j(\mathbf{X}_t, \bt_c) 
  & =  \sum_{m=1}^M \theta_{c,m} ~\psi^{(j)}_m (\mathbf{X}_t) \\ 
 & = \underbrace{ \sum_{h=-H}^H \theta_{c,h}^{(1)} X_{t,j+h} }_{1^{st} order}+
  \underbrace{ \sum_{h_1=-H}^H \sum_{h_2\ge h_1}^H \theta_{c,~(h_1,h_2)}^{(2)} X_{t,j+h_1} X_{t,j+h_2} }_{2^{nd} order}
 \end{array}
\label{eq:cgparticles}
\end{equation}
which consists of first- and second-order interactions over a window of size $H$ with $\bt_c^{(1)}$ and $\bt_c^{(2)}$ denoting the vectors of the corresponding unknown coefficients.  In this case, the total number of unknown coefficients $\bt_c$, is $M=dim(\bt_c)=(2H+1)+(H+1)(2H+1)$ and grows quadratically with the neighborhood-size $H$. \review{Since each of the CG variables $X_{t,j}$ refers to the particle density at bin $j$ (and at time $t$), the neighborhood size $H$ corresponds to the number of bins to the left or to the right of bin $j$ that affect its evolution in time
The feature functions that we generically denote with  $\psi^{(j)}_m$ in \refeq{eq:cgparticles} can also involve
 higher-order interactions or  be  of non-polynomial type.}
  Non-Markovian models could be accommodated as well by accounting for memory terms.  It is obviously impossible to know a priori which feature functions  are relevant in the evolution of the CG states or what types of
interactions are essential (e.g. first, second-order etc). At the same time, and especially in the Small Data regime considered, employing a large vocabulary of feature functions can lead to {\em overfitting}, lack of interpretability and poor predictions, particularly under ``extrapolative" conditions. This highly-important {\em model selection} issue  has been of concern in several coarse-graining studies \citep{noid_perspective:_2013}. We propose of automatically addressing this within the Bayesian framework advocated by employing appropriate sparsity-inducing priors for $\bt_c$ \citep{felsberger_physics-constrained_2019}. In particular, we make use of the Automatic Relevance Determination (ARD, \citep{mackay_probable_1995}) model according to which
\begin{equation}
    p(\theta_{c,m}\mid \tau_m) = \mathcal{N}(\theta_{c,m}~\mid~ 0, \tau_m^{-1}), \qquad m=1,2,\ldots ,M =dim(\bt_c).
\end{equation}
The following hyperprior for the  precision hyperparameters $\bs{\tau}=\{\tau_m\}_{m=1}^M$ was used:
\begin{equation}
    p(\tau_k\mid \gamma_0, \delta_0) = Gamma(\tau_k \mid \gamma_0 , \delta_0 )
\end{equation}
The hyperparameters $\gamma_0$ and $\delta_0$ are set to very small values $10^{-9}$ in all ensuing studies \citep{bishop2000variational}. As we demonstrate in the sequel, the hypeprior proposed can give rise to parsimonious solutions for the CG dynamics even in the Small Data setting considered.\\

A discretized version of the CG evolution law   (\refeq{eq:cg} and  \refeq{eq:cgparticles}) with time step $\Delta t$  is considered by employing a forward Euler scheme\footnote{This corresponds to a multistep method in \refeq{eq:cgdiscr} with $K=1$, $a_0=1,a_1=-1,\beta_0=0$ and $\beta_1=-1$.}  which implies the following  residual vector $\bs{R}_l$  at each time-step $l$ (\refeq{eq:cgdiscr}):
 \be
\bs{R}_{l}(\bxx)=  \bxx_{(l+1)\Delta t,j} - \bxx_{l\Delta t,j} - \Delta t ~\bs{F}(\bxx_{l \Delta t,j},\bt_c), \quad \forall ~l
\label{eq:cgdiscr_particle}
\ee
 and the corresponding virtual observables $\hat{\bs{R}}_l$ (\refeq{eq:voresl}).


More importantly, the nature of the CG variables suggests a {\em  conservation of mass} constraint that  has to be fulfilled at each time step $l$. In view of  the discussion of section \ref{sec:transition}, this suggests the scalar constraint   function as in \refeq{eq:constraintl}:
\be
 {c}_l(\bxx_{l\Delta t}) = \sum_{j=1}^{d_c}  X_{l \Delta t,j}~ -1  = 0, \quad \forall ~l
 \label{eq:constraintl_particle}
 \ee
and the corresponding  virtual observables  $\hat{{c}_l}$ (\refeq{eq:vocon}).

\subsubsection{Inference and Learning}
\label{eq:inferenceparticles}
Given the multinomial $p_{cf}$ in \refeq{eq:Coarse_to_fine_particle},  we employed the following procedure for generating training data which consists of  $n$  numerical experiments in which the FG model is randomly initialized and propagated for one coarse time-step $\Delta t$ i.e. for $T=\frac{\Delta t}{\delta t}$ microscopic time-steps. In particular:
\bi
\item For $i=1,\dots,n$, we:
\bi
\item sample CG initial state $\hat{\bxx}_0^{(i)}$ from a density $p_{c,0}(\hat{\bxx}_0^{(i)})$.
\item sample FG initial state $\hat{\bx}_0^{(i)}$ from $p_{cf}(\hat{\bx}_0^{(i)} | \bxx_0^{(i)})$.
\item solve the (discretized) FG model for  $\frac{\Delta t}{\delta t}$ microscopic time-steps and record final state $\hat{\bx}_{\Delta t}^{(i)}$
\ei
\ei
The generated FG data 
 $\{ \hat{\bx}_{\Delta t}^{(i)} \}_{i=1}^n$ over a {\em single} CG time-step are  used subsequently to draw inferences on the CG model states and parameters (section \ref{sec:inference}).
We note that longer time sequences could readily be generated (albeit at an increased cost). The number of samples $n$ is also something that can be selected adaptively since  inferences and predictions can be updated as soon as more data become available.   The density $p_{c,0}(\bxx_0^{(i)})$ from which initial CG states are drawn, can be selected  quite flexibly and some indicative samples are shown in Figure \ref{fig:ad_initial} for the advection-diffusion case, and in Figure \ref{fig:burger_initial} for the inviscid-Burgers' case. 
In summary, the data $\mathcal{D}$ employed, apart from $\{ \hat{\bx}_{\Delta t}^{(i)} \}_{i=1}^n$ above consists of the virtual observables $\{ \hat{\bs{R}}_{0}^{(1:n)}, \hat{\bs{c}}_{1}^{(1:n)} \}$.


As a result of the data employed and the parametrization adopted, we have $\bxx_{\Delta t}^{(1:n)}$ as the sole  latent vector and $\bt_c, \bs{\tau}$ as the unknown (hyper)parameters. Since only a single CG time-step was considered, we omitted the slowness prior (see \refeq{eq:cgprior}). Hence we sought an approximate posterior $q_{\bp}(\bxx_{\Delta t}, \bt_c, \bs{\tau})$ (\refeq{eq:elboqphi}) which we factorized as in \refeq{eq:qfactor} as  follows:
\begin{equation}
    q_{\bp}(\bxx_{\Delta t}^{(1:n)}, \bt_c, \bs{\tau})=\left[ \prod_{i=1}^n q_{\bp} (\bxx^{(i)}_{\Delta t}) \right]
    q(\bt_c  ) q(\bs{\tau})
    \label{eq:particlefactor}
\end{equation}

Upon substitution in  \refeq{eq:ELBOgeneral}, this yields the following ELBO:
\be
\begin{array}{ll}
\mathcal{F}(q_{\bp}( \bxx_{ \Delta t}^{(1:n)}, ~\bt_c, \bs{\tau})) &=\expe_{q_{\bp}} \left[ \log
p( \mathcal{D} | ~\bxx_{ \Delta t}^{(1:n)}, ~\bt_c )\right] +  \expe_{q_{\bp}} \left[ \log p( ~\bt_c \mid \bs{\tau} )\right] \\ &+  \expe_{q_{\bp}} \left[ \log p( \bs{\tau} )\right]- \expe_{q_{\bp}}\left[ \log q_{\bp} \right]
\end{array}
\label{eq:ELBOparticle}
\ee
where:
\be
p(\mathcal{D} | \bxx_{ \Delta t}^{(1:n)}, \bt_c) = p(\hat{\bx}_{\Delta t}^{(1:n)} |  \bxx_{ \Delta t}^{(1:n)}) ~p( \hat{\bs{R}}_{0}^{(1:n)}  |  \bxx_{ \Delta t}^{(1:n)}, \bt_c)~ p(\hat{\bs{c}}_{1}^{(1:n)} |  \bxx_{ \Delta t}^{(1:n)})
\label{eq:likedecomp_par}
\ee

Based on \refeq{eq:ELBOparticle} the optimal variational posterior densities can be obtained as:
\begin{equation}
   \log q^{opt}(\bt_c)= \expe_{q_{\bp}(\bxx_{\Delta t}^{(1:n)})}\left[\log~ p( \hat{\bs{R}}_{0}^{(1:n)}  |  \bxx_{ 0: 1 \Delta t}^{(1:n)}, \bt_c)  \right]+ \expe_{q(\bs{\tau})}\left[\log ~p(\bt_c \mid \bs{\tau}) \right]
\end{equation}
\begin{equation}
   \log q^{opt}(\bs{\tau})= \expe_{q_{\bp}(\bt_c)}\left[\log~p(\bt_c \mid \bs{\tau})\right] +\log~p(\bs{\tau})
\end{equation}
\begin{equation}
\begin{array}{ll}
   \log q_{\bp}^{opt}(X_{\Delta t}^{(i)}) & =\log~p_{cf}(\bx^i_{\Delta t}\mid \bxx^i_{\Delta t})+\expe_{q_{\bp}(\bt_c)}\left[\log ~p( \hat{\bs{R}}_{0}^{(i)}  |  \bxx_{ 0: 1 \Delta t}^{(i)}, \bt_c) \right] \\
   & + \log~p(\hat{\bs{c}}_{1}^{(i)} |  \bxx_{ \Delta t}^{(i)})
   \end{array}
\end{equation}

The equations above are coupled and a closed-form solution can be obtained only for the first two.
In particular, the optimal posterior approximation for $\bt_c$ is a multivariate normal with mean $\mu_{\bt_c}$ and covariance $\bs{S}_{\bt_c}$.\\
\begin{equation}
\mathbf{S}^{-1}_{\bt_c}= \sigma_{R}^{-2} \sum_{i=1}^n \sum_{j=1}^{d_c} \expe_{q_{\bp}(\bxx^{(i)}_{\Delta t}) } \left[  \bs{\psi}^{(j)} (\bxx^{(i)}_{\Delta t}) \left( \bs{\psi}^{(j)} (\bxx^{(i)}_{\Delta t}) \right)^T \right] + \expe_{q_{\bp}(\bs{\tau})}[diag(\bs{\tau})]
\label{eq:theta1}
\end{equation}
\begin{equation}
\mathbf{S}^{-1}_{\bt_c} \bs{\mu}_{\bt_c} = \sigma_{R}^{-2} \sum_{i=1}^n \sum_{j=1}^{d_c} \expe_{q_{\bp}(\bxx^{(i)}_{\Delta t}) } \left[ \bs{\psi}^{(j)} (\bxx^{(i)}_{\Delta t})\right] 
\label{eq:theta2}
\end{equation}
\review{where the vector $\bs{\psi}^{(j)}$ consists of the $M$  feature functions $ \psi_m^{(j)}$ in \refeq{eq:cgparticles}.}
The optimal posterior approximation for the vector $\bs{\tau}$ of the  hyperparameters $\{ \tau_m\}_{m=1}^M$ reduces to a product of independent Gamma-densities \citep{bishop2000variational} with  parameters $\gamma_m$ and $\delta_m$ which are given by:
\begin{equation}
    \gamma_m=\gamma_0+0.5, \qquad \delta_m=\delta_0+\frac{1}{2}~\left( {\mu}_{\bt_c, m}+ {S}_{\bt_c, (m,m)} \right), \quad m=0,1,\ldots,M=dim(\bt_c)
    \label{eq:tau1}
\end{equation}

Finally and since closed-form updates for the optimal posterior $q_{\bp}^{opt}(X_{\Delta t}^{(i)})$ are impossible, we employed Stochastic Variational Inference (SVI) as detailed in section \ref{sec:inference} by assuming a multivariate lognormal (in order to ensure positivity of $X_{\Delta t,j}$) with parameters $\bp=\{ \bs{\mu}_i, \bs{S}_i\}_{i=1}^n$\footnote{Diagonal covariances $\bs{S}_i$ were employed.}. Noisy gradients with respect to the parameters $\bp$ were estimated with Monte Carlo and  the reparametrization trick \citep{kingma_auto-encoding_2014} and $\bp$ were updated using stochastic gradient ascent (the ADAM algorithm of \citep{kingma2014adam} in particular).
The inference steps are  summarized in  Algorithm \ref{alg:particles}.



\begin{algorithm}[!t]
\SetAlgoLined
\KwResult{$\{q_{\bp}(\bxx^{(i)}_{\Delta t})\}_{i=1}^n$, ~$q(\bt_c)$,~ $q(\bs{\tau})$}
\KwData{$\{ \bxx^{(i)}_{0}, \hat{\bx}^{(i)}_{\Delta t}\}_{i=1}^n$}
 Initialize the parameters for the variational distributions\;
 Set iteration counter $w$ to zero\;
 Set convergence limit $\epsilon$\;
 \While{$||parameters_{w}-parameters_{w-1}||^2 > \epsilon$}{
  \For{$i\gets1$ \KwTo $n$}{
    Update  $q_{\bp}(\bxx^{(i)}_{\Delta t})$ by maximizing the ELBO (see \refeq{eq:ELBOparticle}) }
  update  $q(\bt_c)$ according to \refeq{eq:theta1} and \refeq{eq:theta2} \;
  update  $q(\bs{\tau})$ according to \refeq{eq:tau1}  \;
  update the iteration counter by one \;
 }
 \caption{ Inference algorithm for particle systems}
 \label{alg:particles}
\end{algorithm}

\review{
\begin{table}[!ht]
    \centering
    \begin{tabular}{l|c|c|c|c|}
      & $d_f=dim(\bx)$  & $d_c=dim(\bxx)$ & FG time-step $\delta t$ & CG time-step $\Delta t$  \\
       \hline \hline
     Advection-Diffusion & $250\times 10^3$  & $\le 64$ & $2.5 \times 10^{-3}$ & $2$ \\
     \hline
     inviscid Burgers & $250\times 10^3$ & $\le 128$&$2.5 \times 10^{-3}$ & $4$ \\
  \hline
    \end{tabular}
    \caption{FG/CG state-space dimensions and FG/CG time-steps for particle systems investigated.}
    \label{tab:particles}
\end{table}
}

%% file: ad.tex
For the simulations presented in this section $d_f=250 \times 10^3$ particles were used, which, at each microscopic time step $\delta t=2.5 \times 10^{-3}$  performed random, non-interacting,  jumps of size $\delta s=\frac{1}{640}$, either  to the left with probability $p_{left}=0.1875$ or to the right with probability $p_{right}=0.2125$. The positions were restricted in $[-1,1]$ with periodic boundary conditions.  It is well-known \citep{cottet_vortex_2000} that in the limit (i.e.  $d_f \to \infty$) 
 the particle density $\rho(s,t)$ can be modeled with an advection-diffusion PDE with diffusion constant $D=(p_{left}+p_{right})\frac{\delta s^2} {2\delta t}$ and velocity $v=(p_{right}-p_{left})\frac{\delta s}{\delta t}$:
  \be
  \cfrac{\pa \rho }{\pa t} +v \cfrac{\pa \rho}{\pa s}=D \frac{\pa^2 \rho}{\pa s^2}, \qquad s \in (-1,1)..
  \label{eq:addensity}
  \ee

 
For the CG description, $64$ bins were employed i.e. $d_c=64$ and a time step  $\Delta t=2$ (see Table \ref{tab:particles}). 
Furthermore we employed first- and second-order feature function as in \refeq{eq:cgparticles} with a neighborhood size $H=5$ which implies a total of $M=77$ unknown parameters $\bt_c$.
We incorporate virtual observables pertaining to the residuals $\hat{\bs{R}}_0$ with $\sigma_R^2=10^{-6}$  (\refeq{eq:vlikeres})\footnote{A very interesting possibility which is not explored here would be to learn $\sigma_R^2$ i.e. the strength of the enforcement of the CG evolution law from the data. This would increase the flexibility of the model in cases where the vocabulary of the  feature functions selected in the right-hand side of the CG dynamics is not rich enough.} and the virtual observables $\hat{\bs{c}}_1$ pertaining to the  conservation-of-mass constraint with $\sigma_c^2=10^{-10}$ (\refeq{eq:vlikecon}).

\begin{figure}[h]
 \includegraphics[width=0.89\textwidth]{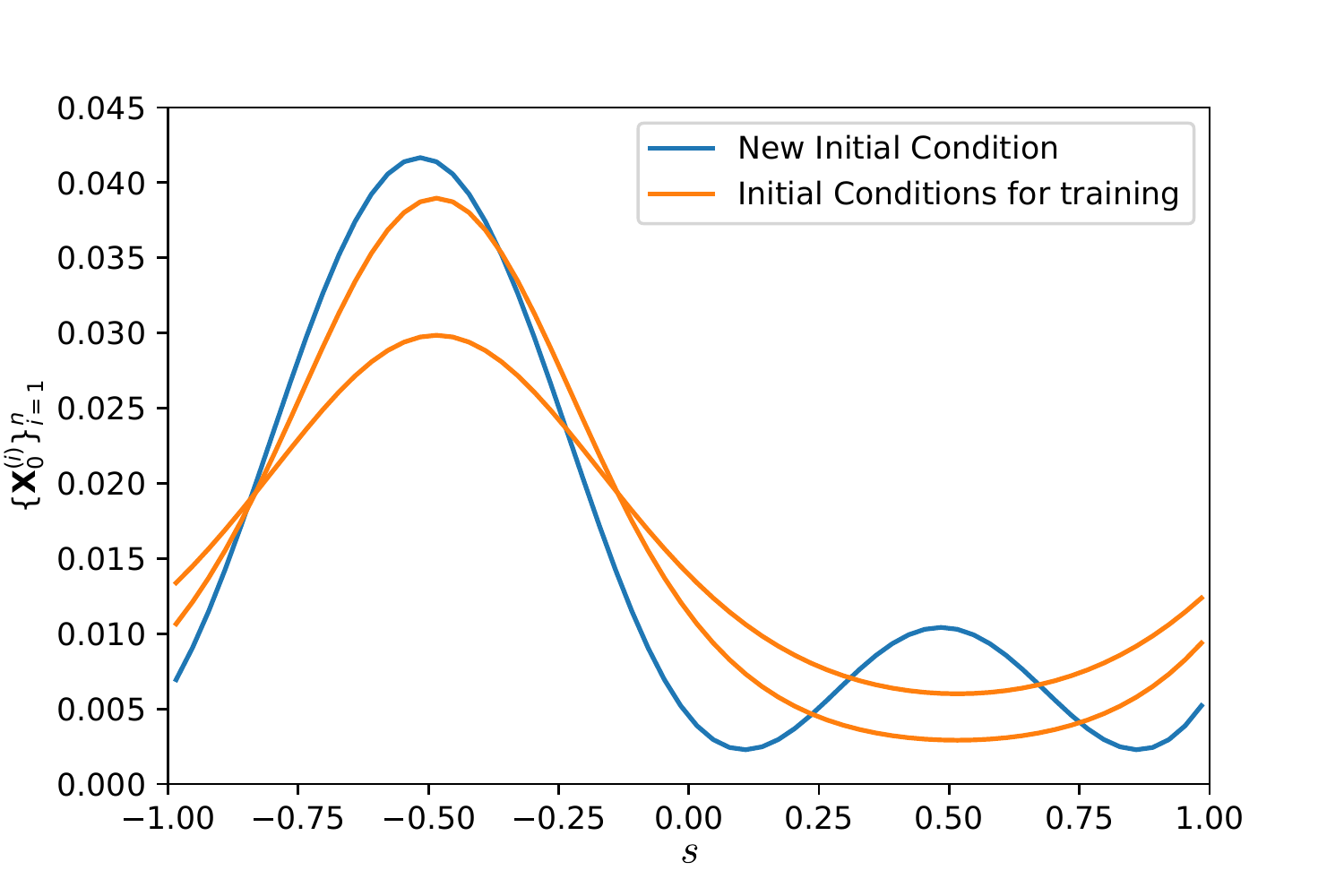}
 \caption{Sample initial conditions $\left\{\bxx_0^{(i)} \right\}_{i=1}^n$ for the Advection-Diffusion problem (orange) and an initial condition (blue) used for``extrapolative" predictions.}
 \label{fig:ad_initial}
\end{figure}

We employed $n=32$ and $n=64$ time sequences for training that were generated as detailed in section \ref{eq:inferenceparticles}  with initial conditions $\{\bxx_0^{(i)} \}_{i=1}^n$ such as the ones seen in  Figure \ref{fig:ad_initial}. The initial conditions were generated by sampling the amplitude of a $sine$ function, which was shifted up to ensure all values are positive and then normalized. 

\begin{figure}
 \includegraphics[width=0.89\textwidth]{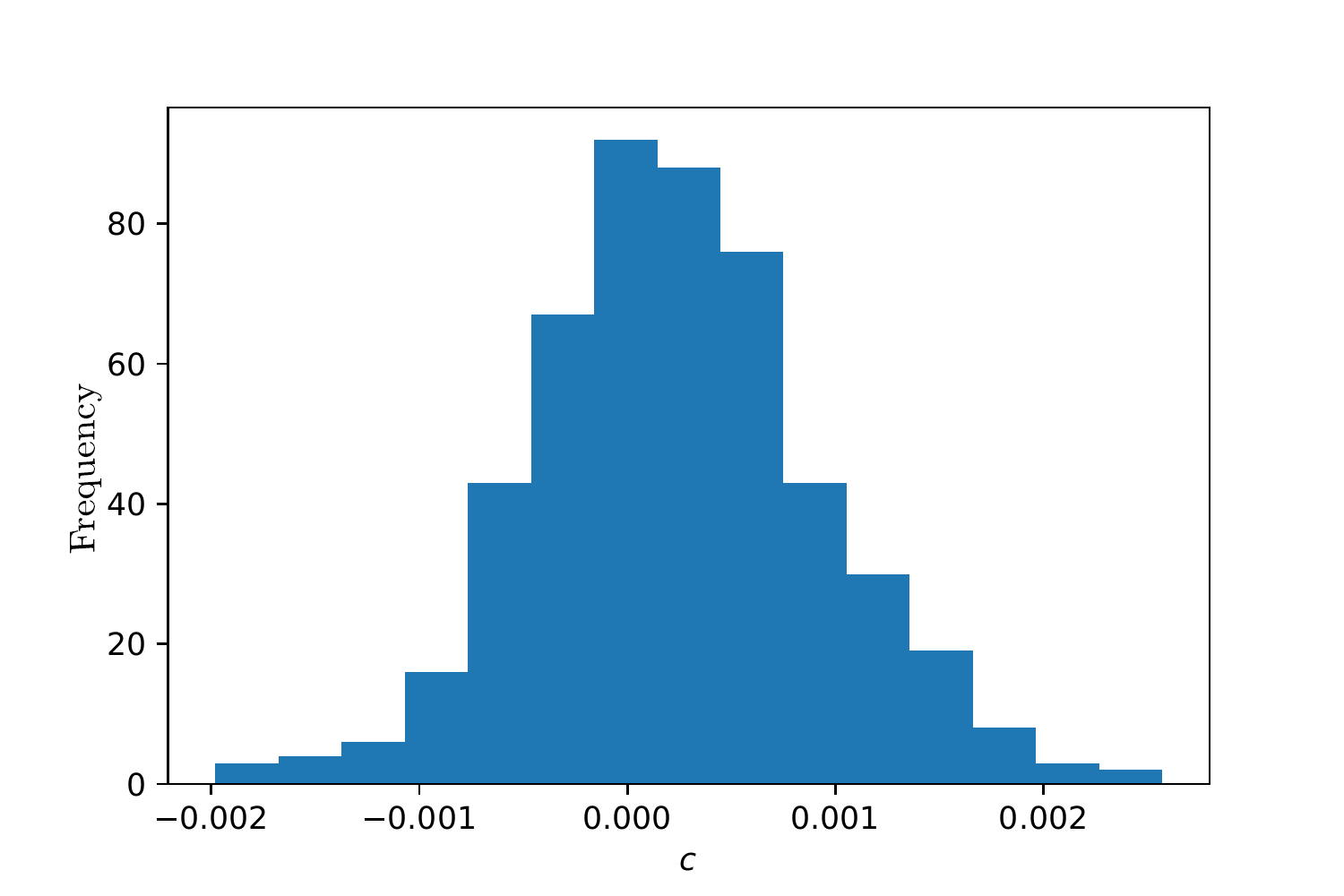}
 \caption{Histogram of the mass constraint $c_{1}$}
 \label{fig:ad_constraint}
\end{figure}

\begin{figure}
 \includegraphics[width=0.89\textwidth]{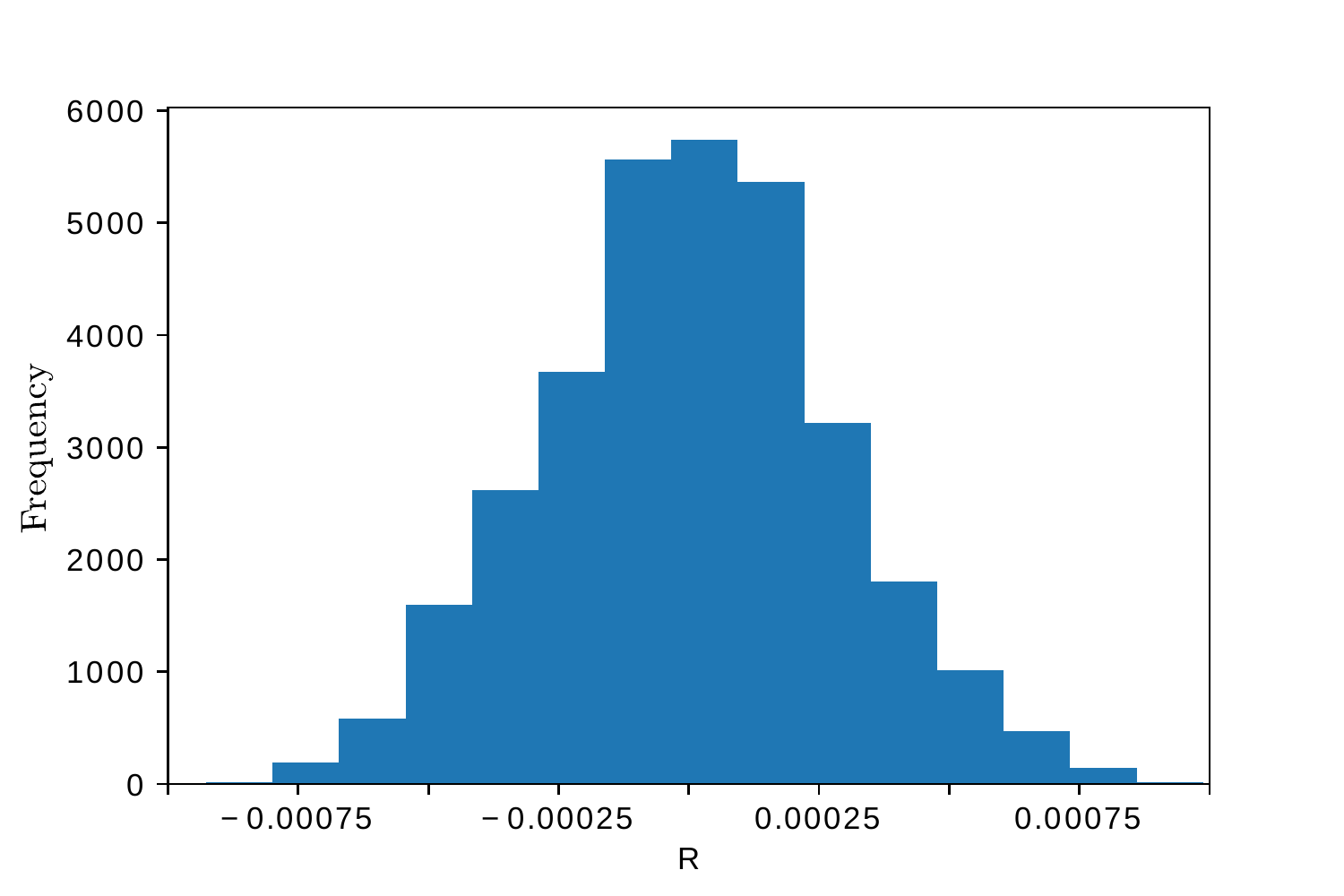}
 \caption{Histogram of the norm of the  residual  $\bs{R}_0$}
 \label{fig:ad_residual}
\end{figure}

\begin{figure}
 \includegraphics[width=0.89\textwidth]{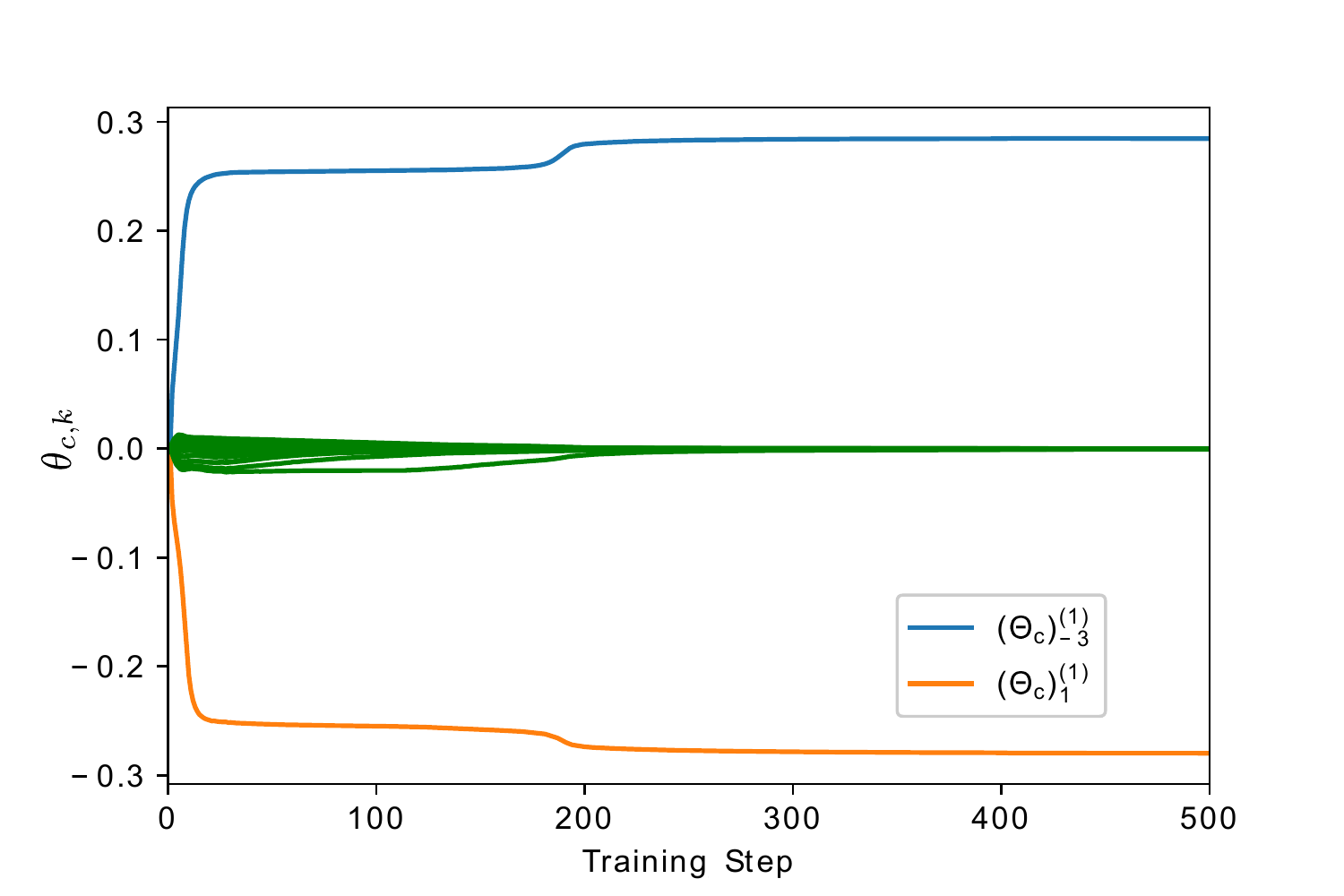}
 \caption{Evolution of a  subset of $\bt_c$ parameters with respect to the iterations of the SVI for $n=64$.}
 \label{fig:ad_coeff}
\end{figure}

\begin{figure}
 \includegraphics[width=0.49\textwidth]{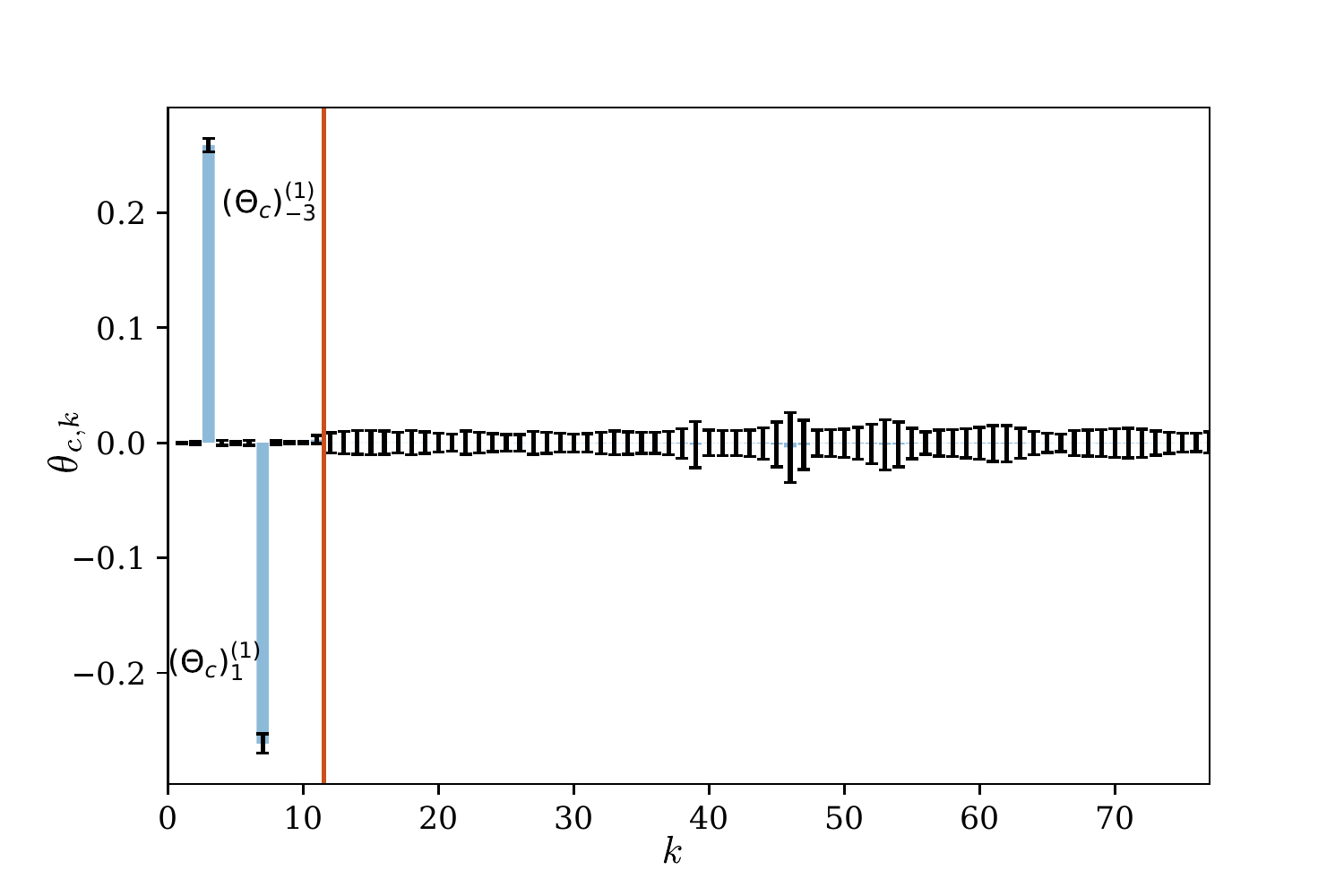} \includegraphics[width=0.49\textwidth]{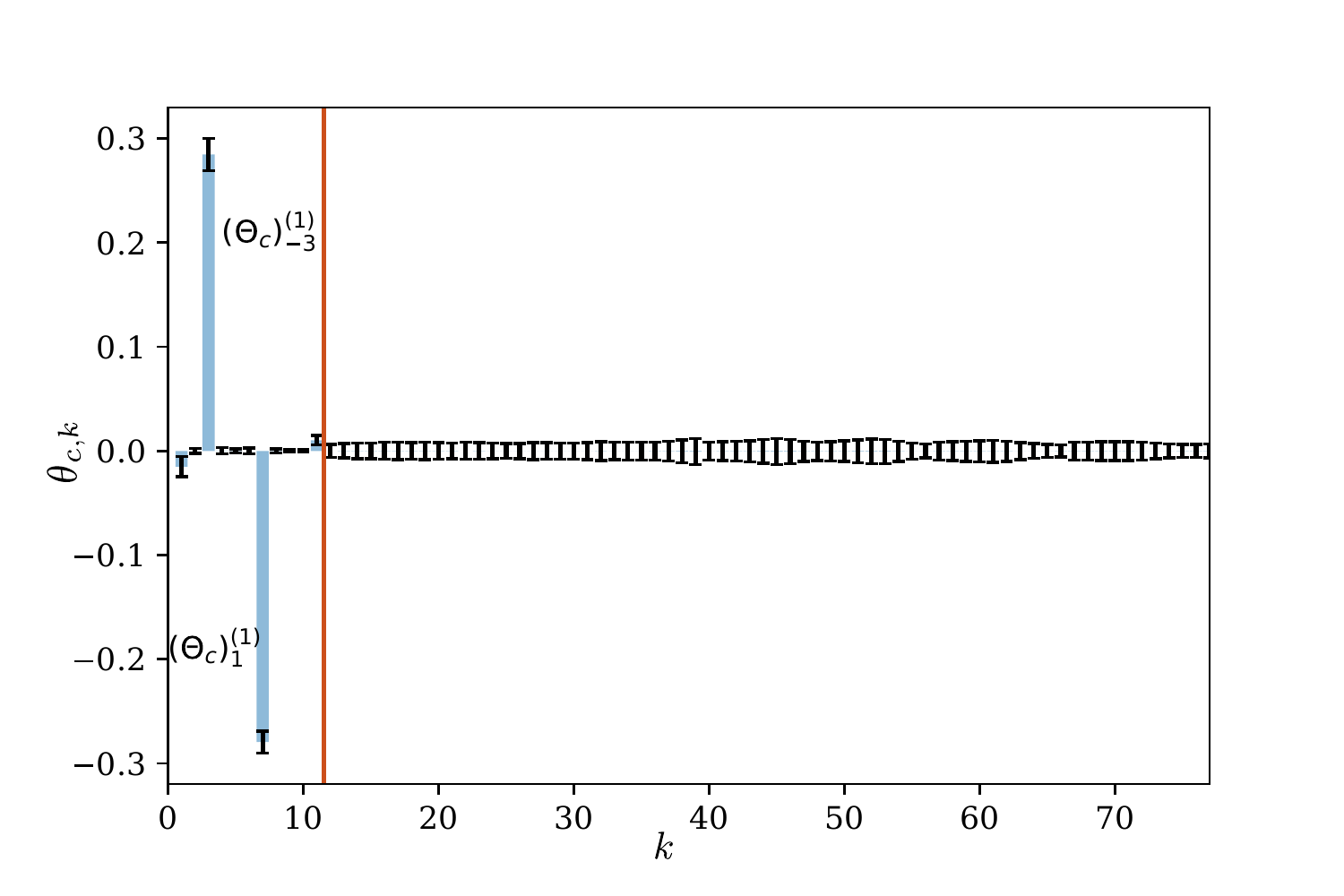}
 \caption{Comparison of the inferred parameters $\bt_c$ for $n=32$ (left) and $n=64$ (right) training data sequences. The black bars indicate +/- 1  standard deviation. The red vertical line separates first- from second-order coefficients.}
 \label{fig:ad_theta}
\end{figure}

Figure \ref{fig:ad_constraint}  provides a histogram of the function values of the conservation-of-mass constraint $\left\{c_1(\bxx_{\Delta t}^{(i)}) \right\}_{i=1}^n$ upon convergence. The small values suggest that this has been softly incorporated in the CG states. A similar histogram for the norm of the residuals $\left\{\bs{R}_{0}(\bxx^{(i)} )\right\}_{i=1}^n$ is depicted in Figure \ref{fig:ad_residual} which also suggests  enforcement of the CG evolution with the parameters $\bt_c$ learned from the data. The evolution of the posterior mean $\bs{\mu}_{\bt_c}$ (\refeq{eq:theta2}) of (a subset of) these parameters over the iterations of the SVI is depicted in  Figure \ref{fig:ad_coeff}. Therein, and more clearly in Figure \ref{fig:ad_theta}, one can observe the ability of the ARD prior  to deactivate the vast majority of the right-hand-side feature functions and reveal a small subset of non-zero, salient terms. 
Both with $n=32$ and $n=64$ training data sequences, only parameters $\bt_c$ associated with first-order-interactions (\refeq{eq:cgparticles}) are activated. In particular, these are  $\theta_{c,-3}^{(1)}$ and $\theta_{c,1}^{(1)}$ which are associated with the feature functions $X_{t,j-3}$ and $X_{t,j+1}$ respectively in \refeq{eq:cgparticles}. This shares similarities with a finite-difference discretization scheme for the advection-diffusion and could be considered as an upwind scheme. The two identified coefficients do not form a centered difference operator but the center of the operator is shifted to the left and therefore takes into account the direction of the particle movement. As the value of the coefficients is not exactly the same the diffusive part is also captured.\\


\begin{figure}
 \includegraphics[width=0.89\textwidth]{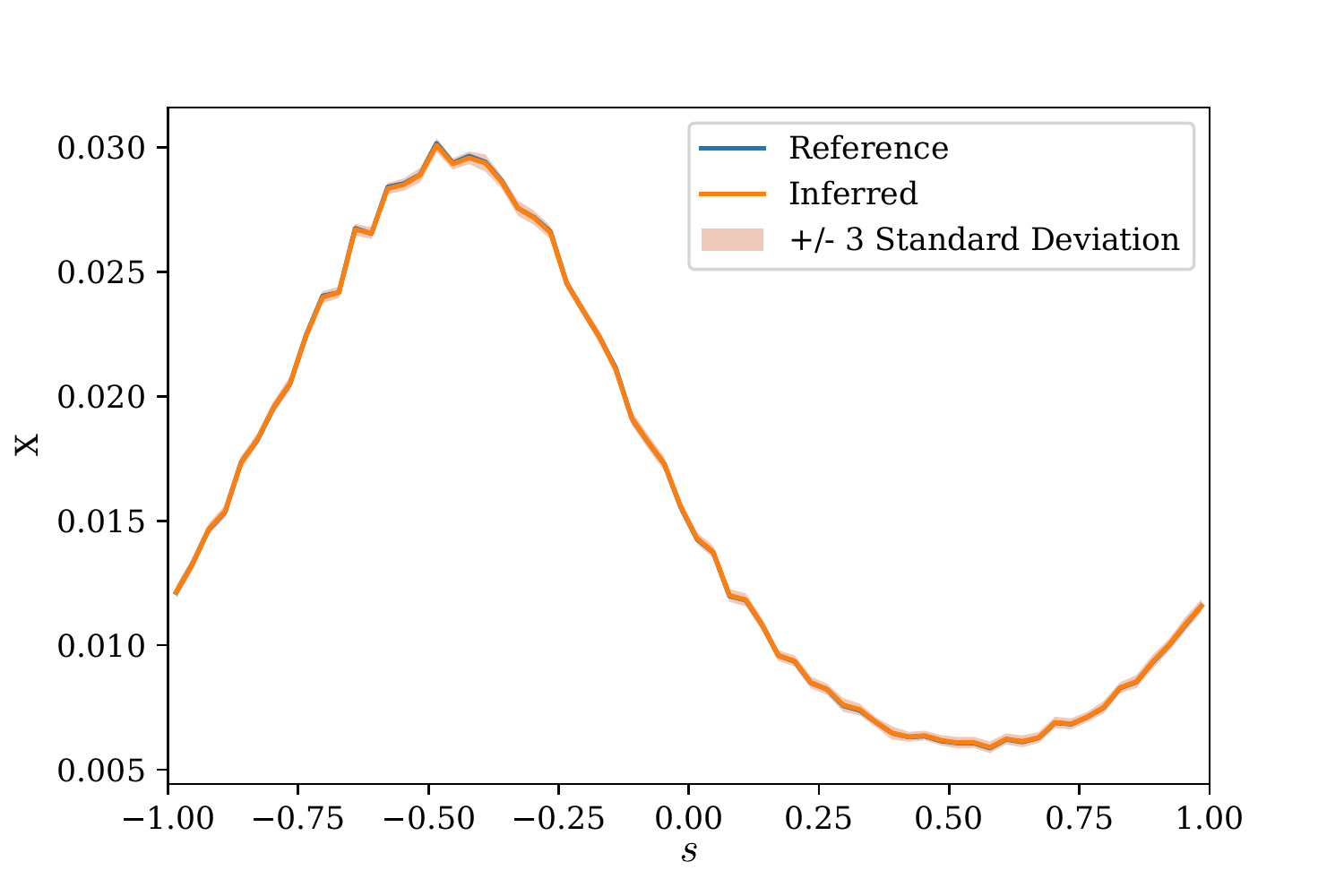}
 \caption{Inferred CG state $\bxx_{\Delta t}^{(i)}$ for a data sequence $i$. Reference is obtained by sorting the particles into bins according to their position.}
 \label{fig:ad_inf}
\end{figure}

\begin{figure}[h]
 \includegraphics[width=.99\textwidth]{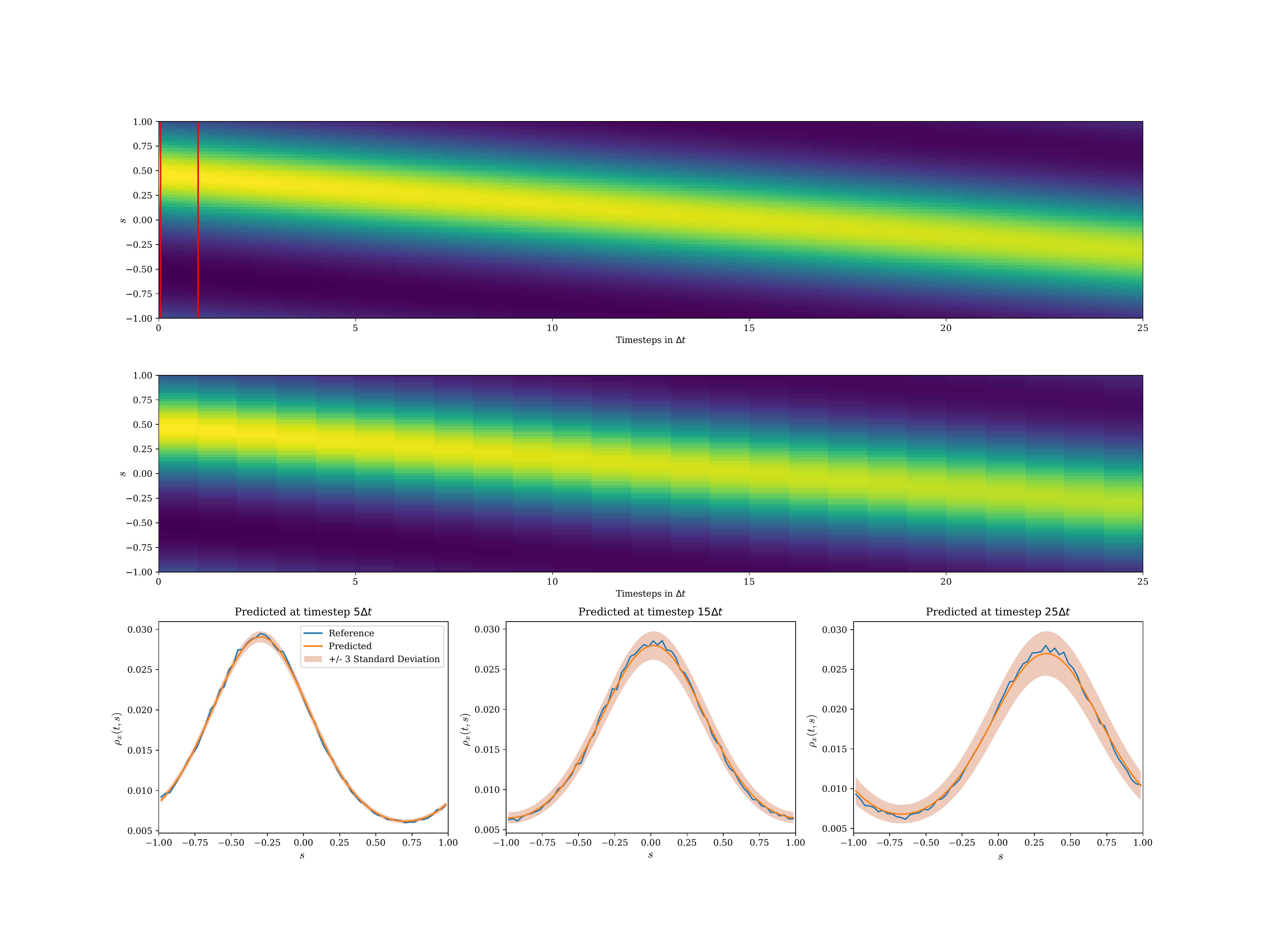}
 \caption{Prediction based on an initial condition contained in the training data.  Top: Reference data (the vertical lines indicate the time instances with given data), Middle: Predictive posterior mean, Bottom: snapshots at three different time instances.}
 \label{fig:ad_train}
\end{figure}

\begin{figure}
 \includegraphics[width=0.49\textwidth]{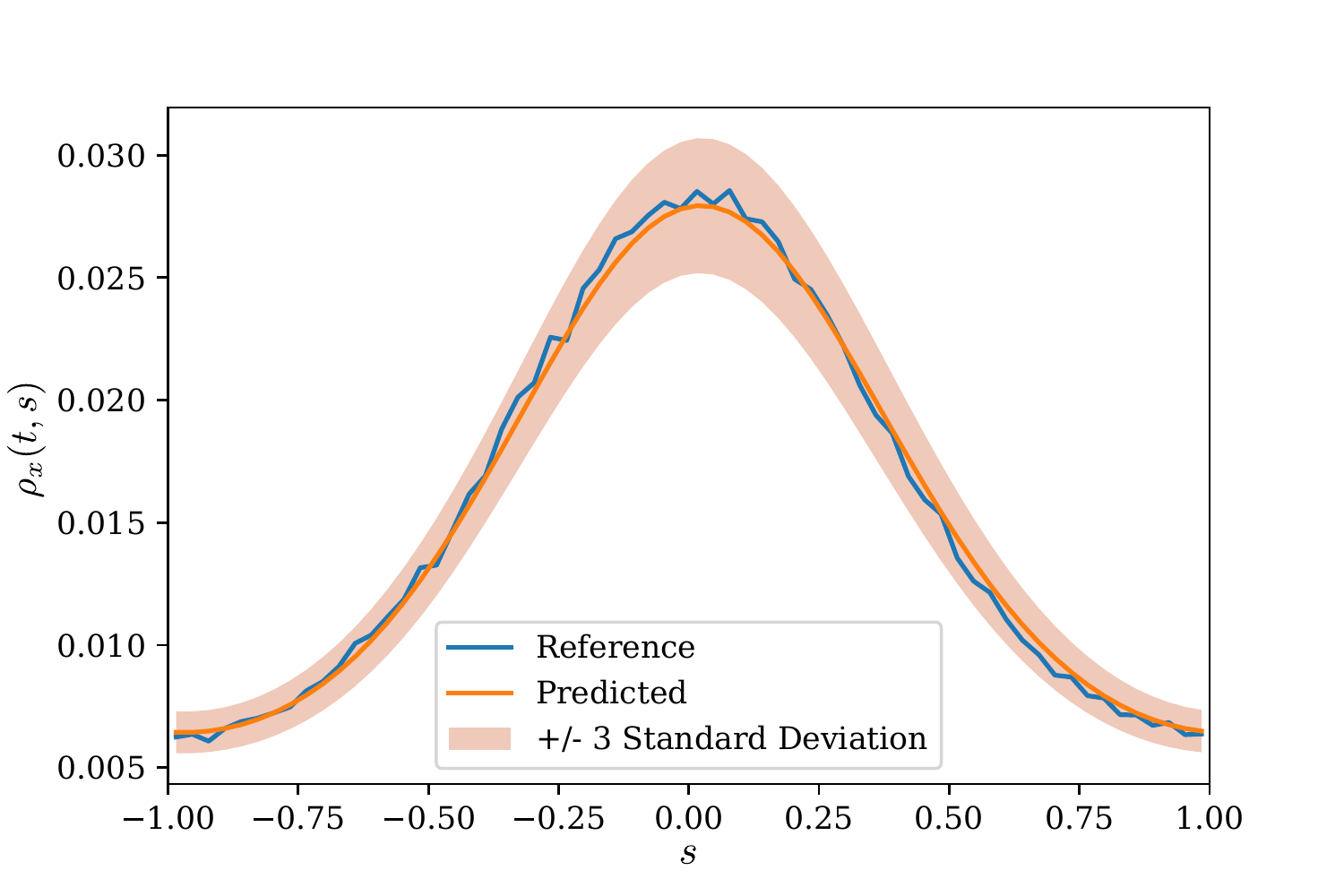} \includegraphics[width=0.49\textwidth]{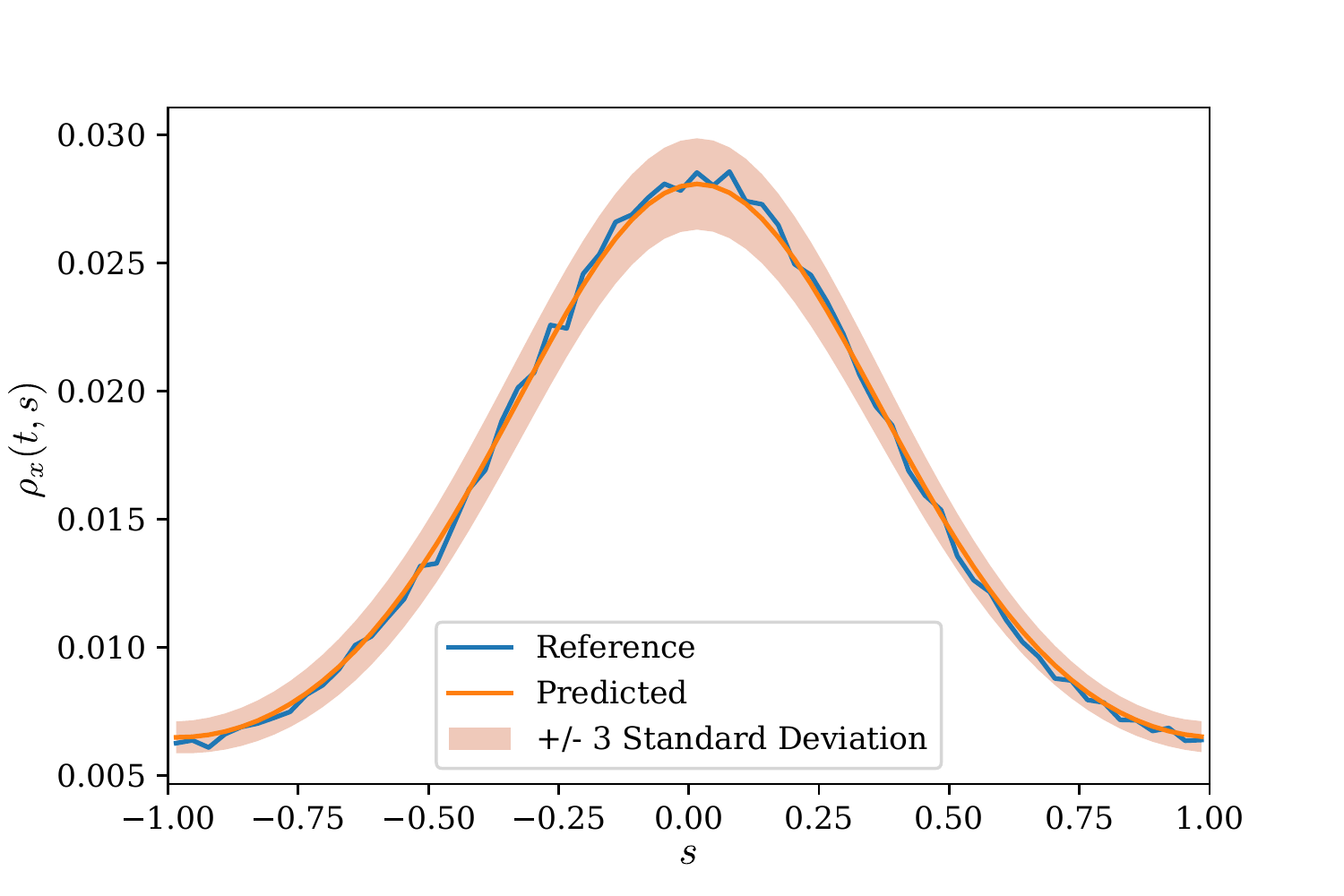}\\
 \includegraphics[width=0.49\textwidth]{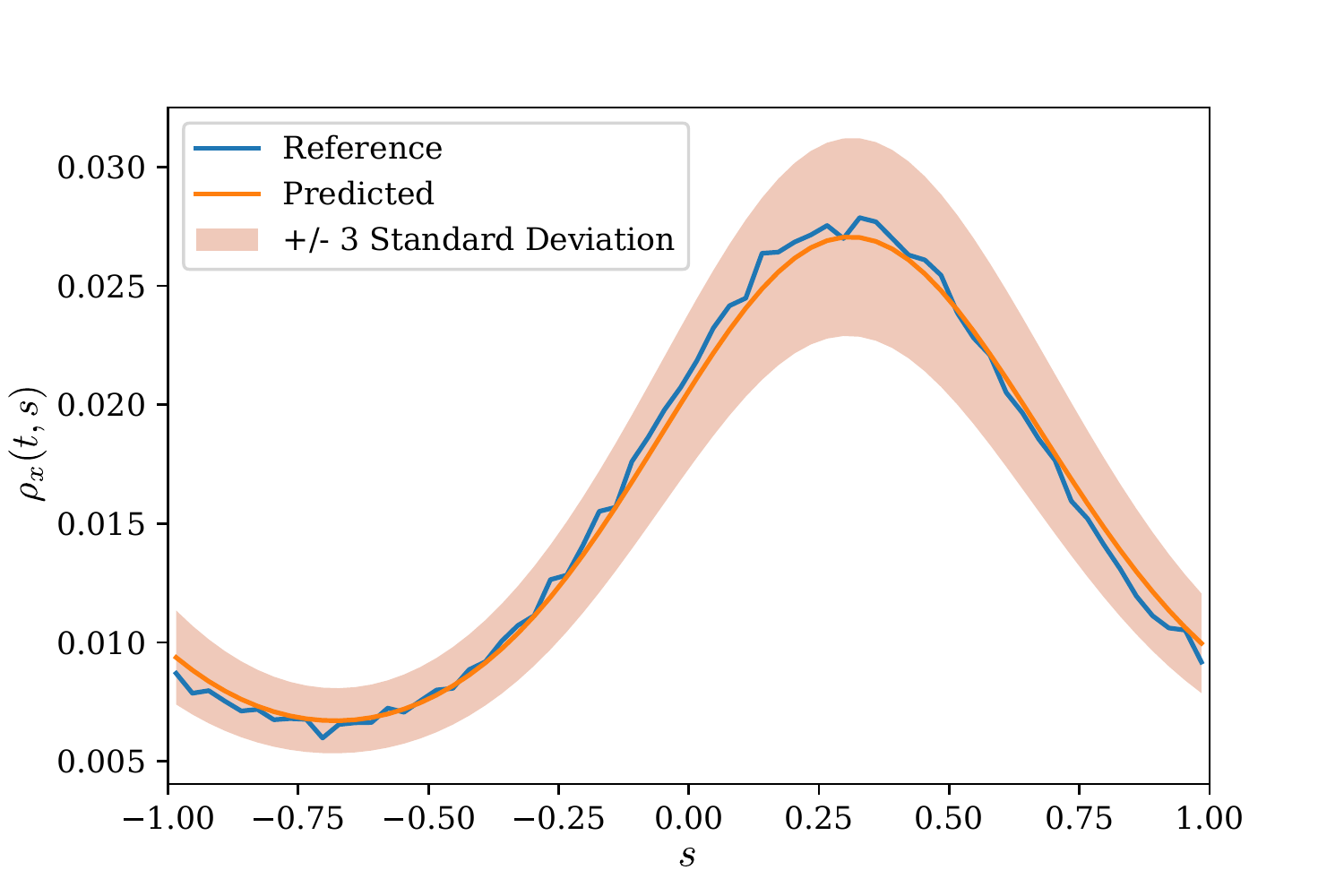} \includegraphics[width=0.49\textwidth]{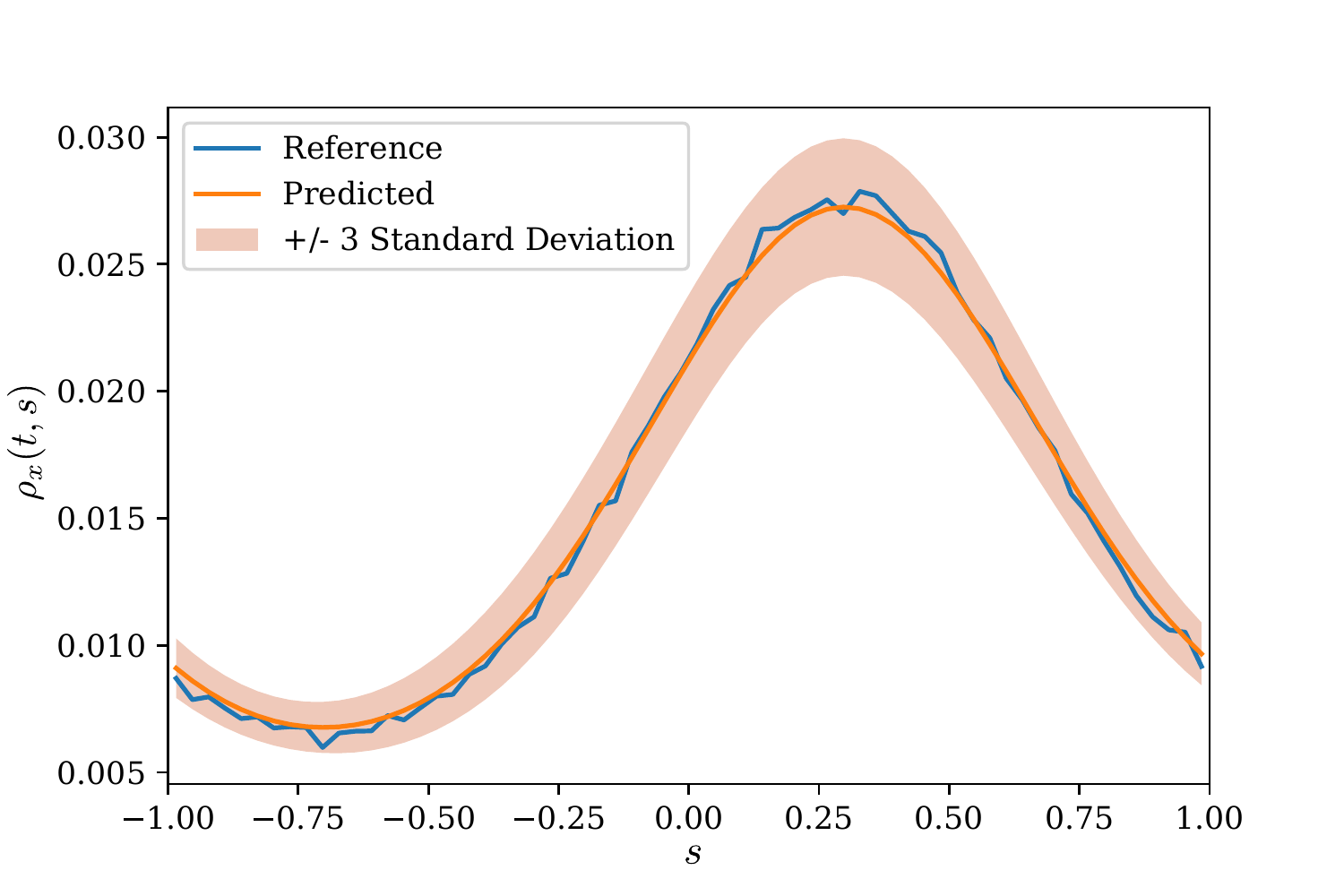}\\
 \caption{Comparison of the predictions for $n=32$ (left) and $n=64$ (right) at $15\Delta t$ (top) and $25 \Delta t$ (bottom). }
 \label{fig:ad_data}
\end{figure}

Figure \ref{fig:ad_inf} depicts one of the inferred CG states $\bxx_{\Delta t}^{(i)}$ as well as the associated posterior uncertainty. Once the CG evolution law is learned, this state can be propagated into the future as detailed in section \ref{sec:predictions} in order to generate predictions. 
%
%
%
%
%
Indicative predictions (under ``interpolative" conditions) can be seen in Figure \ref{fig:ad_train} where the particle density $\rho_x(t,s)$ up to $25 \Delta t$ into the future is drawn. The latter as well as the associated uncertainty bounds are  estimated directly from  the reconstructed FG states. As one would expect, the predictive uncertainty grows, the further into the future one tries to predict. 
Figure  \ref{fig:ad_data} compares the predictive performance as a function of the training data used i.e. $n=32$ or $n=64$. In both cases, the ground truth is envelopped and as one would expect, more training data lead to smaller uncertainty bounds.\\

\begin{figure}[h]
 \includegraphics[width=.95\textwidth]{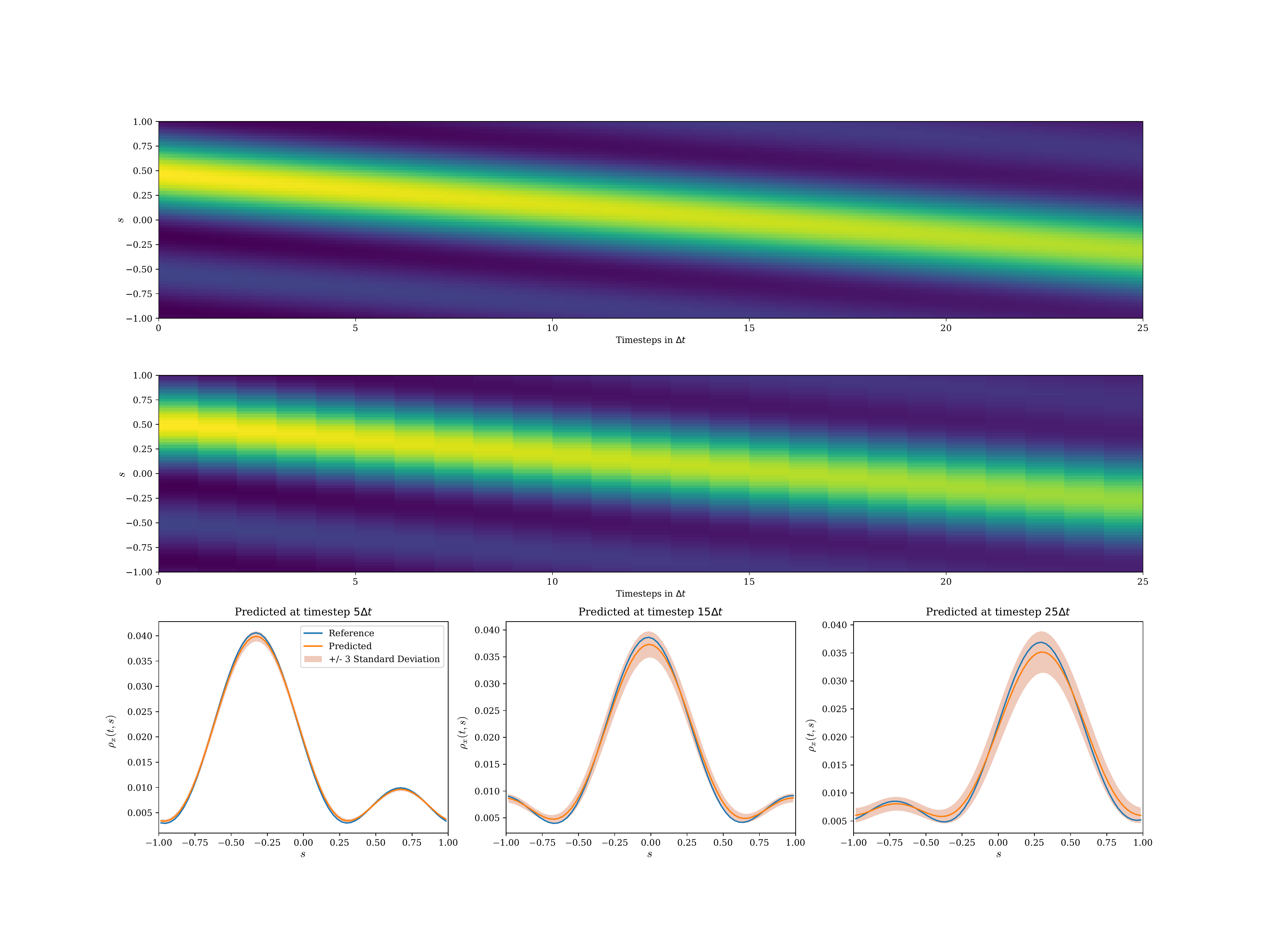}
 \caption{Prediction based on an initial condition NOT contained in the training data.  Top: Reference data, Middle: Predictive posterior mean, Bottom: snapshots at three different time instances}
 \label{fig:ad_prednew}
\end{figure}

We also tested the trained model (on $n=64$) under ``extrapolative" conditions i.e. for a different initial condition than the ones  included in the training data (Figure \ref{fig:ad_initial}). The predictive estimates in Figure \ref{fig:ad_prednew}  show very good agreement with the reference solution. It is important  to point out that the model can correctly advect and diffuse the particle-bump initially introduced around $s=0.5$ which suggests that the CG dynamics learned reflect the most important features of the problem. 

\begin{figure}
\includegraphics[width=0.95\textwidth]{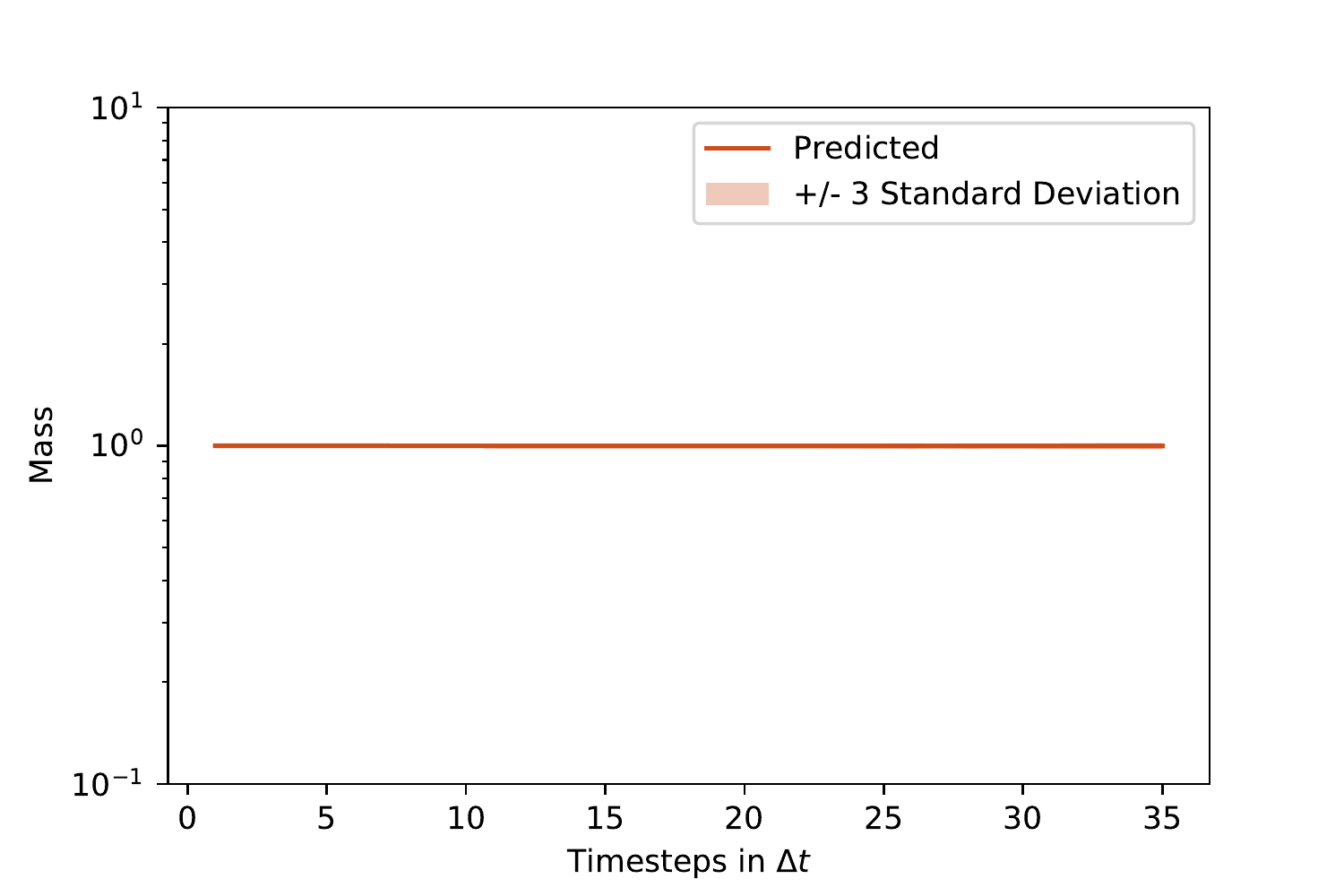}
 \caption{Evolution of the mass constraint (target value is $1$) in time including future time-instants. "Predicted" corresponds to the posterior mean.}
 \label{fig:mass_ad}
\end{figure}

Finally, in Figure \ref{fig:mass_ad}, the evolution of the mass constraint into the future is depicted  and good agreement with the target value is observed.

%% file: burgers.tex
The second test-case involved an FG system of  $d_f=500 \times 10^3$ particles  which perform {\em interactive} random walks i.e. the jump performed at each fine-scale time-step $\delta t=2.5 \times 10^{-3}$ depends on the positions of the other walkers. In particular we adopted interactions as described in   \cite{roberts_convergence_1989,chertock_particle_2001,li_deciding_2007}
 so as, in the limit (i.e. when $d_f \to \infty, \delta t \to 0, \delta s \to 0$), the particle density $\rho(s,t)$ follows the inviscid Burgers' equation:
 \be
  \cfrac{\pa \rho }{\pa t} +\frac{1}{2} \cfrac{\pa \rho^2}{\pa s}=0, \qquad s \in (-1,1).
  \label{eq:inviscidbu}
  \ee

For the CG description, $128$ bins  were employed i.e. $d_c= 128$  and a time step $\Delta t= 4$ (see Table \ref{tab:particles}).
As compared with the previous case, we enlarged the neighborhood size $H$ in the first- and second-order interactions to $H=8$, which yielded $M=170$ right-hand-side terms in  \refeq{eq:cgparticles}. We incorporate virtual observables pertaining to the residuals $\hat{\bs{R}}_0$ with $\sigma_R^2=10^{-7}$  (\refeq{eq:vlikeres}) and the virtual observables $\hat{\bs{c}}_1$ pertaining to conservation-of-mass constraint with $\sigma_c^2=10^{-10}$ (\refeq{eq:vlikecon}).

\begin{figure}[h]
 \includegraphics[width=0.89\textwidth]{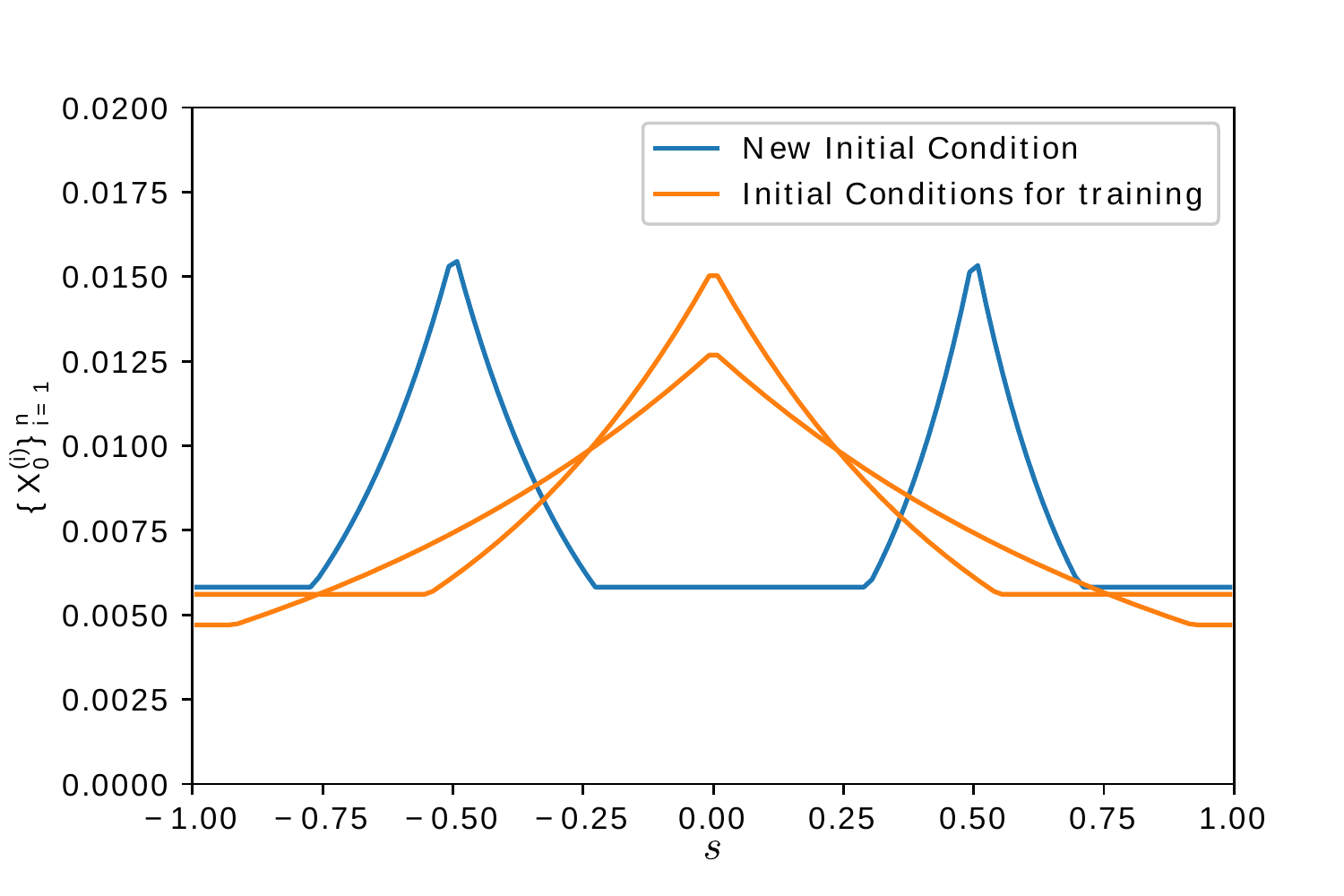}
 \caption{Sample initial conditions $\left\{\bxx_0^{(i)} \right\}_{i=1}^n$ for the Burgers' problem (orange) and an initial condition (blue) used for``extrapolative" predictions.}
 \label{fig:burger_initial}
\end{figure}

We employed $n=32$, $n=64$  and $n=128$ time sequences for training that were generated as detailed in section \ref{eq:inferenceparticles}  with initial conditions $\{\bxx_0^{(i)} \}_{i=1}^n$ such as the ones seen in  Figure \ref{fig:burger_initial}. They were generated by randomizing the width and height of a triangular profile.

\begin{figure}
 \includegraphics[width=0.89\textwidth]{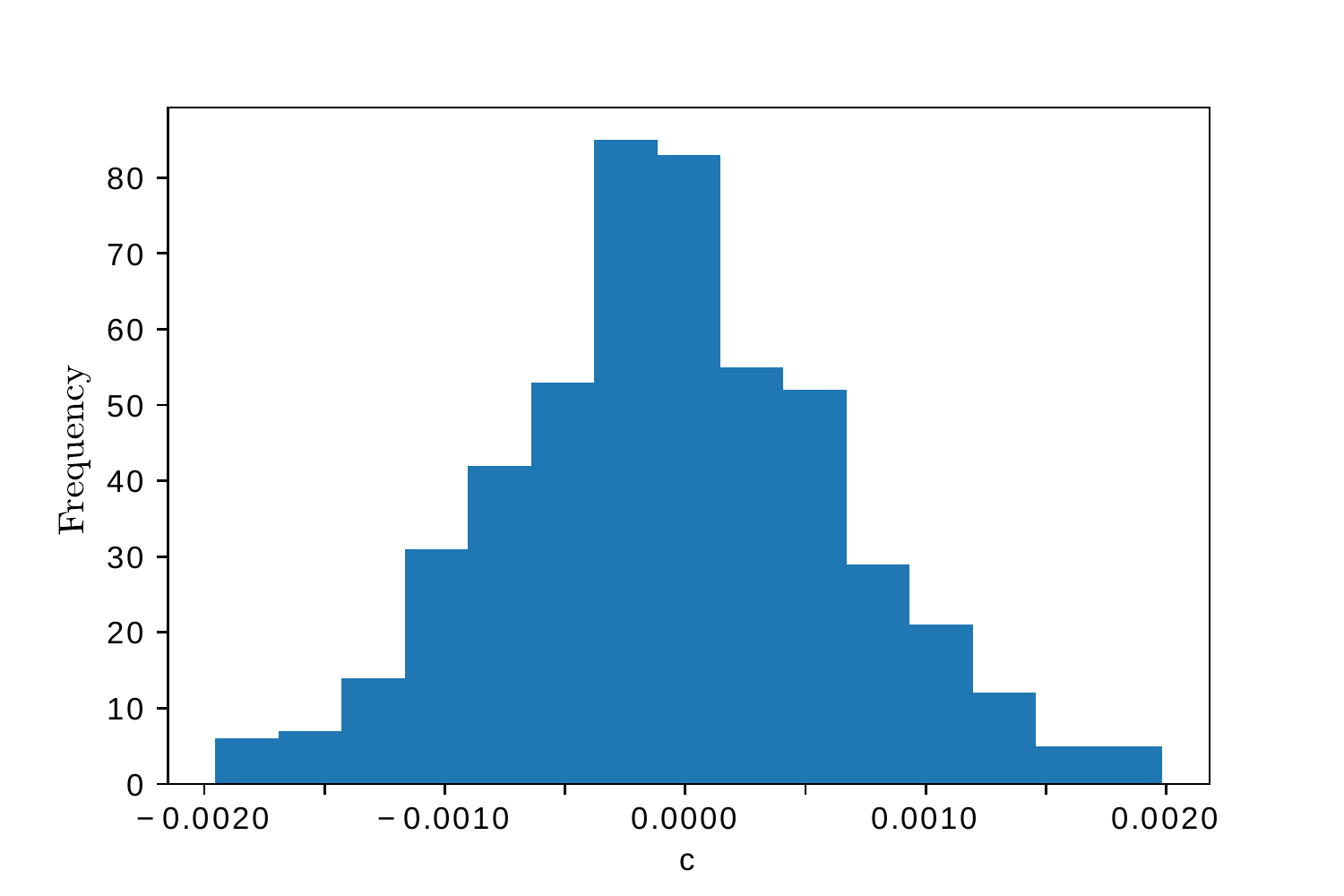}
 \caption{Histogram of the mass constraint  $c_{1}$}
 \label{fig:burger_constraint}
\end{figure}

\begin{figure}
 \includegraphics[width=0.89\textwidth]{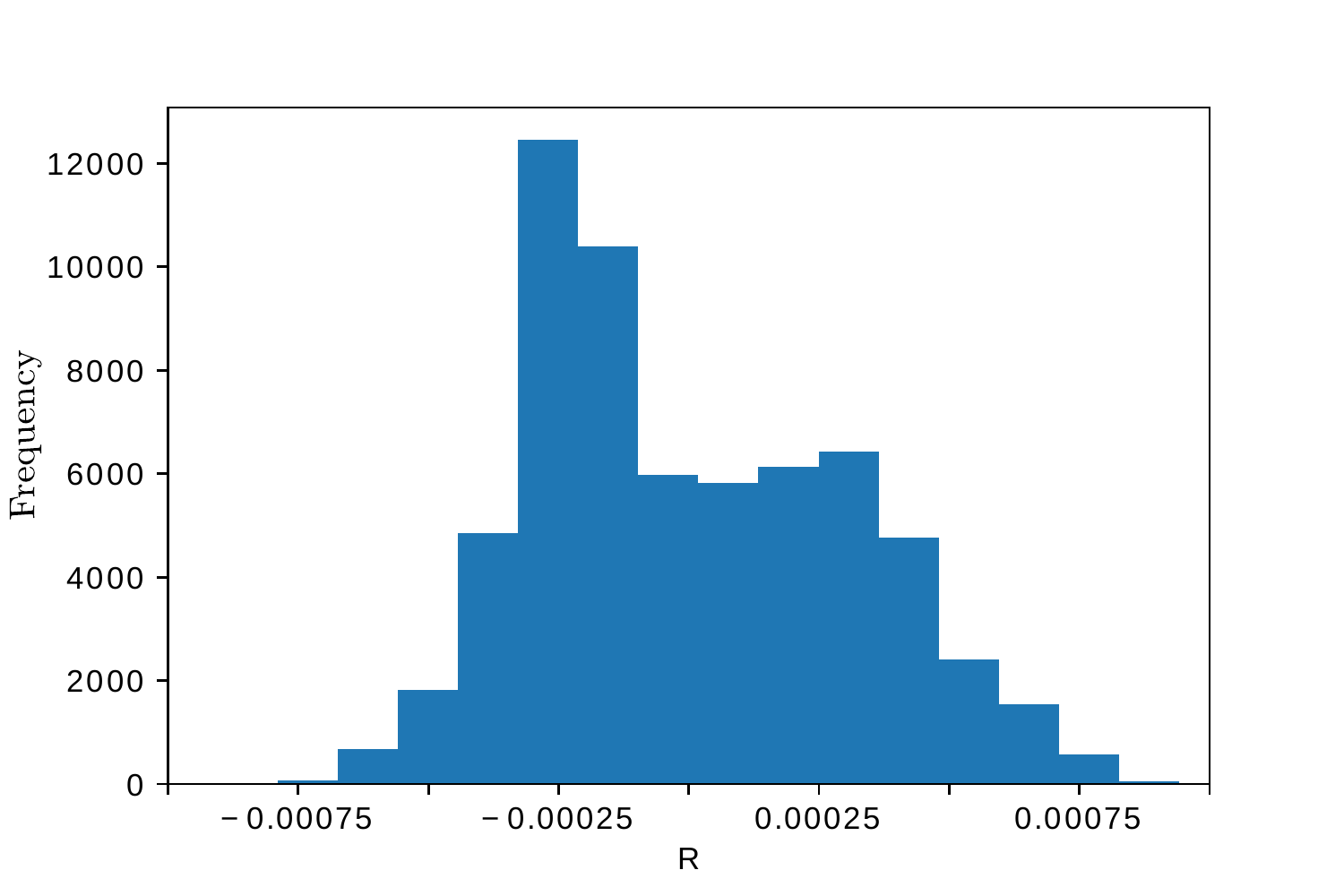}
 \caption{Histogram of the norm of the residual constraint  $\bs{R}_{0}$}
 \label{fig:burger_residual}
\end{figure}

\begin{figure}
 \includegraphics[width=0.89\textwidth]{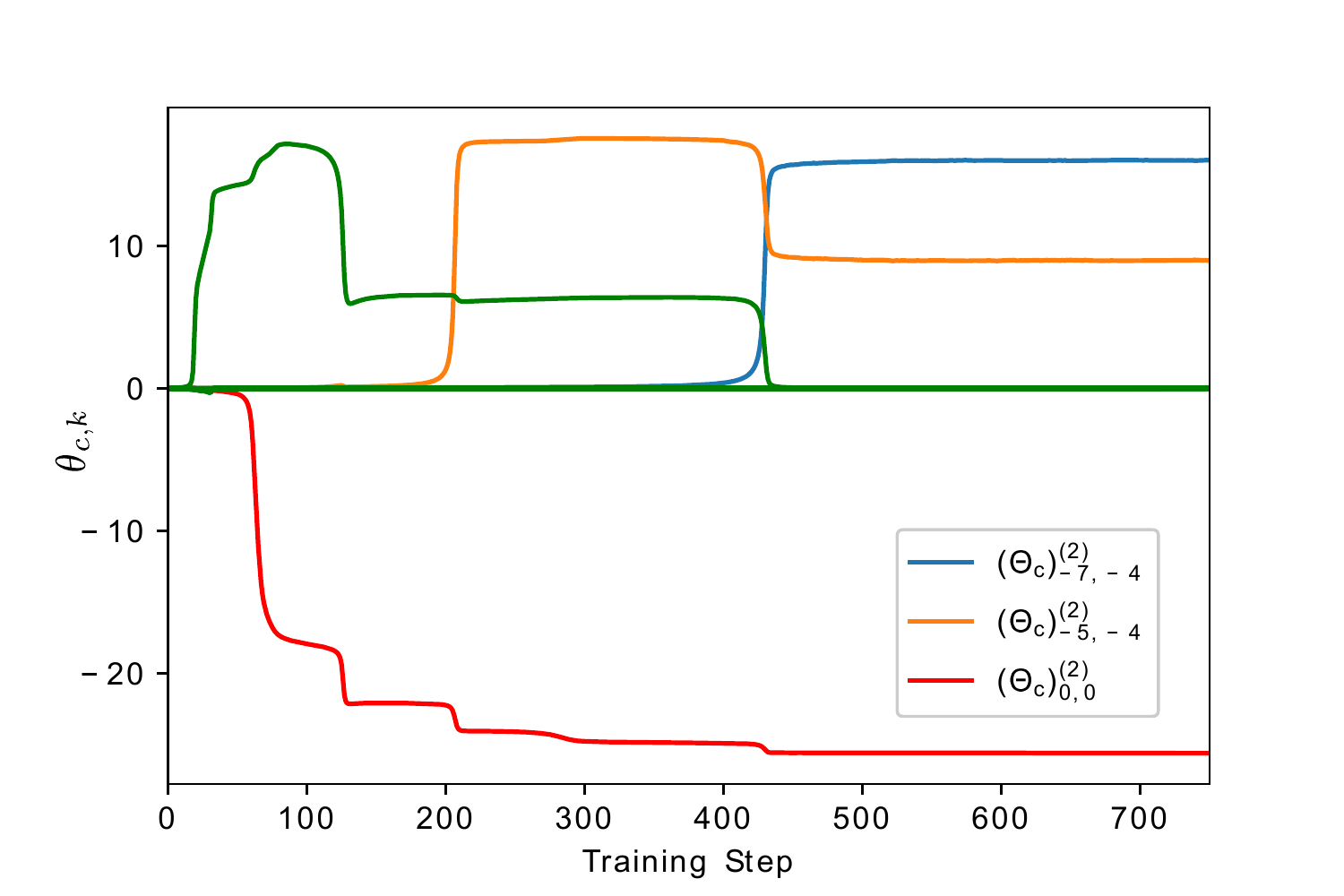}
 \caption{Evolution of a  subset of $\bt_c$ parameters with respect to the iterations of the SVI for $n=64$.}
 \label{fig:burger_coeff}
\end{figure}

\begin{figure}
 \includegraphics[width=0.49\textwidth]{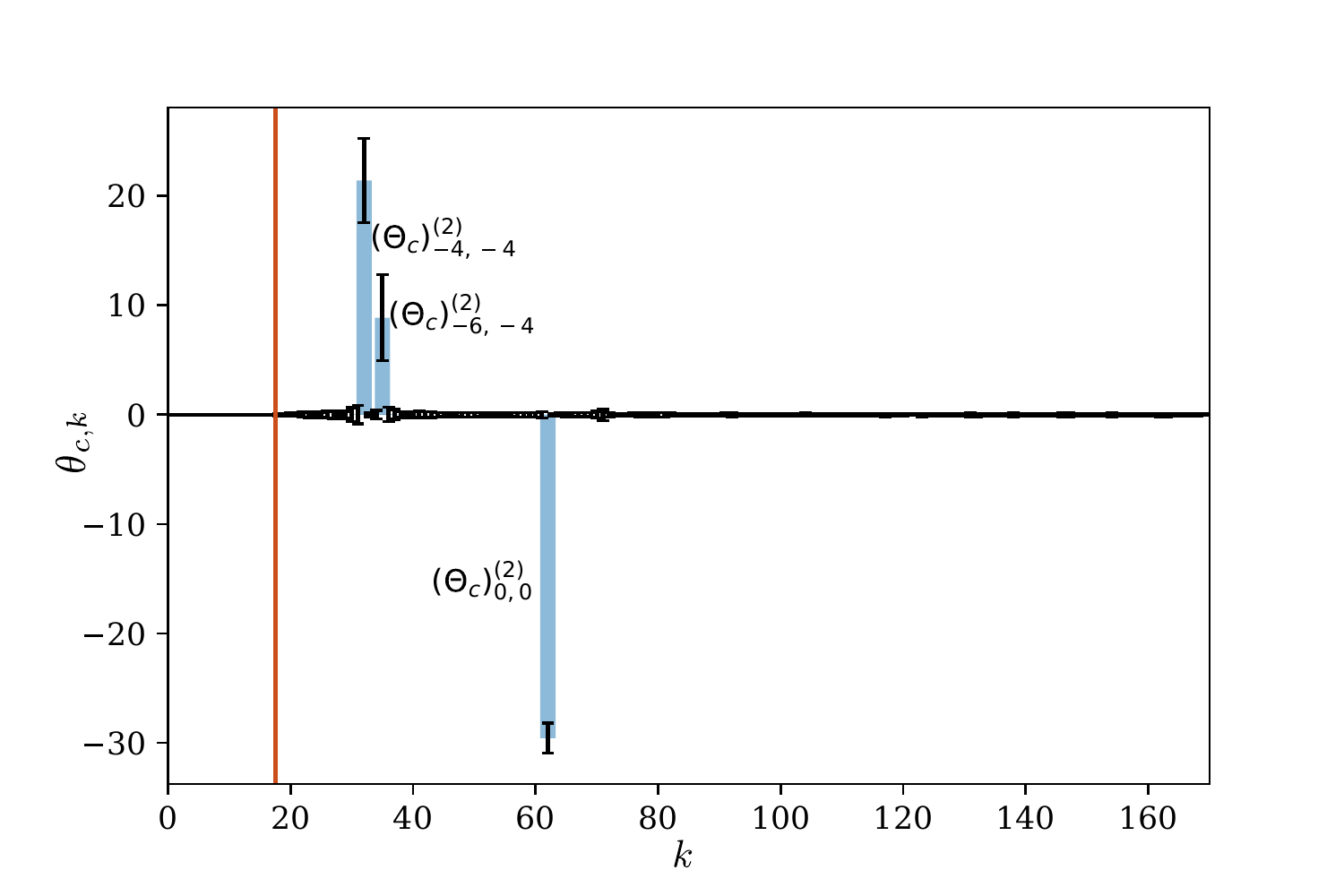} \includegraphics[width=0.49\textwidth]{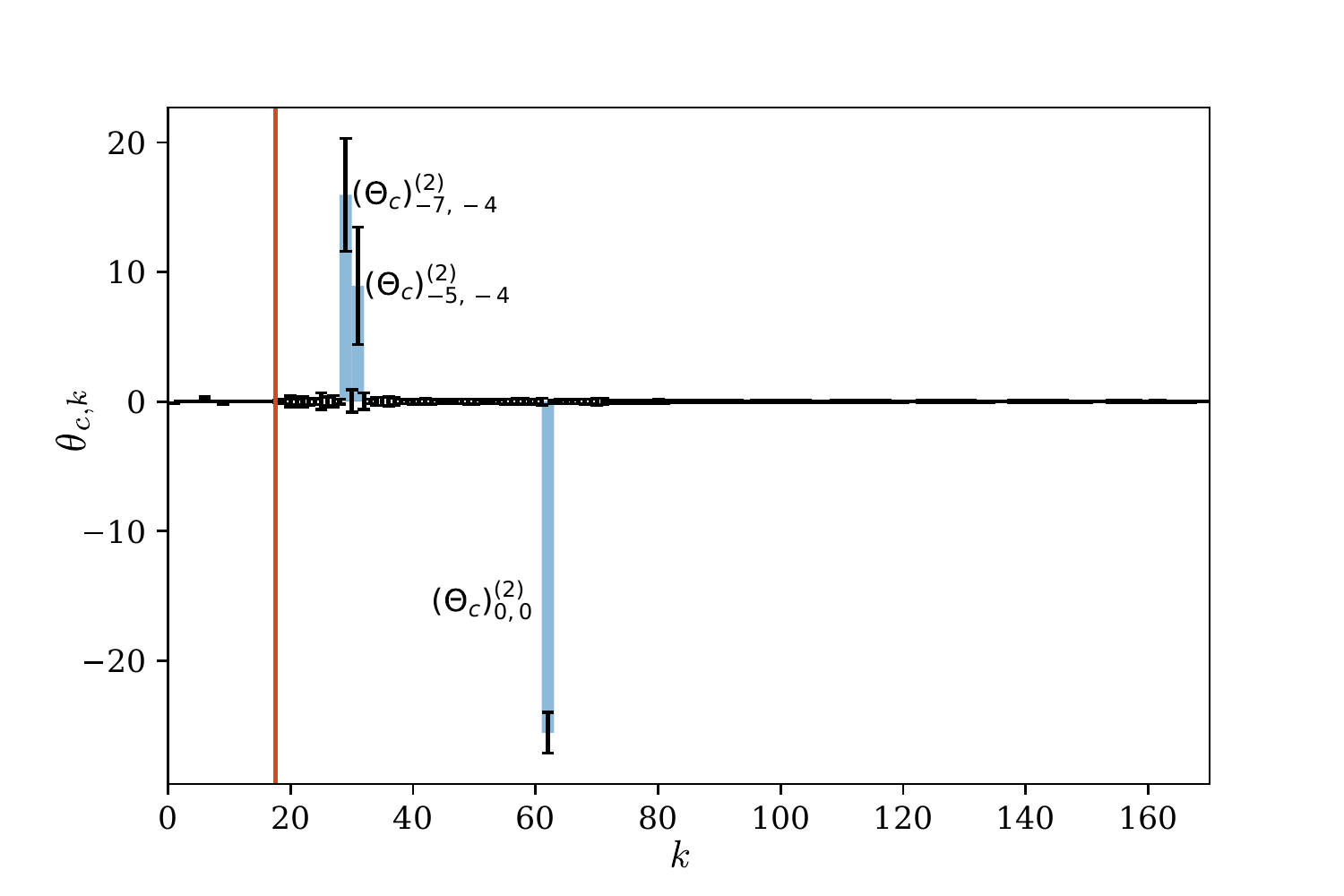}\\
 \includegraphics[width=0.49\textwidth]{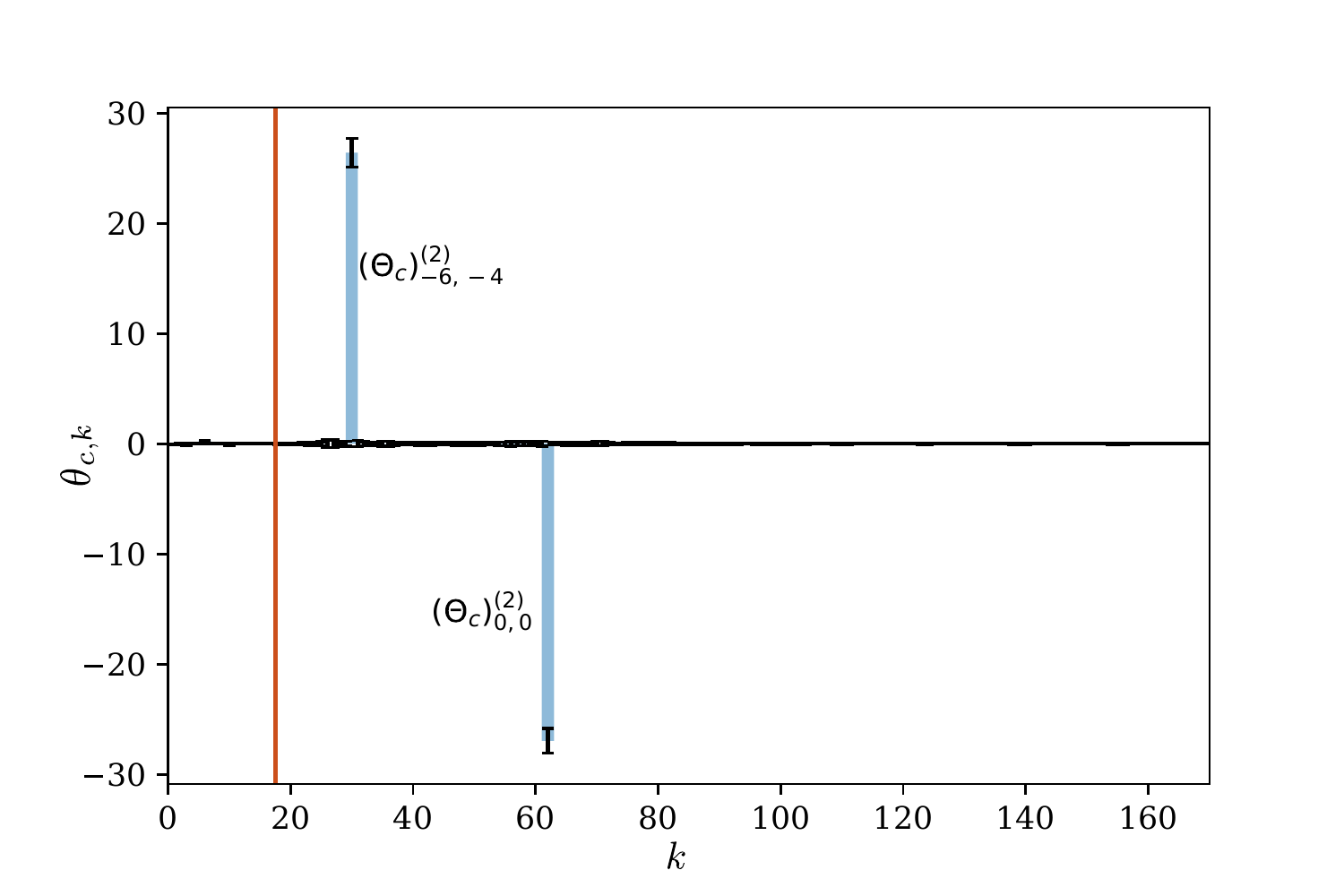}
 \caption{Comparison of the inferred parameters $\bt_c$ for $n=32$ (top-left),  $n=64$ (top-right) and $n=128$ (bottom-left) training data. The black bars indicate +/- 1 standard deviation. The red vertical line separates first- from second-order coefficients.}
 \label{fig:burger_theta}
\end{figure}

Figure \ref{fig:burger_constraint}  provides a histogram of the function values of the conservation-of-mass constraint $\left\{c_1(\bxx_{\Delta t}^{(i)}) \right\}_{i=1}^n$ upon convergence. The small values suggest that this has been softly incorporated in the CG states. A similar histogram for the norm of the residuals $\left\{\bs{R}_{0}(\bxx^{(i)} )\right\}_{i=1}^n$ is depicted in Figure \ref{fig:burger_residual} which also suggests  enforcement of the CG evolution with the parameters $\bt_c$ learned from the data. The evolution of the posterior mean $\bs{\mu}_{\bt_c}$ (\refeq{eq:theta2}) of (a subset of) these parameters over the iterations of the SVI is depicted in Figure  \ref{fig:burger_coeff}. As in the previous example,   in Figure \ref{fig:burger_theta} one can observe the ability of the ARD prior model to yield sparse solutions for the right-hand side of the CG evolution law. 
%
%
%
For all three training datasets with  $n=32,64, 128$ time-sequences, only parameters $\bt_c$ associated with second-order-interactions (\refeq{eq:cgparticles}) are activated. In particular, these are the negative coefficient $\bt_{c,(0,0)}^{(2)}$ (in all three cases) as well as different  second-order coefficients. 
In the cases of $n=32$ and $n=64$ two coefficients are found with positive mean and  high posterior uncertainty, but they also have negative posterior correlation (correlation coefficient of $-0.88$). As all activated coefficients pertain to feature-functions involving    the actual bin or bins to the left, the learned evolution law could be interpreted as an upwind scheme, which takes the direction of the Burgers' flow into account. Such schemes are considered advantageous  for numerical simulations of fluid flows. 

 \begin{figure}
 \includegraphics[width=0.89\textwidth]{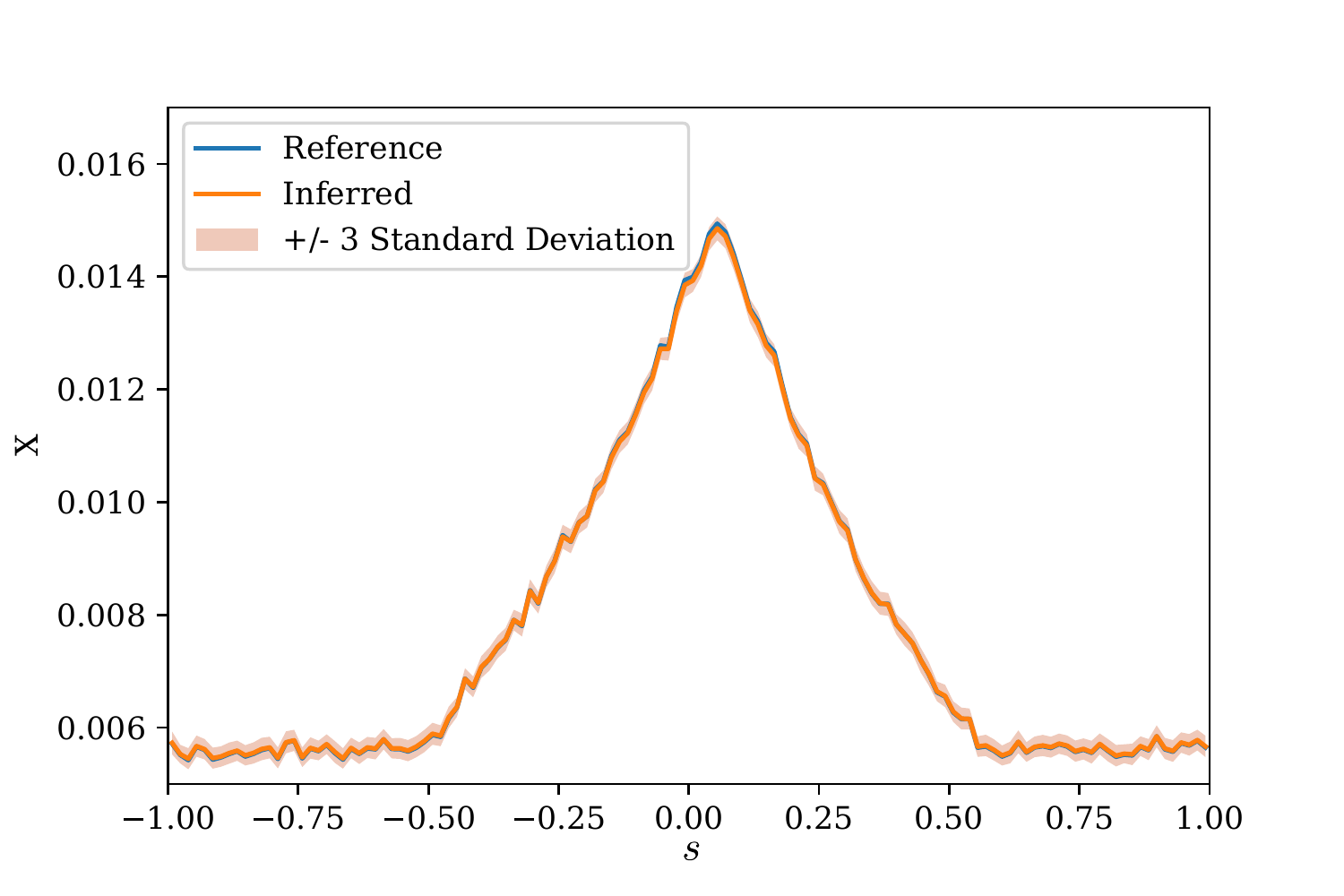}
 \caption{Example of inferred CG state $\bxx_{\Delta t}^{(i)}$ for data sequence $i$.}
 \label{fig:burger_inf}
\end{figure}

\begin{figure}[h]
 \includegraphics[width=.95\textwidth]{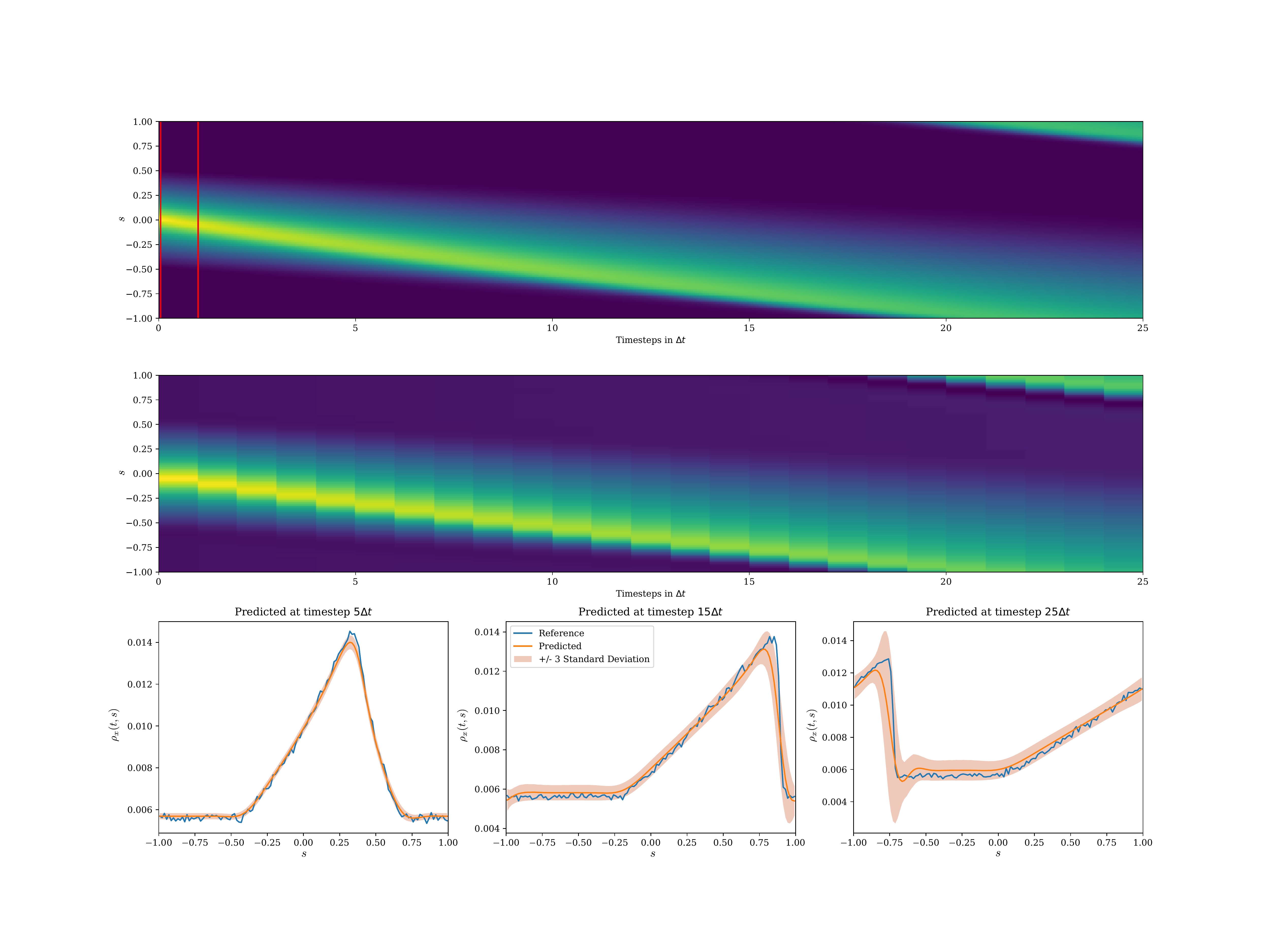}
 \caption{Prediction based on an initial condition contained in the training data. Top: Reference data (the vertical lines indicate the time instances with given data), Middle: Predictive posterior mean, Bottom: snapshots at three different time instances}
 \label{fig:pred_train_burger}
\end{figure}

\begin{figure}
 \includegraphics[width=0.49\textwidth]{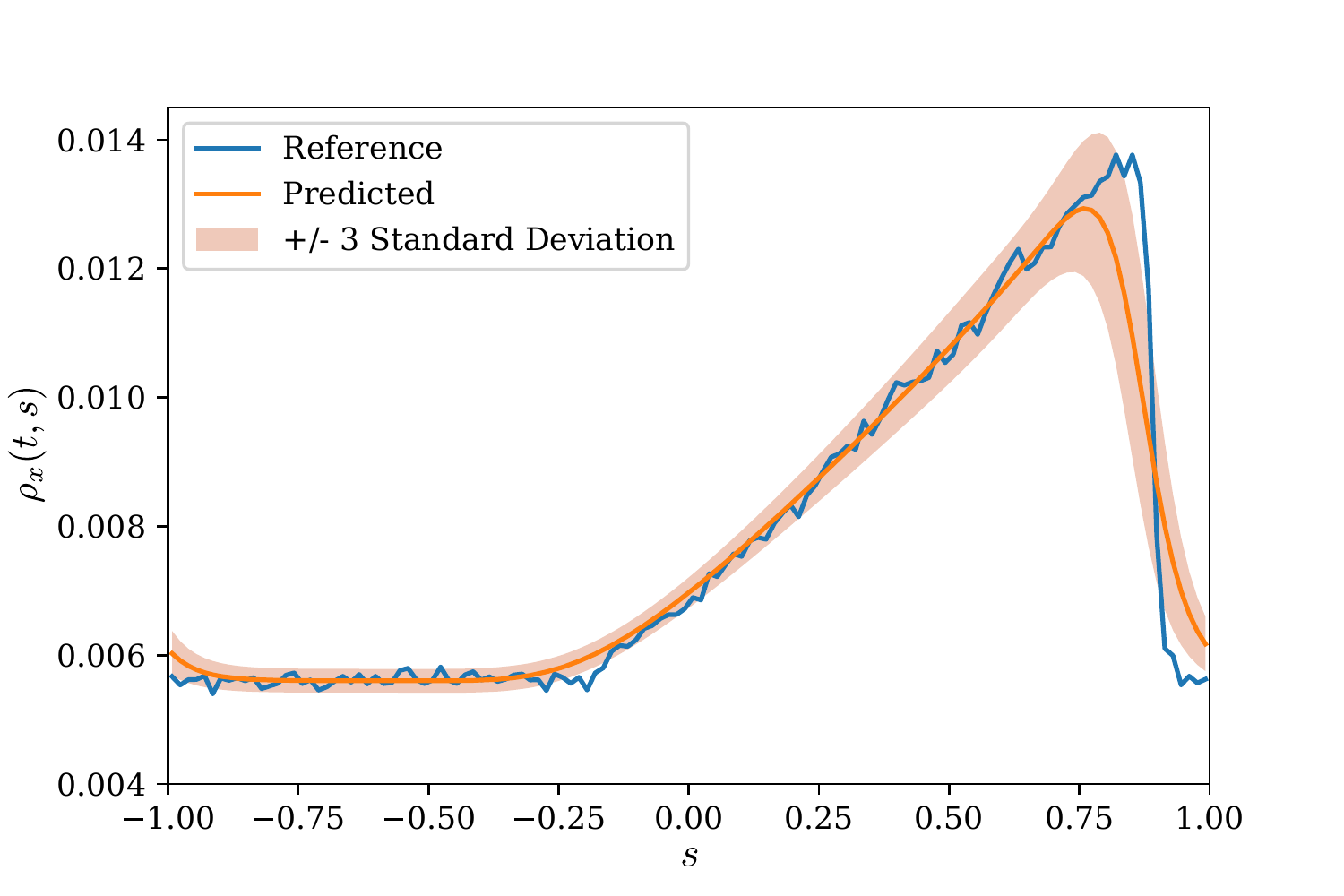} \includegraphics[width=0.49\textwidth]{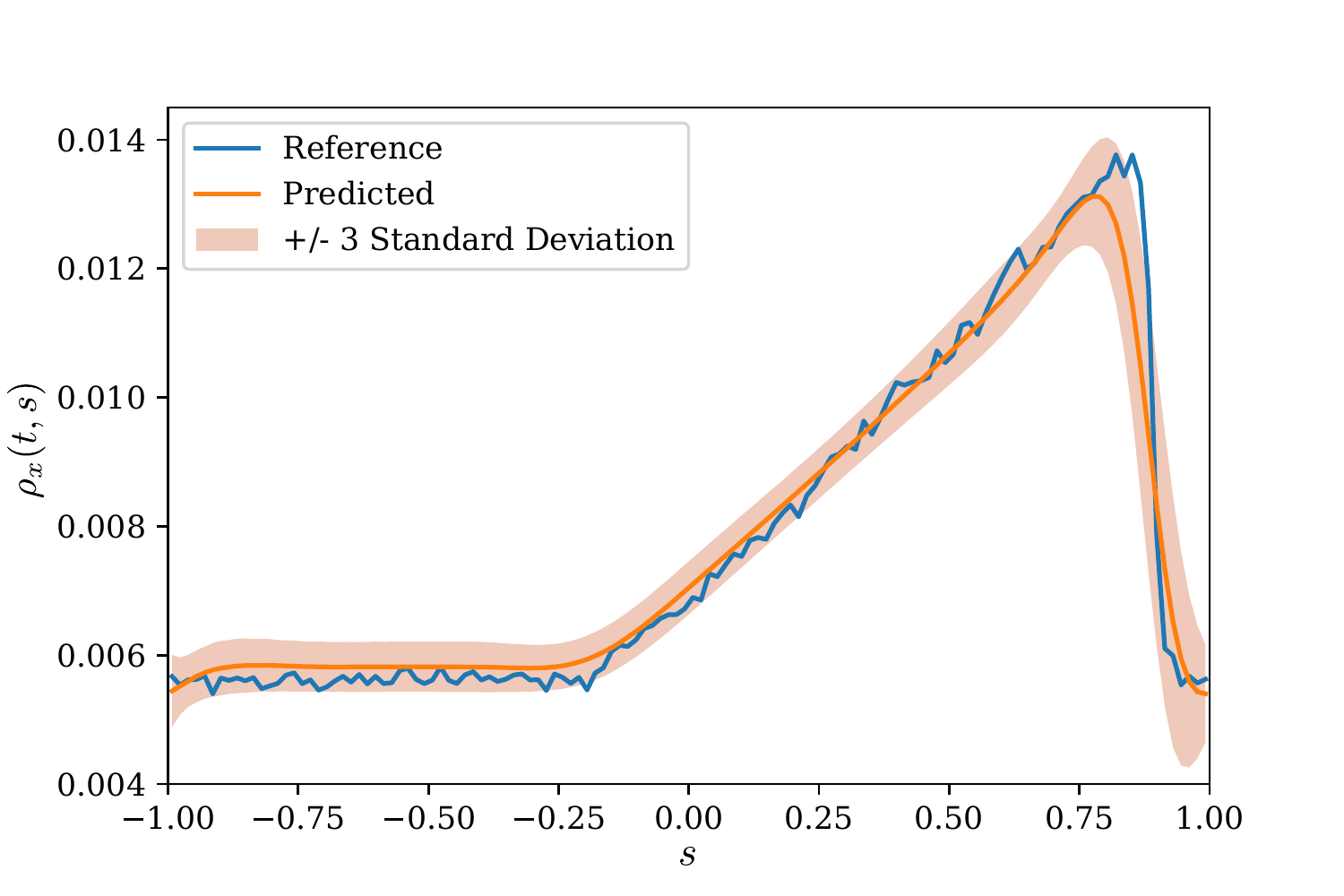}\\
 \includegraphics[width=0.49\textwidth]{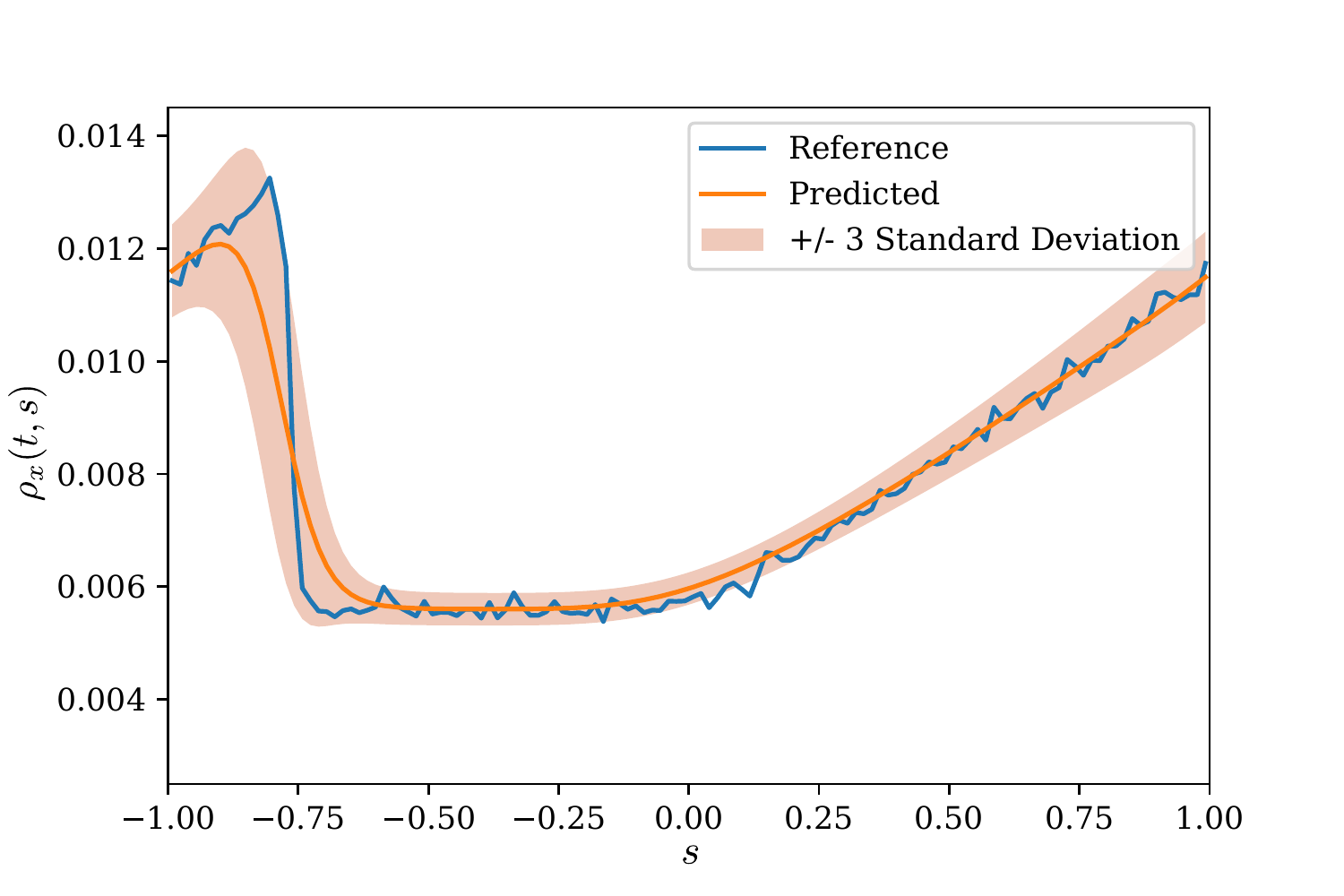} \includegraphics[width=0.49\textwidth]{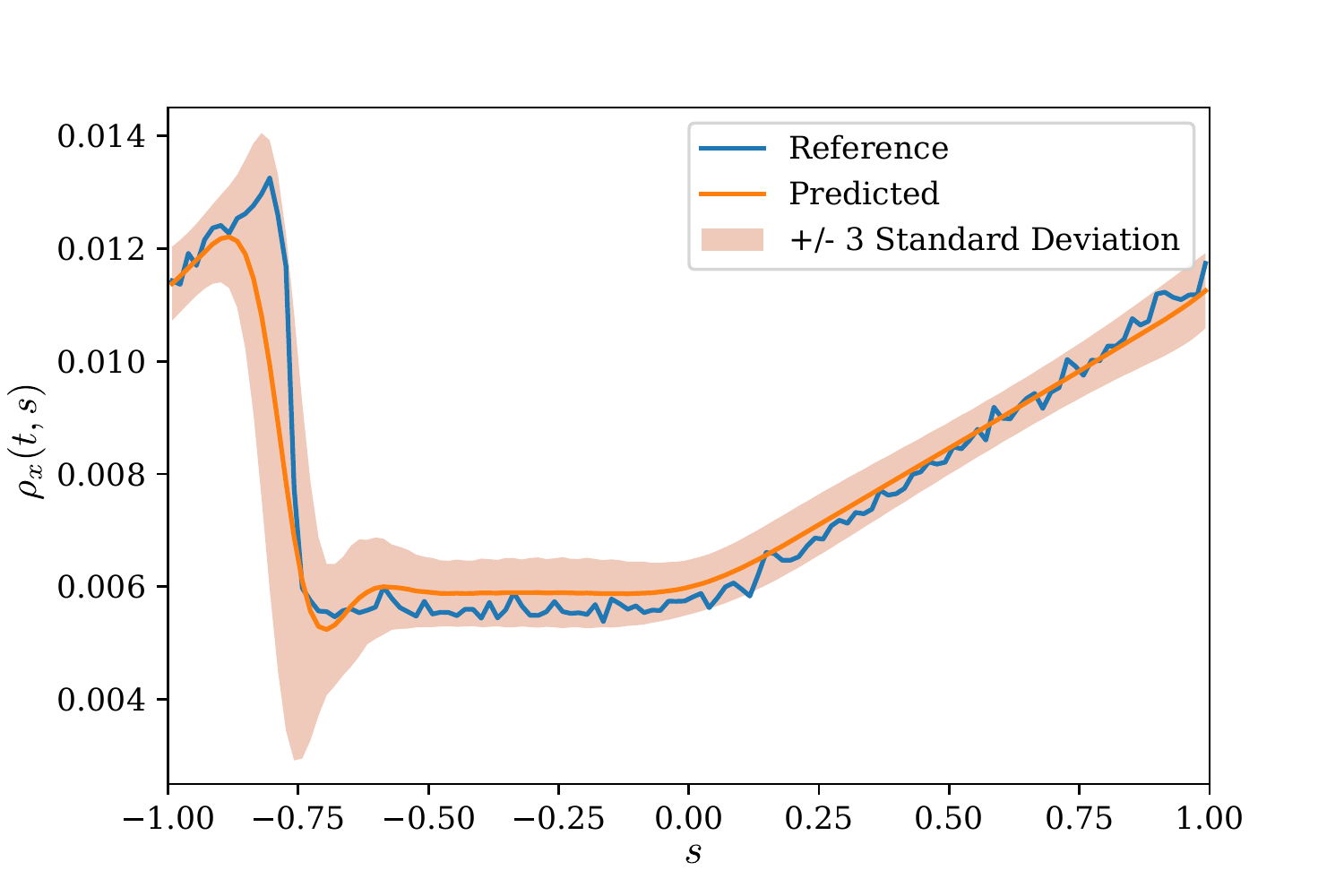}\\
 \caption{Comparison of the predictions for $n=32$ (left) and $n=64$ (right) training data at $15\Delta t$ (top) and $25 \Delta t$ (bottom).}
 \label{fig:burger_data}
\end{figure}

Figure \ref{fig:burger_inf} depicts one of the inferred CG states $\bxx_{\Delta t}^{(i)}$ as well as the associated posterior uncertainty. Given the learned CG dynamics, this state can be propagated into the future as detailed in section \ref{sec:predictions} in order to generate predictions.  
Indicative predictions (under ``interpolative" conditions) can be seen in Figure \ref{fig:pred_train_burger} where the particle density up to $25 \Delta t$ into the future is drawn. The latter as well as the associated uncertainty bounds are  estimated directly from the reconstructed FG states. As in the previous example,  the predictive uncertainty grows, the further into the future one tries to predict. 
Figure  \ref{fig:burger_data} compares the predictive performance as a function of the training data used i.e. $n=32$ or $n=64$. The increase in data leads for this example to a better fit of the posterior mean to the reference, which captures the location of the shock more precisely. The predictive uncertainty bounds are particularly large at the location  of the shock which is the most challenging component in such systems.\\

\begin{figure}[h]
 \includegraphics[width=.95\textwidth]{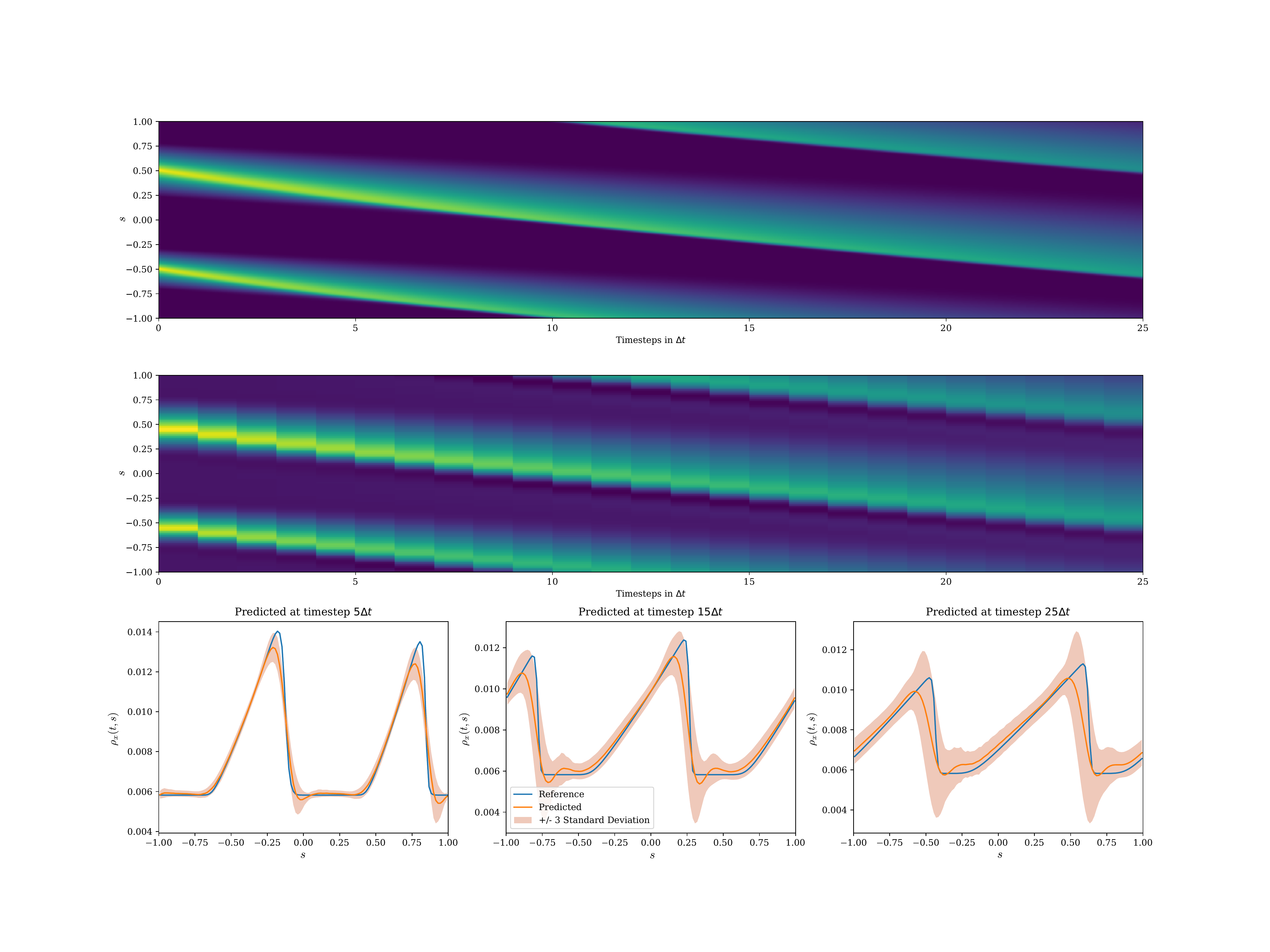}
 \caption{Prediction based on an initial condition NOT contained in the training data. Top: Reference data, Middle: Predictive posterior mean, Bottom: snapshots at three different time instances}
 \label{fig:pred_new_burger}
\end{figure}

\begin{figure}
 \includegraphics[width=0.95\textwidth]{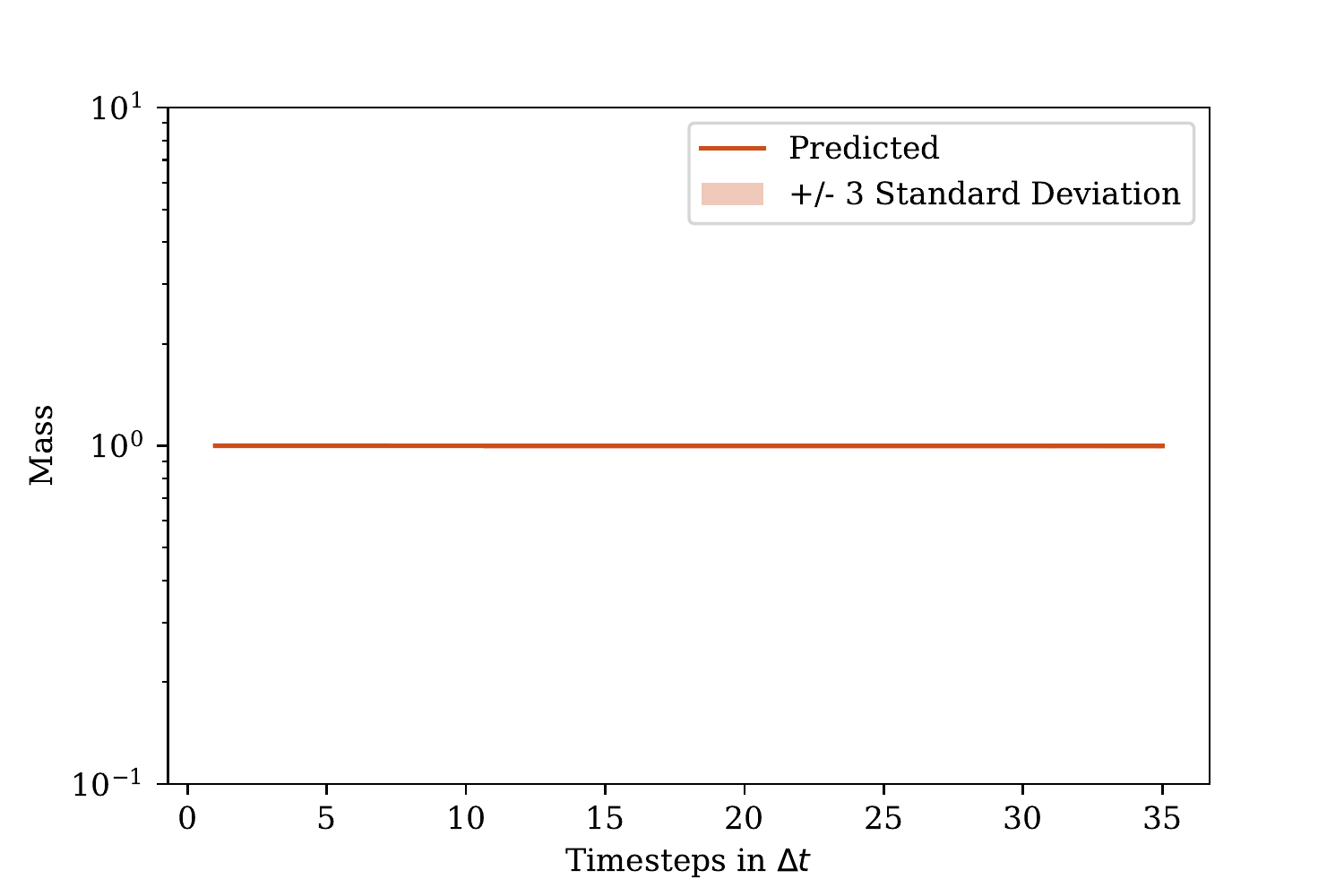}
 \caption{ Evolution of the mass constraint (target value is $1$) in time including future time-instants. "Predicted" corresponds to the posterior mean.}
 \label{fig:burger_mass}
\end{figure}

We also test the trained model (on $n=64$) under ``extrapolative" conditions i.e. for a ``bimodal" initial condition which was quite different from the ones  included in the training data (Figure \ref{fig:burger_initial}). The predictive estimates in Figure \ref{fig:pred_new_burger} show very good agreement with the reference solution. We want to point out that the trained model is capable of capturing the development, the position as well as the propagation of a shock front. 
Finally, in Figure \ref{fig:burger_mass}, the evolution of the mass constraint into the future is depicted  and good agreement with the target value is observed.


%% file: pendulum.tex
In this final example we consider time sequences of images of a nonlinear pendulum in two dimensions as in \citep{champion2019data}. 
%
\subsubsection{FG model}
For the FG data we generate a series of black-and-white images of a moving disc tied on a string and forming a pendulum (see Figure \ref{fig:pendpred}). Each  image consists of  $29 \times 29$ pixels each and each pixel's value was either $1$ (occupied) or $-1$ (unoccupied). Hence $\bx_t$ was a $d_f=29^2=581$-dimensional vector of binary variables. The dynamics of the pendulum  can be fully described by the rotation angle $y_t$ which follows a nonlinear, second-order ODE of the form:
\be
\ddot{y}_t+\sin(y_t)=0
\label{eq:refpendulum}
\ee
The primary goal is to identify the right CG variables as well as CG dynamics solely from image data i.e. binary vectors $\{ \hat{\bx}_{0:T \Delta t}^{(i)}\}_{i=1}^{n}$ collected over $T$ time-steps as the pendulum is initialized from $n$  states/positions.
The length of time sequences in the following numerical results was $T=74$ and the CG time-step $\Delta t=0.05$\footnote{For the generation of images a {\em microscopic} time-step $\delta t=0.01$  for the integration of \refeq{eq:refpendulum} was used.}.\review{ We also considered the effect of missing data i.e. only observing a subset of the $T+1$ values in  each sequence  and present respective results in Section \ref{sec:Appendix}.} 

\subsubsection{CG variables and  coarse-to-fine mapping}
The only knowledge introduced a priori with regards to the CG variables $\bxx_t$ is that $dim(\bxx)=d_c=2$. We intend to investigate procedures that can automatically identify $d_c$ i.e. the number of CG variables. We note at this stage that such efforts could be guided by  the ELBO $\mathcal{F}$ (e.g. \refeq{eq:ELBOgeneral}) which  approximates the model evidence and therefore provides a natural Bayesian score for comparing models with different numbers of CG variables.
%

\begin{figure}[!ht]
 \includegraphics[trim=20 400 0 00 ,clip,width=1.0\textwidth]{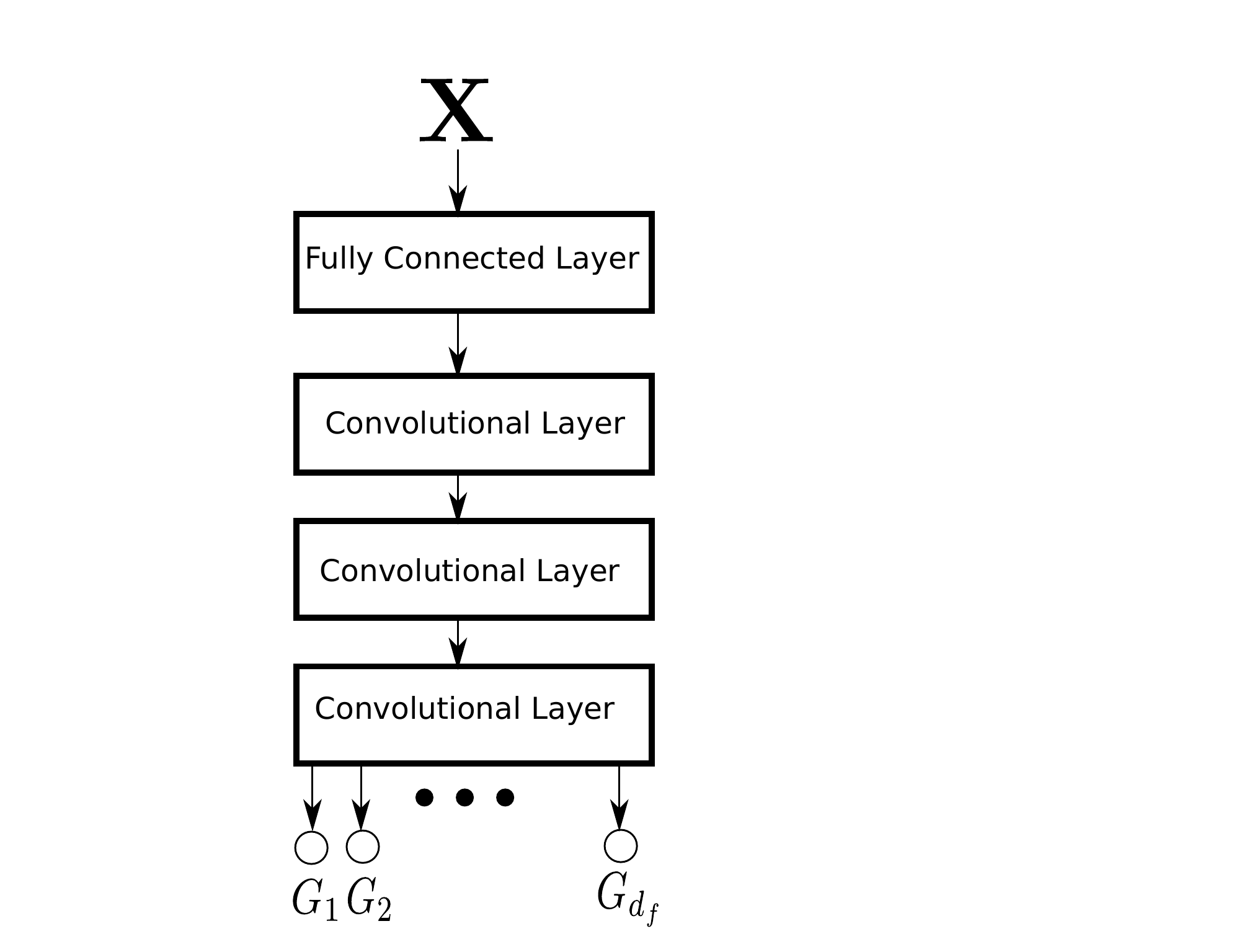}
 \caption{Deep neural net employed for the link functions $G_s$ (\refeq{eq:pcfpendulum}).  After one dense layer which $32\cdot7\cdot7$ nodes and rectified linear unit activation function (ReLU), two two-dimensional transposed convolutional layers with 32 filters and a kernel size of 3 as well as a ReLU activation function are applied followed by one-last two-dimensional transposed convolutional layers with one filter, kernel size 3 and without activation to generate the functions $G_{s}$ .}
 \label{fig:nn}
\end{figure}

The other pertinent model component is the coarse-to-fine map which is enabled by the $p_{cf}(\bx_t | \bxx_t)$ (section \ref{sec:emission}). To that end, we employed the following logistic model\footnote{We omit the time-index $t$ for clarity.}:
\begin{equation}
    p_{cf}(\bx | \bxx) = \prod_{s=1}^{d_f} p_{cf}(x_s | \bxx)
    \label{eq:pcfpendulum}
\end{equation}
with
\begin{equation}
    p_{cf}(x_s | \bxx) = \begin{cases}
     \cfrac{1}{1+\exp(-G_{s}(\bxx; \bt_{cf}))} \;\; \text{for} \; x_s=1 \\
      \cfrac{1}{1+\exp(+G_{s}(\bxx; \bt_{cf}))} \;\; \text{for} \; x_s=0 \\
    \end{cases}
    \label{eq:pcfpendulum_detail}
\end{equation}
where $x_s$ is the value ($1,0$) of each of the pixels $s=1,\ldots,d_f$. 
For the link functions $\{ G_{s} \}_{s=1}^{d_f}$, we employed a deep neural net with weights $\bt_{cf}$, the details of which are shown in Figure \ref{fig:nn}.
One fully connected layer followed by two transposed convolutional layers were found to be flexible enough to accurately represent the functions $G_{s}$. The CNNs were specifically chosen because of their ability to extract/map features from/to images.


\subsubsection{The CG evolution law and the virtual observables}

With regards to the evolution law of the CG states $\bxx_t=\{X_{t,1},X_{t,2}\}$, we postulate the following form:
\be
\begin{array}{ll}
\dot{X}_{t,1} & =F_1(\bxx_t,\bt_c)=X_{t,2} \\
\dot{X}_{t,2} & = F_2(\mathbf{X}_t, \bt_c) = \mathbf{\bt_c}^T \bs{\psi}(X_{t,1})=\sum_{m=0}^M \theta_{c,m} ~\psi_m (X_{t,1})
\end{array}
\label{eq:cgpendulum}
\ee
where $\bt_c$ denote the associated parameters. In total we employed $M=101$ feature functions of the following type:
\be
\psi_m(X) =\left\{ \begin{array}{ll} 1, &m=0 \\ \sin(m X), & m =1,\ldots , M/2=50  \\  \cos( (m-50) X), & m =51,\ldots , M =100\end{array} \right.
 \label{eq:featurepend}
 \ee
 The form of \refeq{eq:cgpendulum} implies a second-order ODE where the second CG variable plays the role of the velocity.
 With regards to the parameters $\bt_c$, the sparsity-inducing ARD prior detailed in section \ref{eq:parcg} was employed.
 
 To enforce the associated dynamics, we made use of the sympletic Euler time-discretization scheme, which is a first-order integrator, that is explicit in the first  variable ($X_{t,1}$) and {\em implicit} in the other ($X_{t,2}$)\footnote{This corresponds to a multistep method in \refeq{eq:cgdiscr} with $K=1$, $a_0=1,a_1=-1,\beta_0=0$ and $\beta_1=-1$ for the explicit part and $K=1$, $a_0=1,a_1=-1,\beta_0=-1$ and $\beta_1=0$ for the implicit part. }. The associated virtual observables (see \refeq{eq:voresl}) were enforced with $\sigma_R^2=10^{-5}$. 


\subsubsection{Inference and Learning}
As in the previous examples (\refeq{eq:particlefactor}), the approximate posterior was factorized as:

\begin{equation}
    q_{\bp}(\bxx_{0: T \Delta t}^{(1:n)}, \bt_c, \bs{\tau})= \left[ \prod_{i=1}^n q_{\bp} (\bxx^{(i)}_{0:T\Delta t})\right] 
    q(\bt_c  ) q(\bs{\tau})
    \label{eq:pendfactor}
\end{equation}
and  closed-form updates were used for $ q(\bt_c)$ (see Equations (\ref{eq:theta1}) and (\ref{eq:theta2}))  and  $q(\bs{\tau})$ (see \refeq{eq:tau1}).  

\begin{figure}[!t]
 \includegraphics[trim=20 450 0 00 ,clip,width=1.0\textwidth]{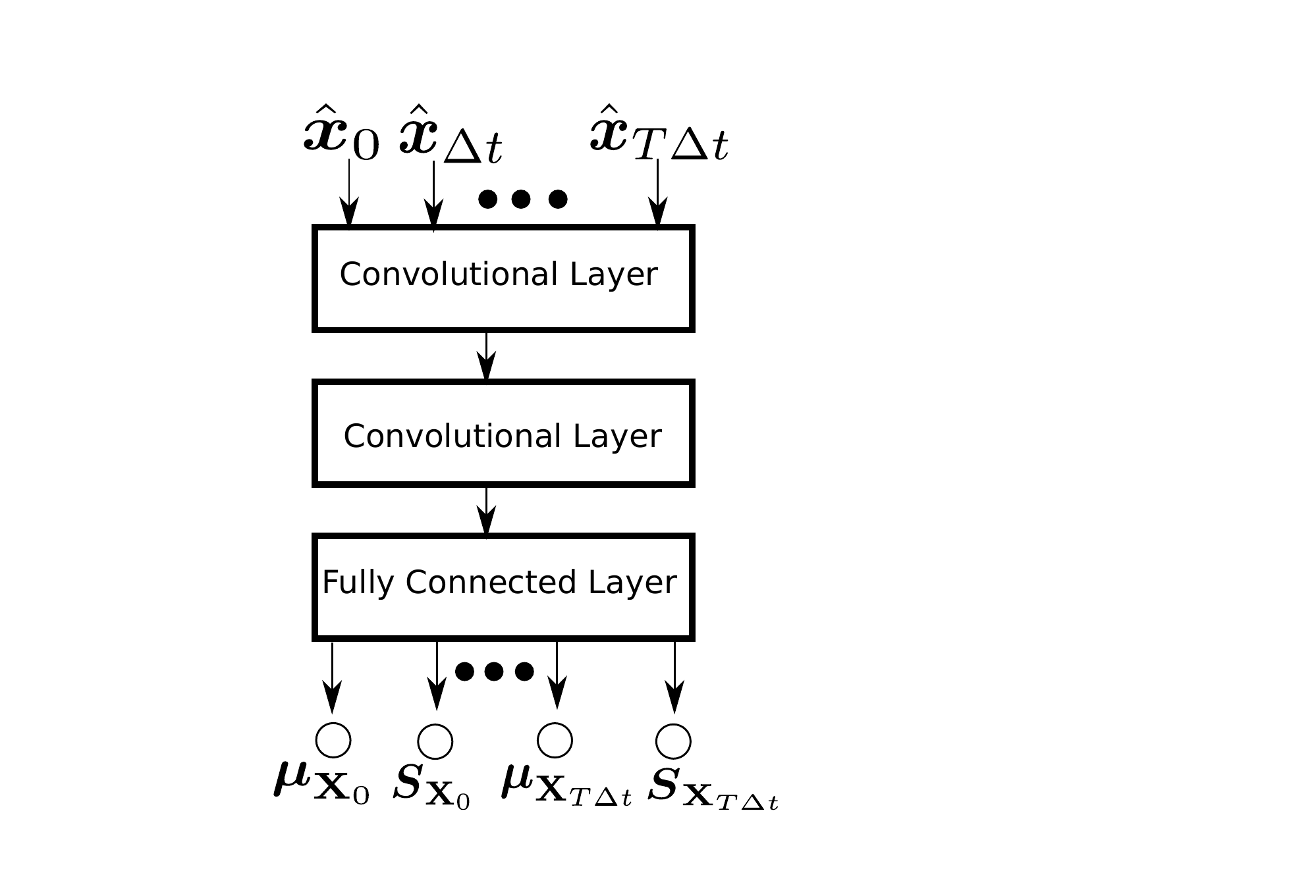}
 \caption{DNN architecture for approximate posterior $q_{\bp}$. The input consists of a time series of pictures of the pendulum and can therefore considered to be three-dimensional, where the first and second dimension are the number of pixels and the third dimension is the number of time steps available for training. This input is given to a three-dimensional convolutional layer with kernel size $(3,3,2)$, 32 filters and a ReLU activation followod by another three-dimensional convolutional layer with kernel size 2 in each dimension, 64 filters and a ReLU activation. The last layer is a fully connected layer with $2 d_c \cdot T$ nodes and without activation to generate the mean and variance values for each time step of the inferred $\bxx$ coordinates.}
 \label{fig:nn_en}
\end{figure}

SVI was applied for the  posterior densities  $q_{\bp} (\bxx^{(i)}_{0:T\Delta t})$ on the vector of the latent CG states $\bxx^{(i)}_{0:T\Delta t}$ which we  approximated with multivariate Gaussians. Since the posterior reveals  the fine-to-coarse map which apart from insight can be used for predictive purposes as well, we employed an {\em amortized} version of SVI (\citep{kingma_auto-encoding_2014}) i.e. explicitly accounted for the dependence of each $q_{\bp} (\bxx^{(i)}_{0:T\Delta t})$ on the corresponding FG observables $\hat{\bx}^{(i)}_{0:T\Delta t}$ i.e.:
\be
q_{\bp} (\bxx^{(i)}_{0:T~\Delta t})=\mathcal{N} \left( \bs{\mu}_{\bp}(\hat{\bx}^{(i)}_{0:T\Delta t})~,~ \bs{S}_{\bp}(\hat{\bx}^{(i)}_{0:T\Delta t})\right)
\label{eq:qpendulum}
\ee
The parameters $\bp$ were the weights of a deep convolutional neural net, the architecture of which is shown in Figure \ref{fig:nn_en}. This  was chosen because it mirrors the DNN architecture employed  for the coarse-to-fine map in Figure \ref{fig:nn}.

Finally it  should be mentioned that the "slowness" prior was employed on the hidden states $\bxx_{0:T \Delta t}^{(1:n)}$ as described in \refeq{eq:cgprior}\footnote{For the prior distribution $p_{c,0}(\bxx_{0}^{(i)})$ a Gaussian mixture distribution with means +1.5 and -1.5 and standard deviation 1.5 was used.}. Maximum-likelihood estimates for the hyperparameter $\sigma_{\bxx}^2$ were employed which readily arise by differentiating the ELBO $\mathcal{F}$ and which yield the following update equation:
\be
\sigma^2_{\bxx} = \frac{1}{n ~T~ d_c } \sum_{i=1}^n \sum_{l=0}^{T-1}  \expe_{q_{\bp}(\bxx_{0:T\Delta t}^{(i)})}\left[\left| \bxx_{(l+1)~\Delta t}^{(i)} -\bxx_{l~\Delta t}^{(i)}\right|^2 \right]
\label{eq:sigmax}
\ee
Maximum likelihood estimates were also obtained for the parameters $\bt_{cf}$ (\refeq{eq:pcfpendulum}) by numerically differentiating the ELBO $\mathcal{F}$ and performing Stochastic Gradient Ascent (SGA).
A general summary of the steps involved for the inference procedure iscan be found in Algorithm \ref{alg:pend}. For the implementation we made use of the Tensorflow framework \citep{abadi2016tensorflow}.\\
\begin{algorithm}[H]
\SetAlgoLined
\KwResult{$\bp$,$q(\bt_c)$,$q(\bs{\tau})$,$\bt_{cf}$,$\sigma_{\bxx}$}
\KwData{$\hat{\bx}^{(1:n)}_{0:T\Delta t}$}
 Initialize all required parameters\;
 Set iteration counter $w$ to zero\;
 \While{$ || ELBO_{w}-ELBO_{w-1} ||^2 > \epsilon$}{
    Update the parameters $\bt_{cf}$  and $\bp$ by SGA of the ELBO ( \refeq{eq:ELBOgeneral}) \;
update  $q(\bt_c)$ according to \refeq{eq:theta1} and \refeq{eq:theta2} \;
  update  $q(\bs{\tau})$ according to \refeq{eq:tau1}  \;
    update the parameter $\sigma_{\bxx}$ according to \refeq{eq:sigmax}\;
    update the iteration counter by one\;
 }
 \caption{ Algorithm for the Pendulum system}
\label{alg:pend}
\end{algorithm}

\subsubsection{Results}

\begin{figure}[!t]
 \includegraphics[width=0.89\textwidth]{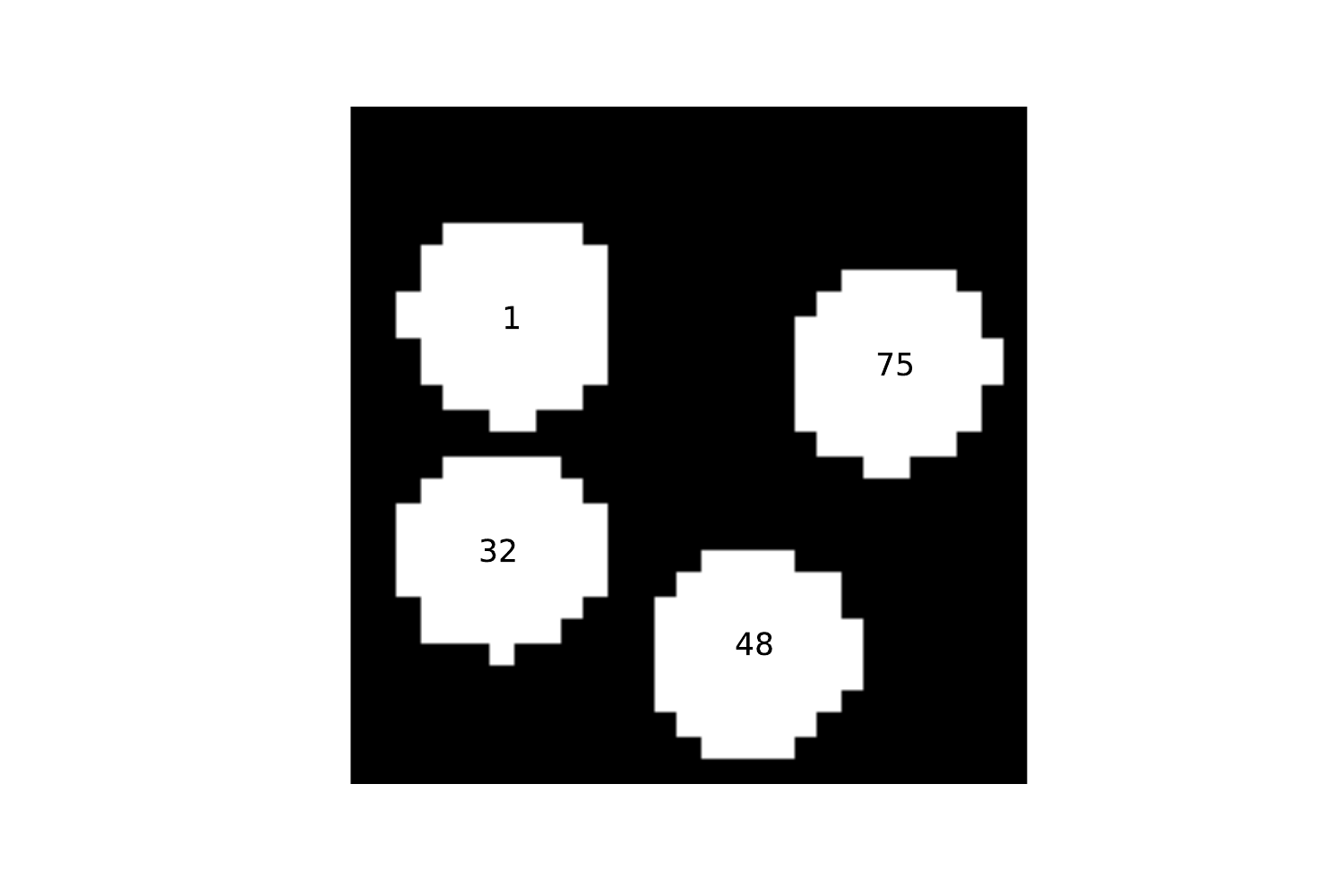}
 \caption{Indicative positions of the pendulum in a data sequence  $\hat{\bx}^{(i)}_{0:T \Delta t}$. The number indicates the corresponding  time-step.}
 \label{fig:pendulum_initial}
\end{figure}

Each data sequence $\hat{\bx}^{(i)}_{0:T\Delta t}$ used consisted of  $75$ images, i.e. $T=74$, generated with a time-step  $\Delta t =0.05$ (Figure \ref{fig:pendulum_initial}). We  investigated two cases for the number of data sequences i.e. $n=16$ and $n=64$. The data generation involved sampling uniformly the initial angle $y_0 \in [-\pi, \pi]$ and assuming zero initial velocity i.e $\dot{y}_0=0$. \review{We emphasize that none of the data sequences contained  a complete oscillation of the pendulum i.e. always partial trajectories were observed.}

\begin{figure}
 \includegraphics[width=0.49\textwidth]{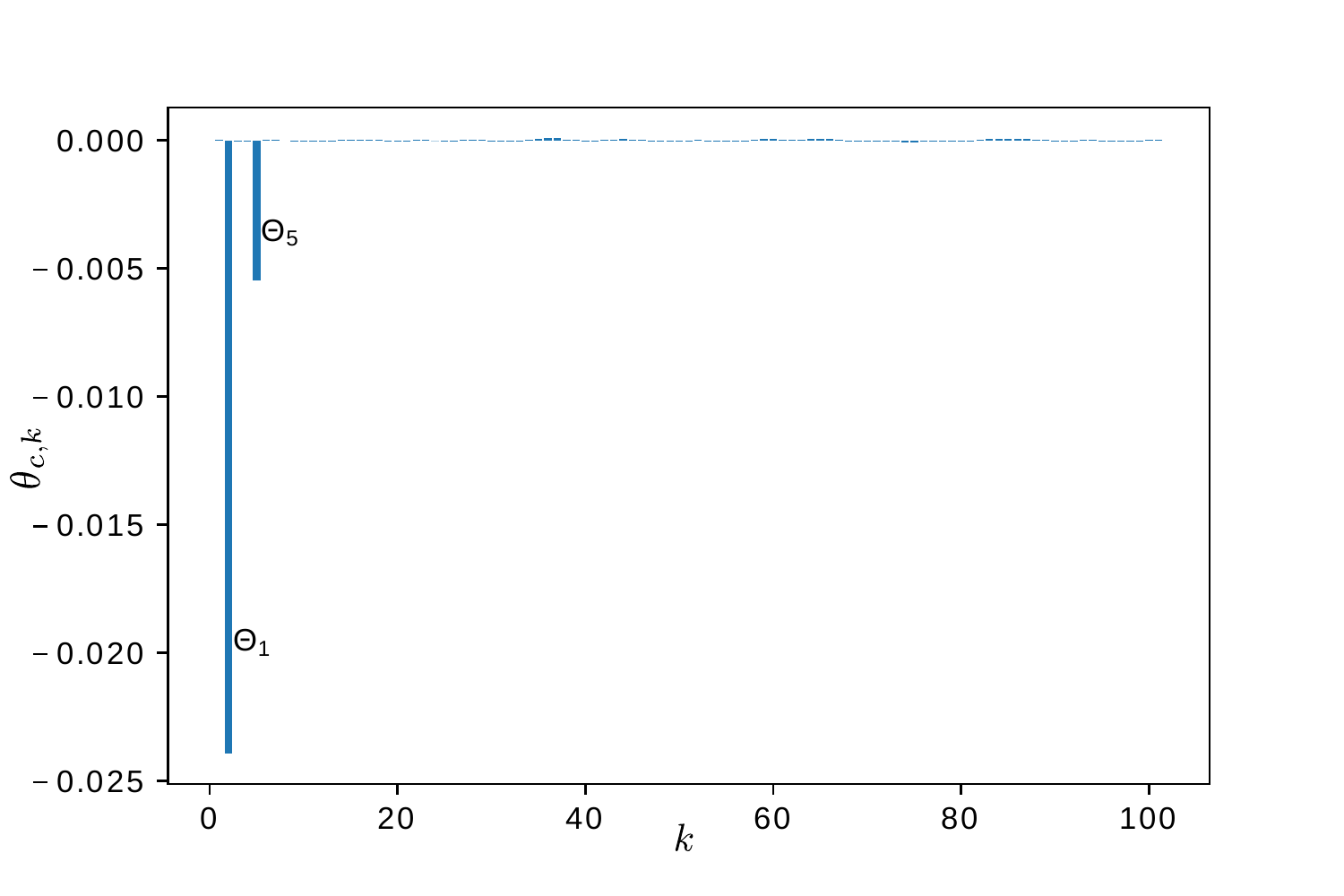} \includegraphics[width=0.49\textwidth]{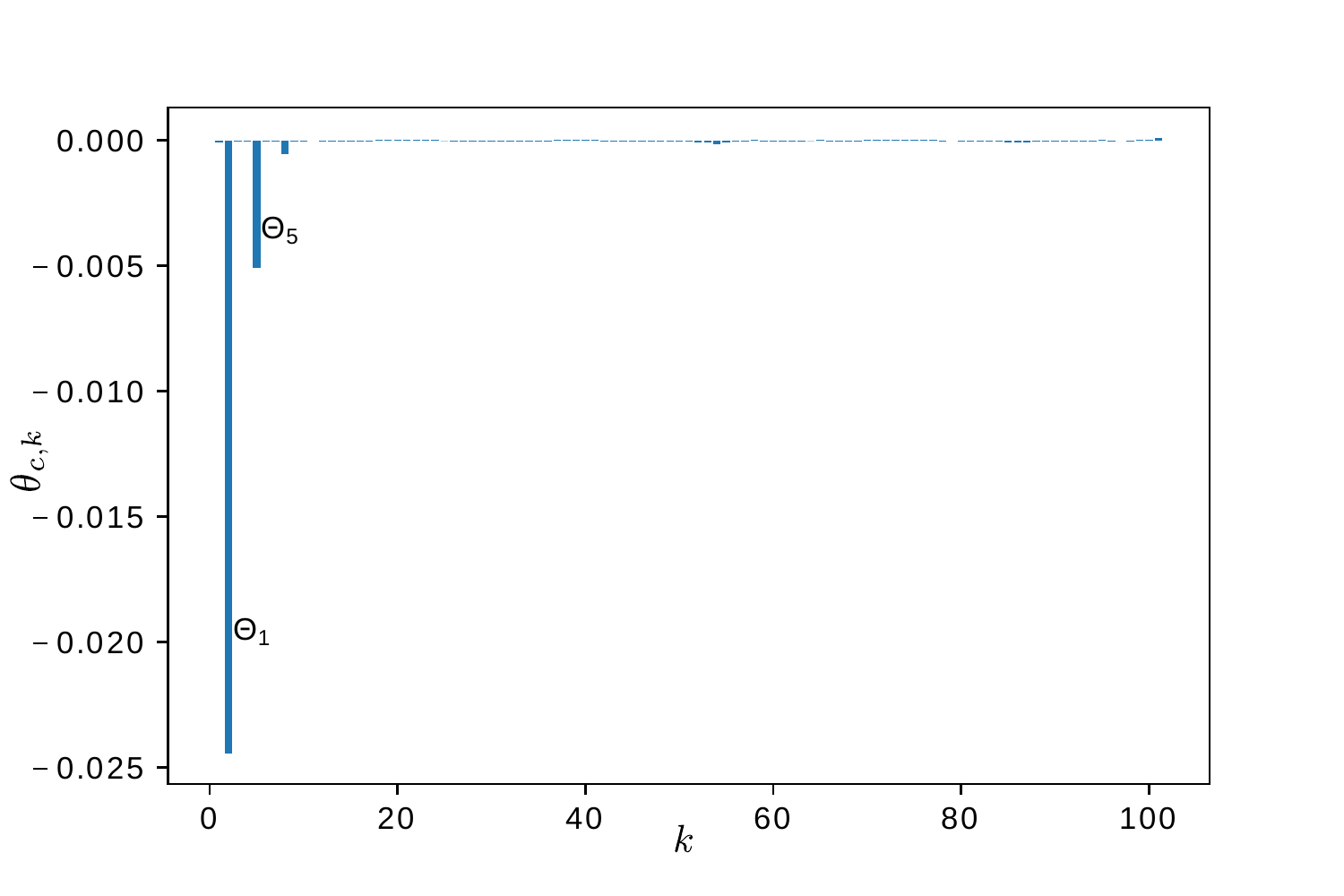}
 \caption{Posterior means of the inferred $\bt_c$ that parametrize the CG evolution law (\refeq{eq:cgpendulum}) for $n=16$ (left) and $n=64$ (right) training data.}
 \label{fig:pendulum_theta}
\end{figure}

Figure \ref{fig:pendulum_theta} indicates the posterior means of the inferred $\bt_c$ that parametrize the CG evolution law (\refeq{eq:cgpendulum}) for $n=16$ and $n=64$. Of the $101$ possible terms, only $2$ are activated due the ARD prior. 

\begin{figure}
 \includegraphics[width=0.49\textwidth]{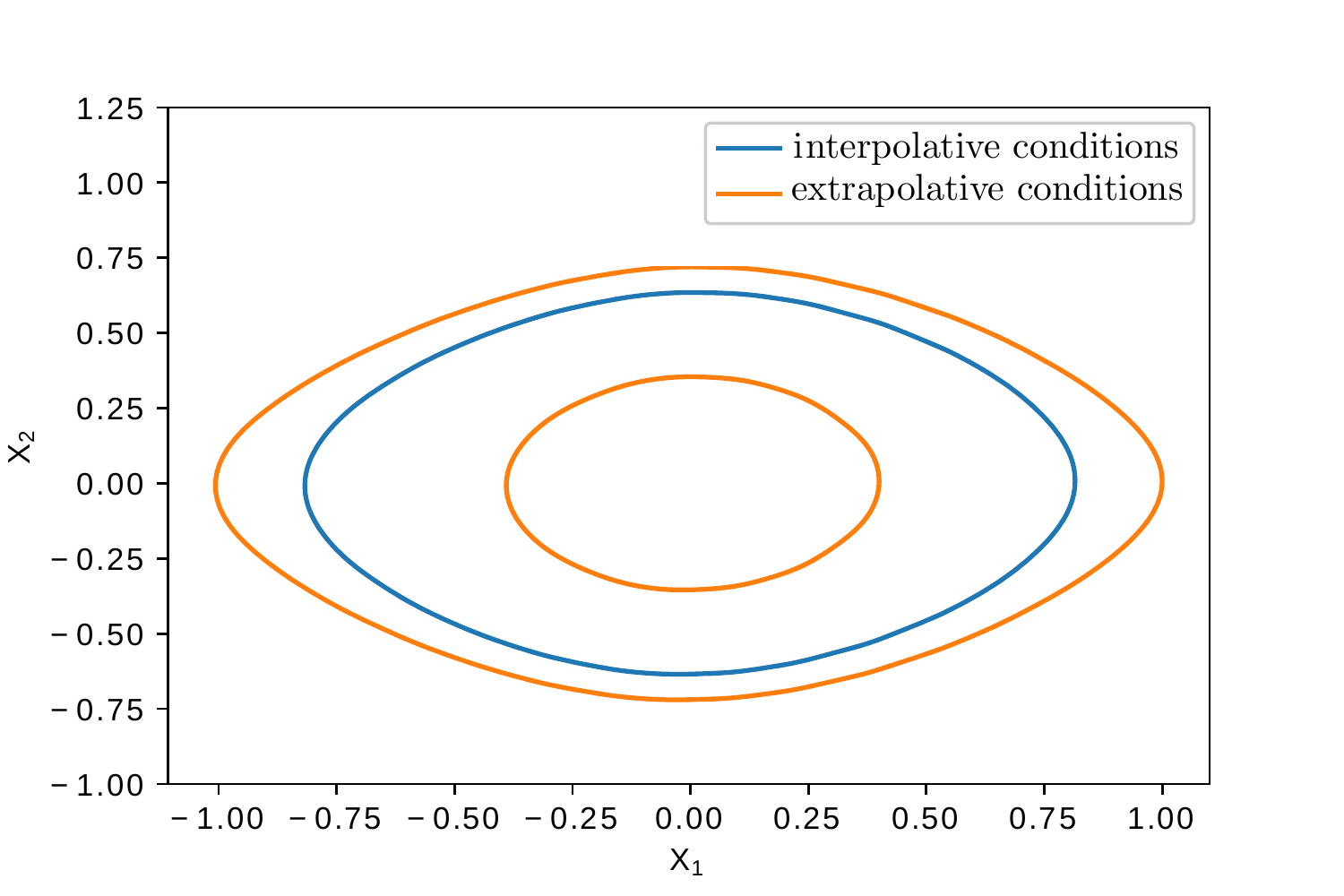} \includegraphics[width=0.49\textwidth]{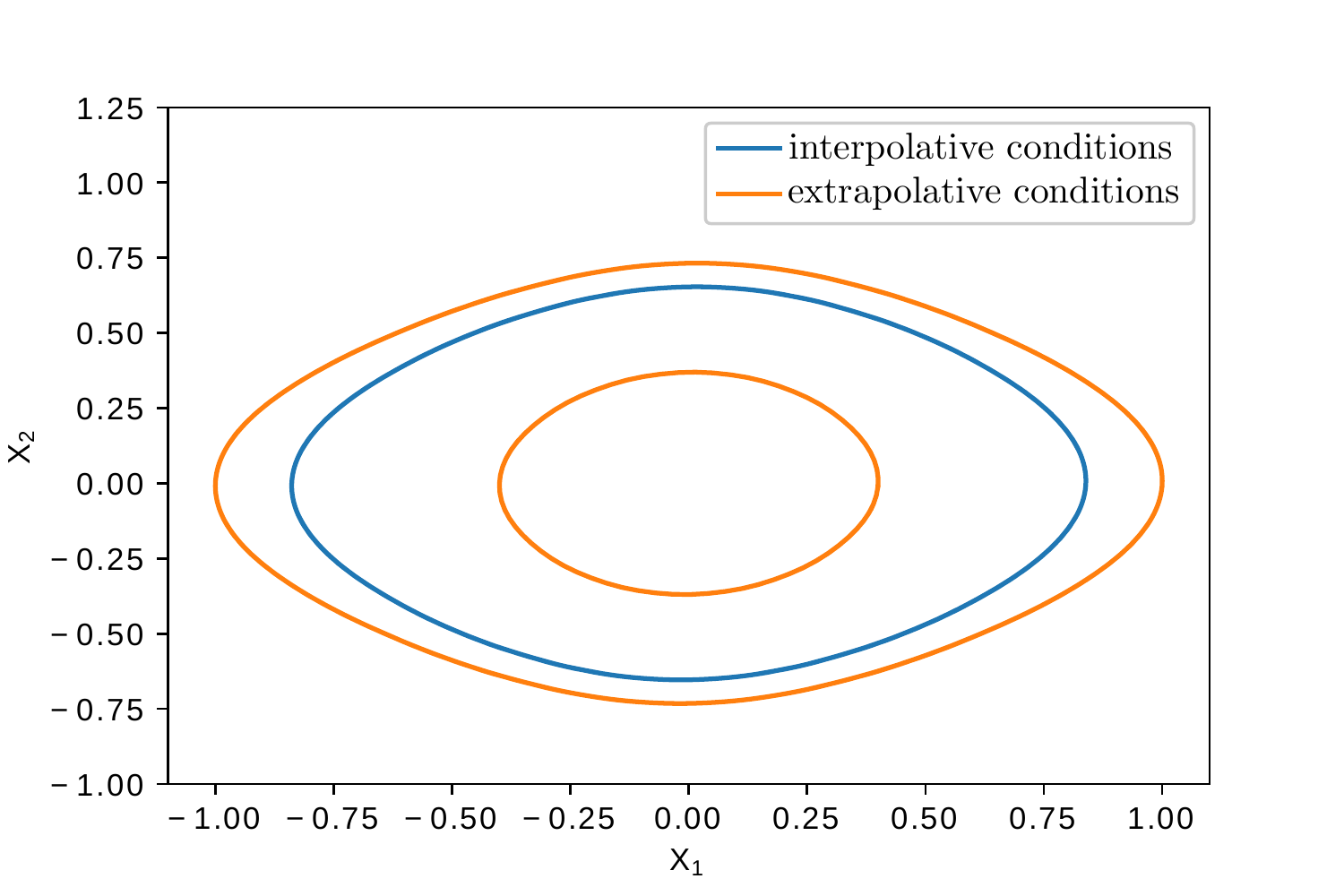}
 \caption{Comparison of trajectories  in state space $\bxx$ of the CG dynamics learned for $n=16$ (left)  and $n=64$ (right) training data.}
 \label{fig:pendulum_phase}
\end{figure}

\begin{figure}
 \includegraphics[width=0.49\textwidth]{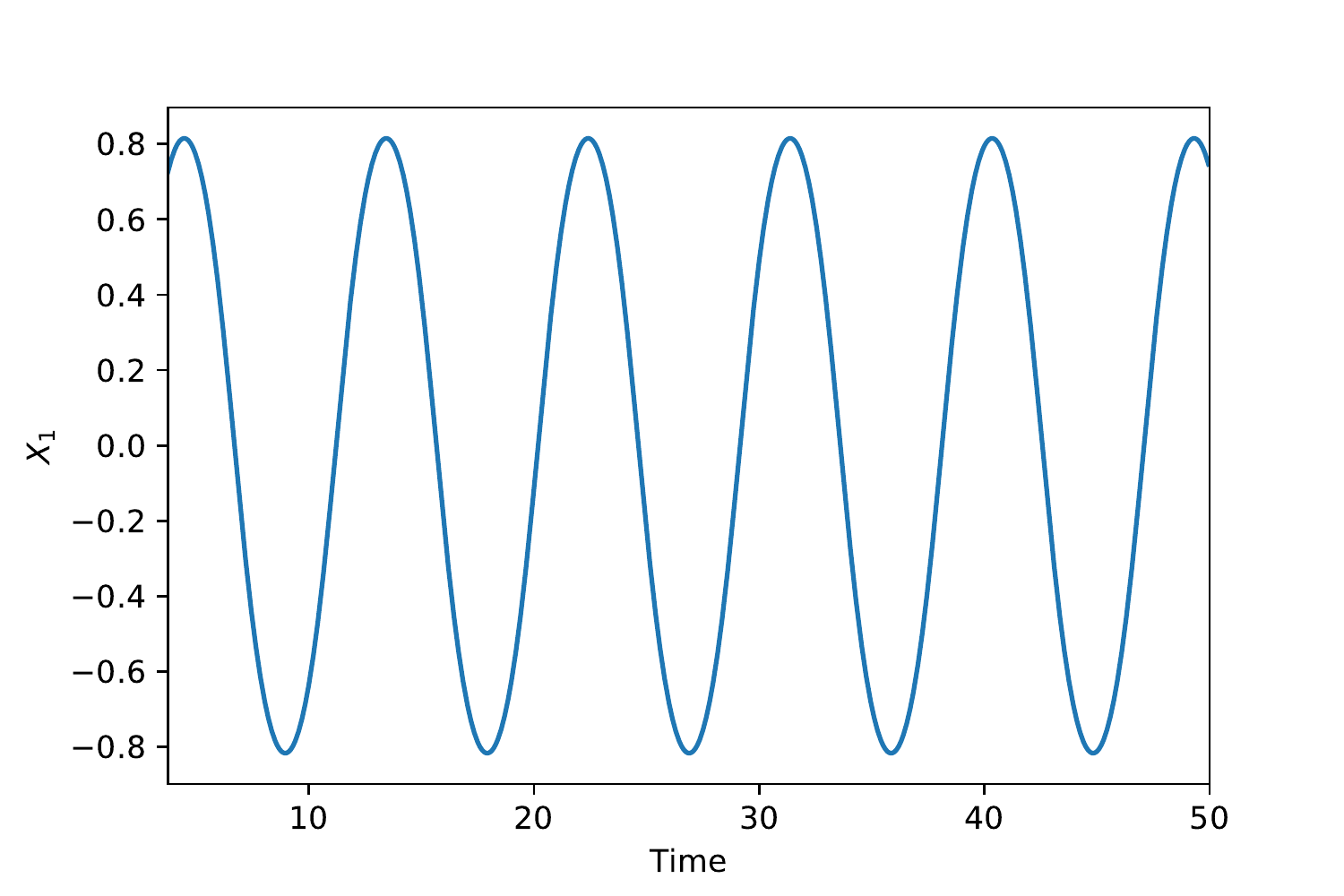} \includegraphics[width=0.49\textwidth]{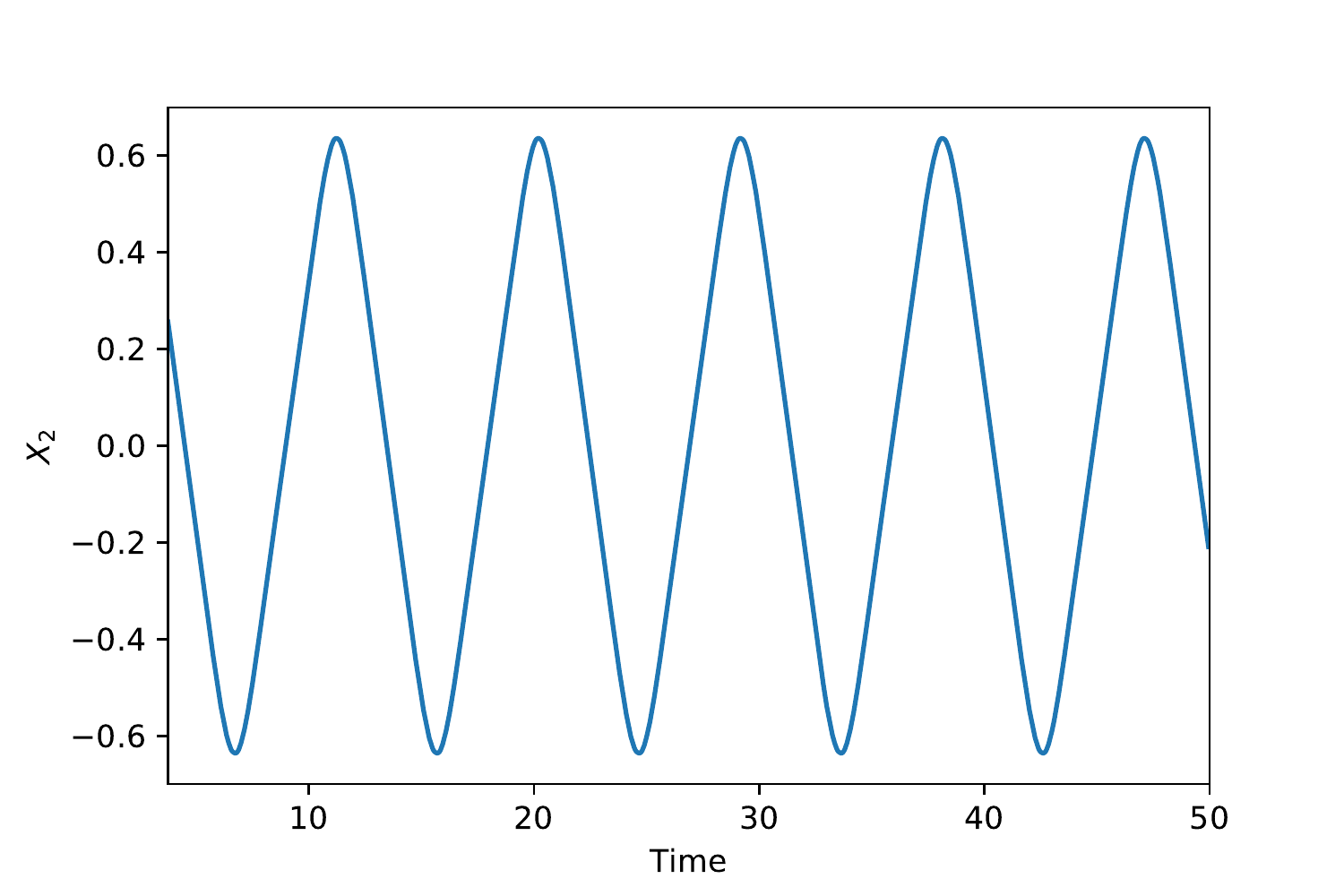}
 \caption{Predicted posterior mean of CG state variables $\bxx_t$}
 \label{fig:pendulum_xv}
\end{figure}

Figure \ref{fig:pendulum_phase} illustrates trajectories in the two-dimensional  CG state-space obtained with various initial conditions for the CG model identified with $n=16$ and $n=64$ data sequences. The blue curves correspond to ``interporlative"  settings i.e. to the CG states of an observed sequence of images, whereas the orange curves to ``extrapolative settings" i.e. to the CG states inferred by initializing the pendulum from an arbitrary position {\em not} contained in the training data. 
In Figure \ref{fig:pendulum_xv} the predicted evolution in time of both coarse-grained variables is shown. The periodic nature of the CG dynamics is obvious, even though the CG state variables implicitly identified do not correspond to the natural ones i.e. $y_t$ and $\dot{y}_t$. 

\begin{figure}
 \includegraphics[width=0.49\textwidth]{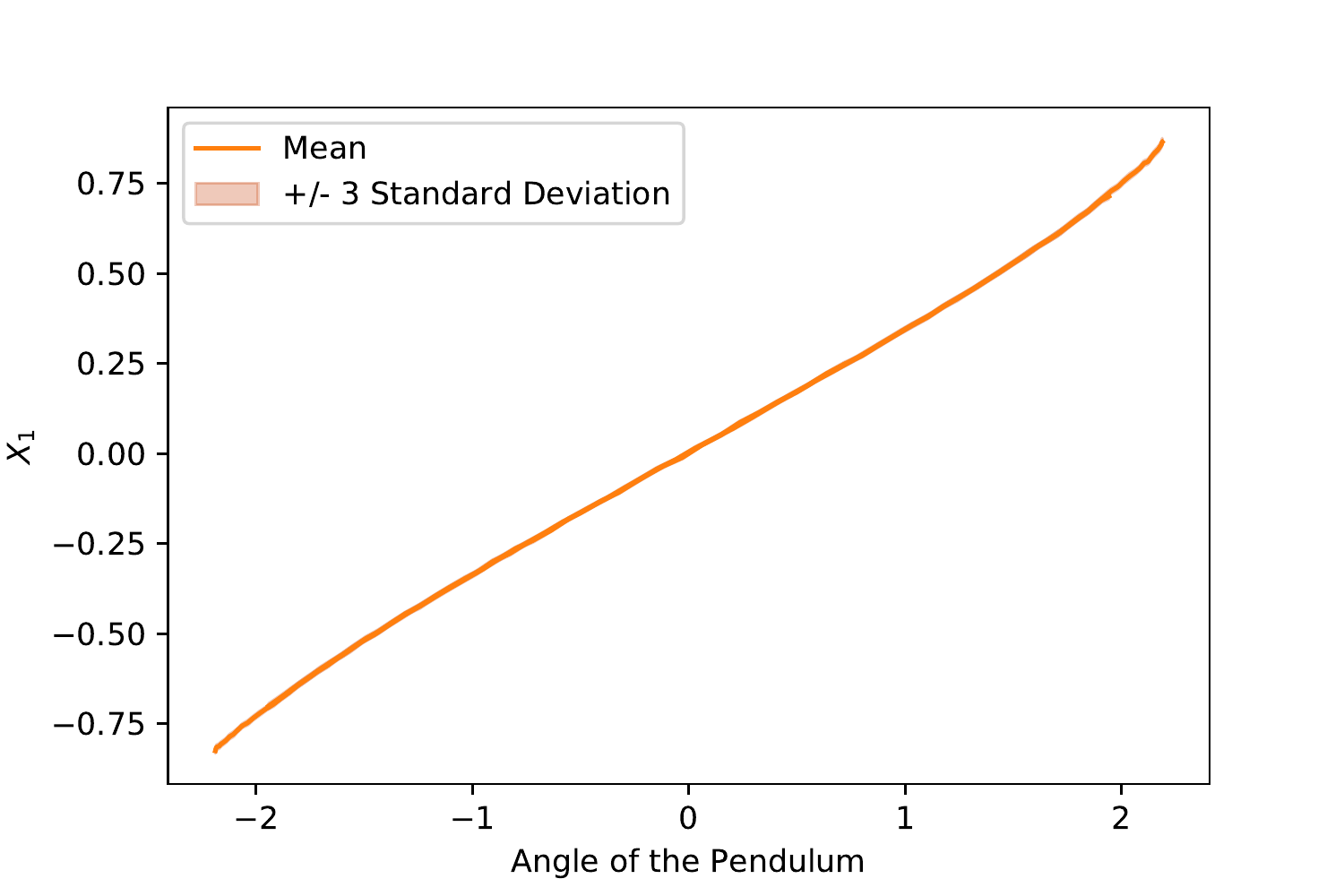} \includegraphics[width=0.49\textwidth]{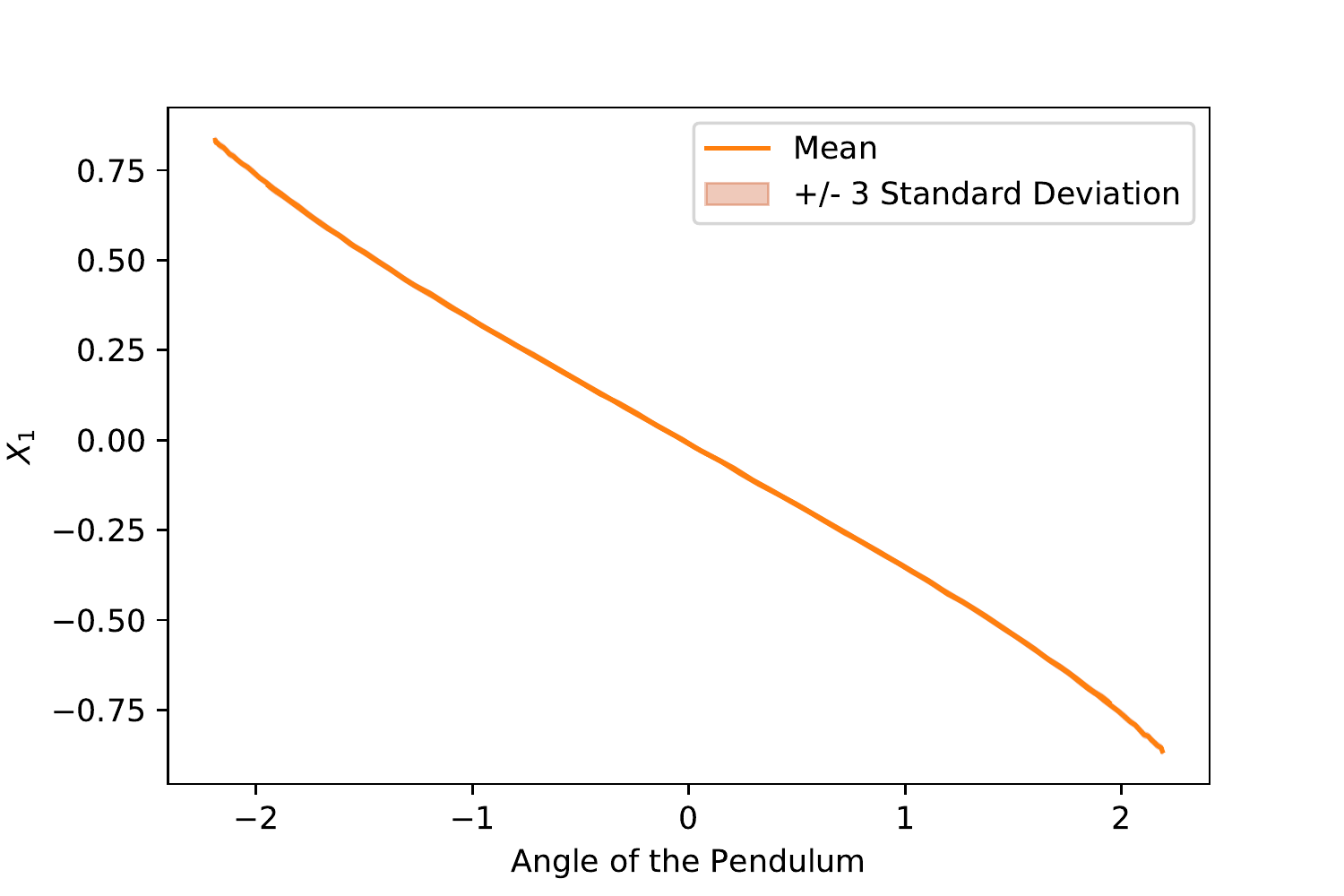}
 \caption{Mapping between the angle of the pendulum and the coarse-grained coordinates for 32 training data and 64 (right) training data.}
 \label{fig:pendulum_map}
\end{figure}

This can be seen in Figure \ref{fig:pendulum_map} where for data-sequences $\bx_{0:T \Delta t}^{(i)}$ (corresponding to the pendulum  at various positions i.e. angles $y_{0:T \Delta t}$), we compute from the approximate posterior $q_{\bp} (\bxx_{0:T\Delta t}^{(i)} | \bx_{0:T \Delta t}^{(i)})$ (\refeq{eq:qpendulum}) the mean of the corresponding CG states $\bxx_{0:T~\Delta t}^{(i)}$ as well as the (in this case negligible) standard deviation. For each time instant $l =0,1,\ldots,T$, we plot the pairs  of $y_{l \Delta t}$ and (the mean of) $X_{l\Delta t, 1}$ (i.e. the first of the CG variables identified) to show the relation between the two variables. While it is obvious from the scales that the first CG variable identified is {\em not} the angle,  it appears  to be isomorphic to $y$. The latter property persists for $n=64$ even though the sign of the relation has  been reversed. 
The difference between the first CG variable identified and the natural angle $y$ explains the difference between the CG evolution law identified (Figure \ref{fig:pendulum_theta}) and the reference one \refeq{eq:refpendulum}.

\begin{figure}
 \includegraphics[width=0.89\textwidth]{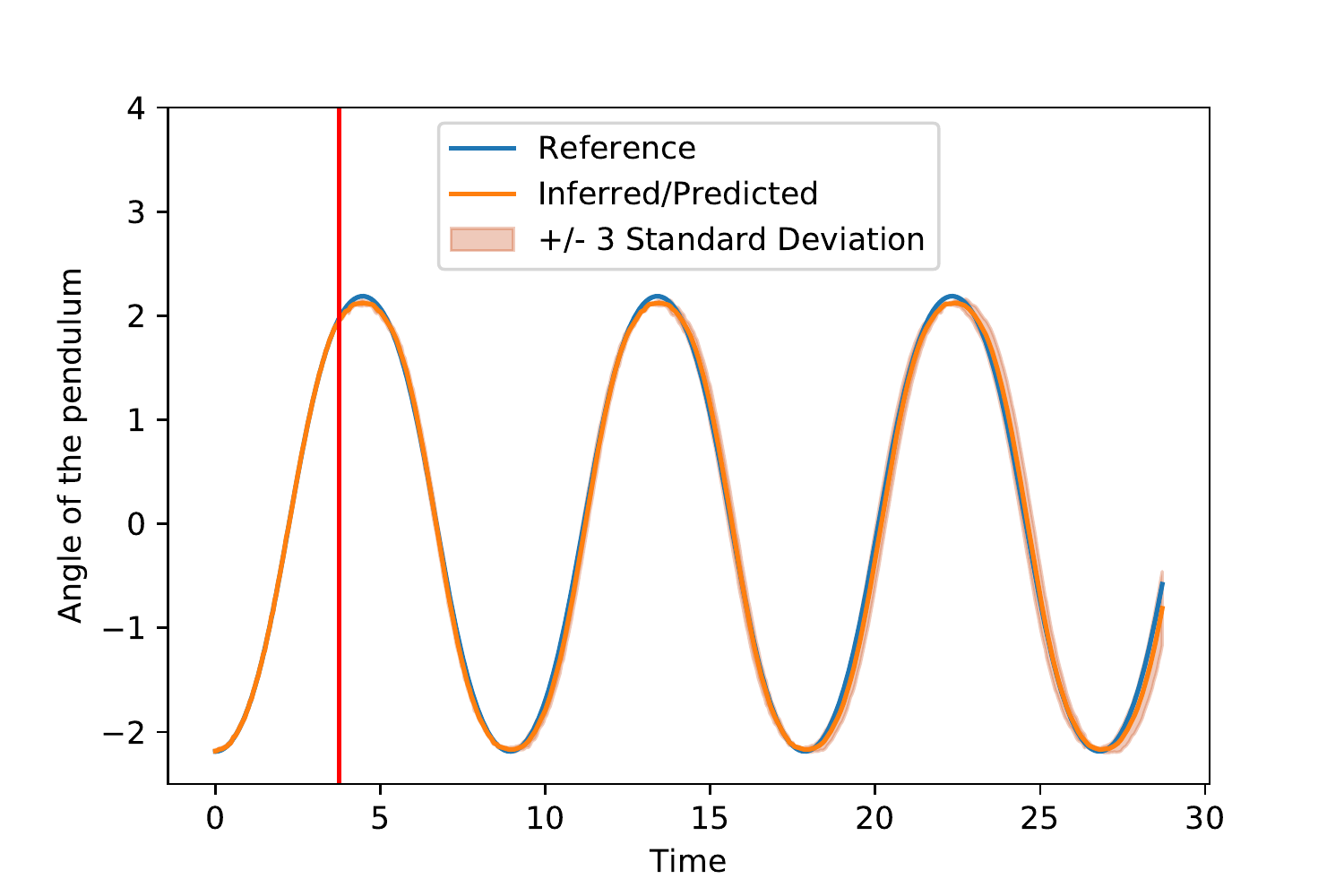} 
 \caption{Inferred/Predicted evolution of the center of mass of the pendulum. The vertical line separates the inferred states from the predictions}
 \label{fig:pendulum_center}
\end{figure}

Figure \ref{fig:pendulum_center} provides predictive estimates of the position of the center of mass in time. These were obtained by propagating the CG variables in time and for each time instant, sampling $p_{cf}$ for corresponding images $\bx$. From the latter, the center of mass was computed from the activated pixels i.e. the pixels with value 1. Naturally, predictive uncertainty arises due the stochasticity in the initial conditions of $\bxx$ as well as in $p_{cf}$. The latter is quantified by the standard deviation and plotted in Figure \ref{fig:pendulum_center}. As in the previous examples, the predictive uncertainty grows, albeit modestly, with time.  

\begin{figure}
\includegraphics[width=0.75\textwidth]{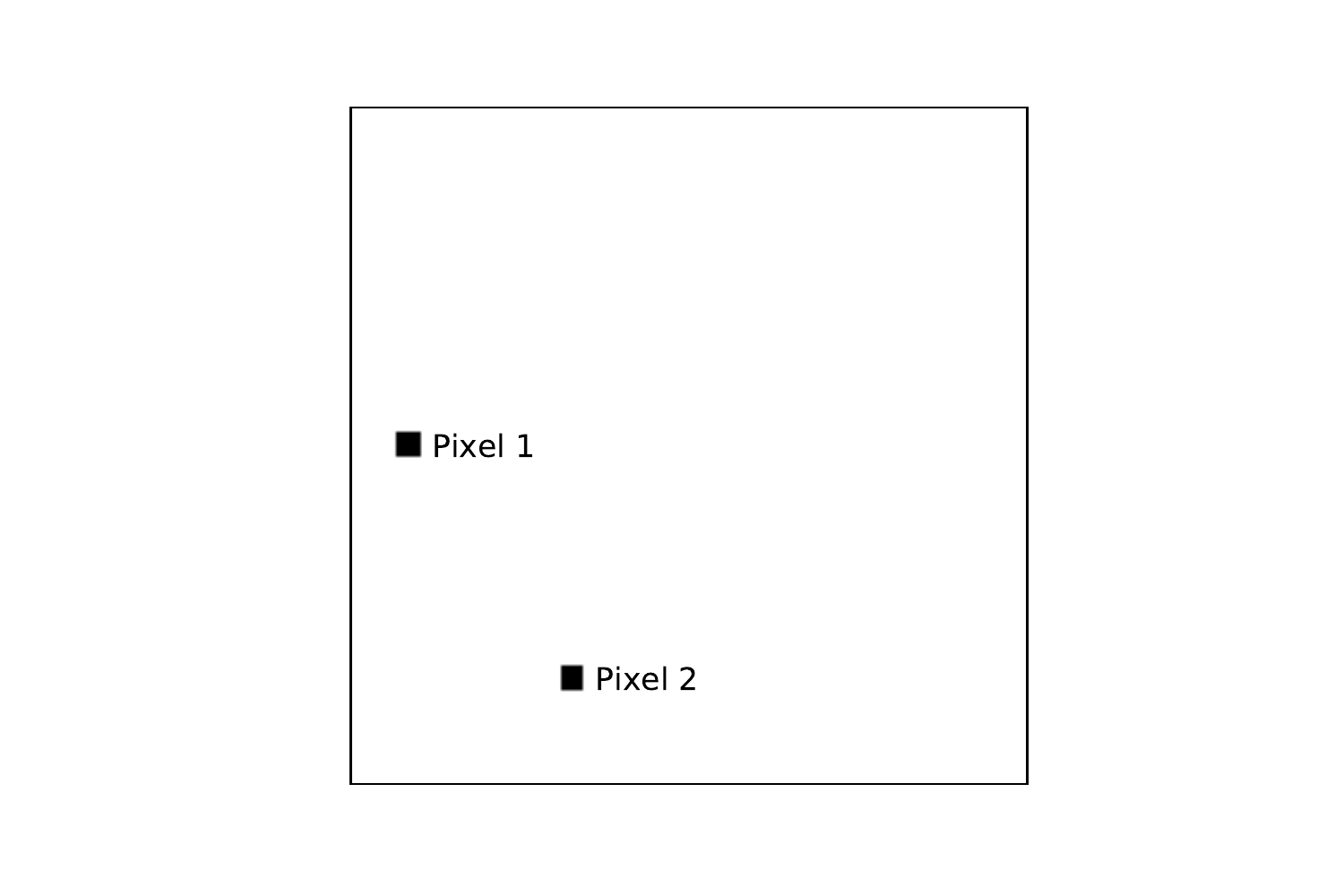} \\
 \includegraphics[width=0.49\textwidth]{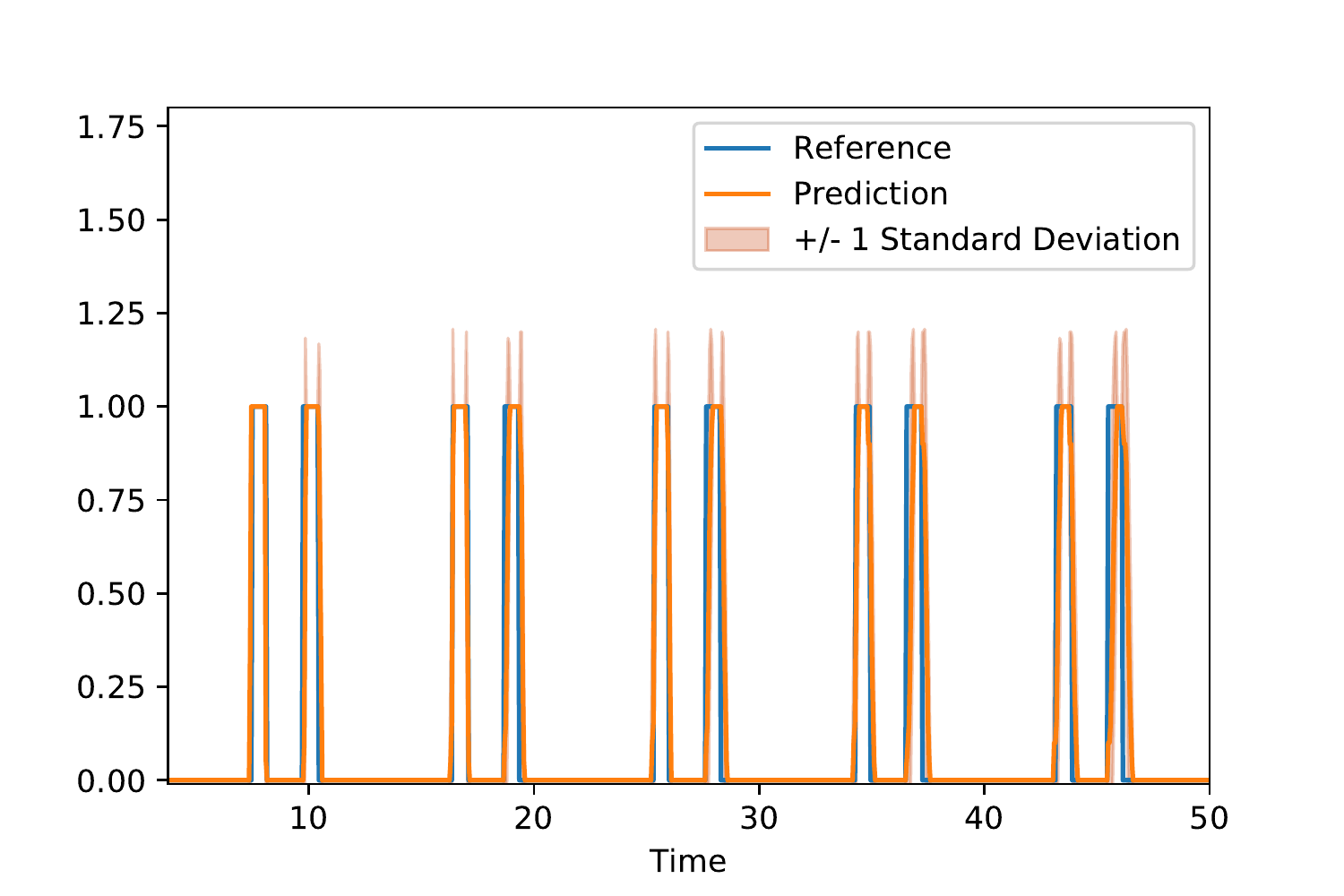}
 \includegraphics[width=0.49\textwidth]{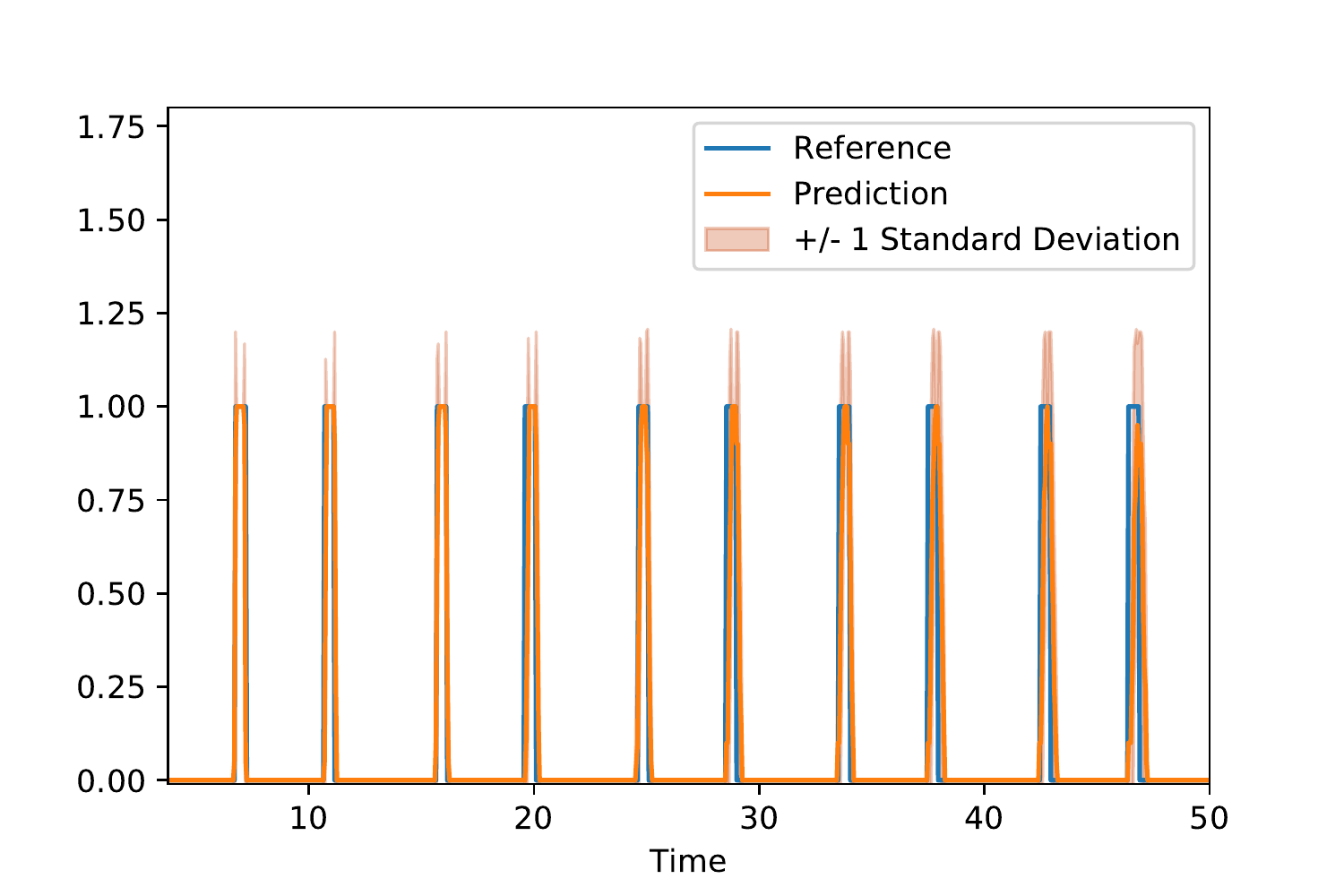}
 \caption{Predicted time history of a single pixel: Pixel 1 (left) and Pixel 2 (right)}
 \label{fig:pendulum_pixel}
\end{figure}

Figure \ref{fig:pendulum_pixel} depicts predictions in time for two pixels in the image. One can clearly distinguish the change-points i.e. when the pendulum crosses the pixel and its value is changed from 0 to 1 as well as the predictive uncertainty which is concentrated at those change-points. This demonstrates one of the strengths of our approach as due to the coarse-to-fine mapping the whole FG state is reconstructed and every observable can be computed together with the associated predictive uncertainty.

\begin{figure}[p]
        \centering
        \begin{subfigure}[b]{0.475\textwidth}
            \centering
            \includegraphics[trim=20 50 0 50 ,clip,width=0.8\textwidth]{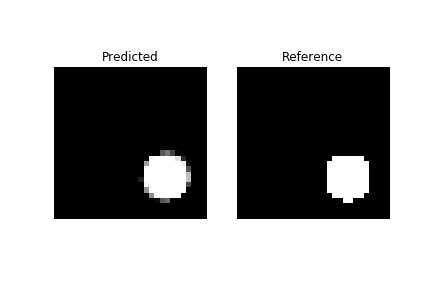}
            \caption[]%
            {{\small Time step 125}}    
        \end{subfigure}
        \hfill
        \begin{subfigure}[b]{0.475\textwidth}  
            \centering 
             \includegraphics[trim=20 50 0 50 ,clip,width=0.8\textwidth]{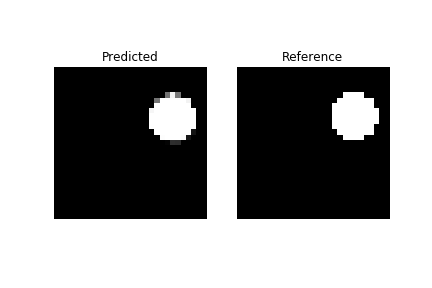}
            \caption[]%
            {{\small Time step 275}}    
        \end{subfigure}
        \vskip\baselineskip
        \begin{subfigure}[b]{0.475\textwidth}   
            \centering 
            \includegraphics[trim=20 50 0 50 ,clip,width=0.8\textwidth]{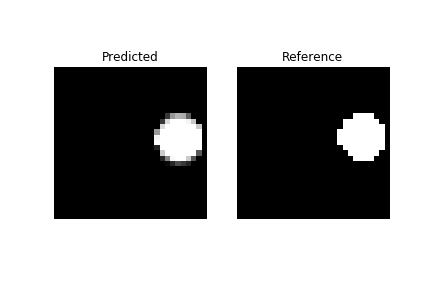}
            \caption[]%
            {{\small Time step 425}}    
        \end{subfigure}
        \quad
        \begin{subfigure}[b]{0.475\textwidth}   
            \centering 
            \includegraphics[trim=20 50 0 50 ,clip,width=0.8\textwidth]{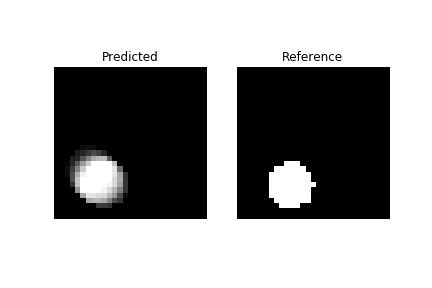}
            \caption[]%
            {{\small Time step 575}}    
        \end{subfigure}
        \vskip\baselineskip
        \begin{subfigure}[b]{0.475\textwidth}   
            \centering 
            \includegraphics[trim=20 50 0 50 ,clip,width=0.8\textwidth]{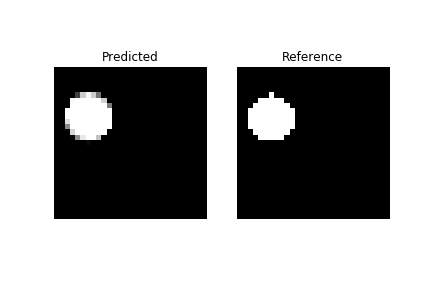}
            \caption[]%
            {{\small Time step 725}}    
        \end{subfigure}
        \quad
        \begin{subfigure}[b]{0.475\textwidth}   
            \centering 
            \includegraphics[trim=20 50 0 50 ,clip,width=0.8\textwidth]{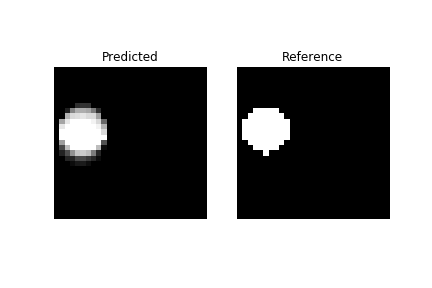}
            \caption[]%
            {{\small Time step 875}}    
        \end{subfigure}
        \caption[ Predicted solutions]
        {Predictive posterior means of images of the pendulum compared to the reference data}
        \label{fig:pendpred}
\end{figure}

Finally, Figure \ref{fig:pendpred} compares actual images obtained by the reference dynamics of the pendulum with the predictive posterior mean obtained by the CG model and $p_{cf}$ trained on the data. Even though these extend up to $875$ time-steps i.e. more than $11$ times longer than the time-window over which observations were available, they  match the reference quite accurately, a strong indication that the right CG variables and CG dynamics have been identified. An animation containing all frames can be found by following  \href{https://github.com/SebastianKaltenbach/PhysicalConstraints_ProbabilisticCG/blob/master/pendulum_animated.gif}{this \underline{link}}.

\review{
\subsubsection{Missing data}
}
\label{sec:Appendix}
\review{
The generative nature of the proposed model makes it highly suitable for handling missing FG data either in the form of partial observations of the FG state vector $\bx_t$ or observations over a portion/subset of the time-sequence considered. We investigate the latter case in this section but note that in both situations the only modification required is removing the likelihood terms corresponding to the missing data from \refeq{eq:fgdatalikelihood}. 
}

\begin{figure}[h]
    \centering
    \includegraphics[width=0.49\textwidth]{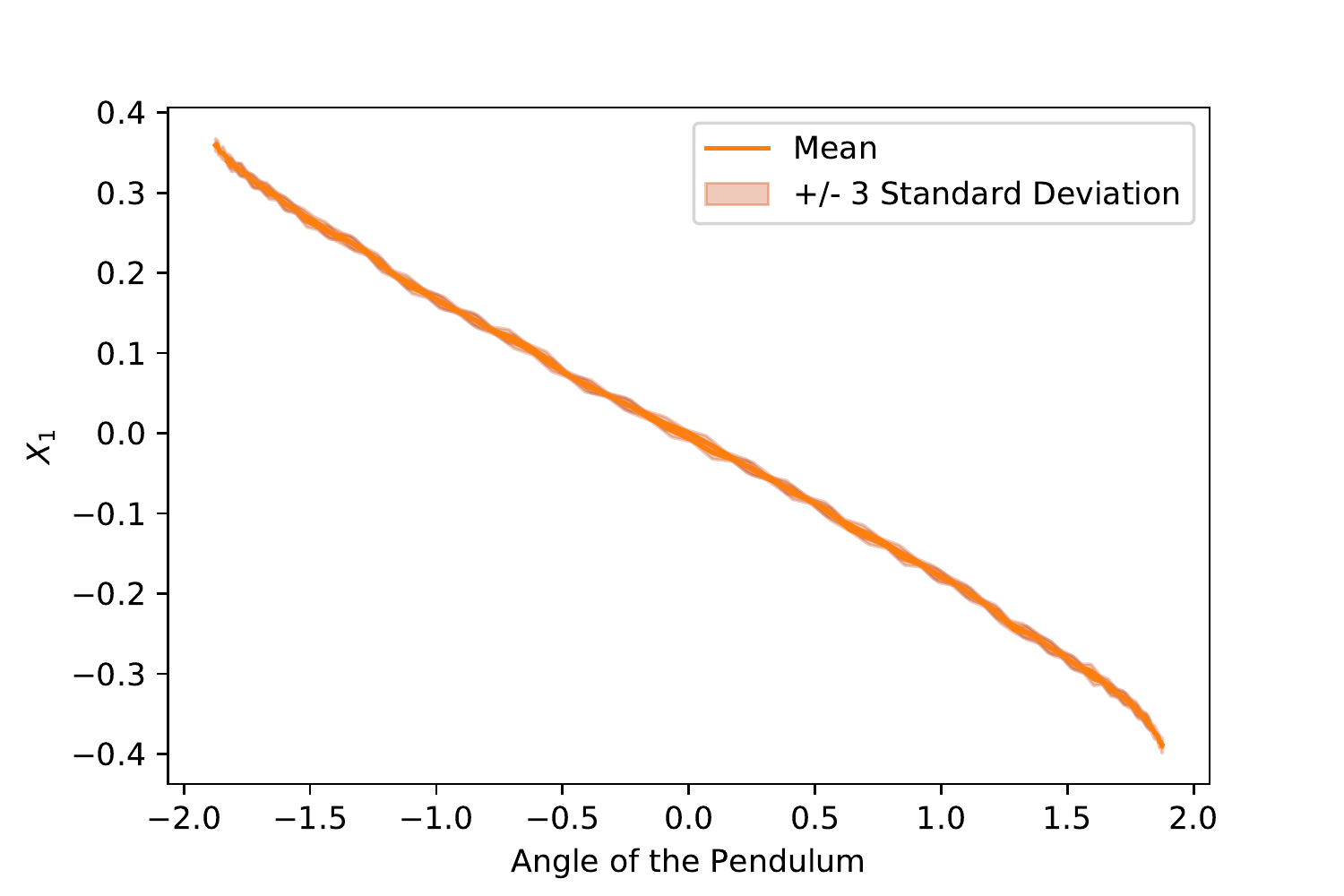}
    \includegraphics[width=0.49\textwidth]{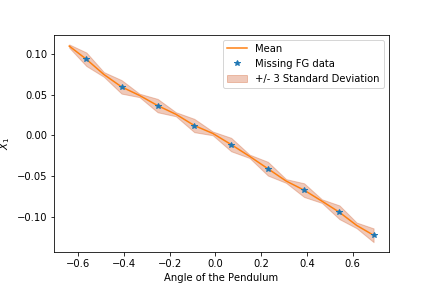}
    \caption{Effect of missing data on the CG variables.The figure on the right is zoomed-in to show the higher uncertainty associated with CG states with missing data}
    \label{fig:md_cg}
\end{figure}

\begin{figure}
    \centering
    \includegraphics[width=0.49\textwidth]{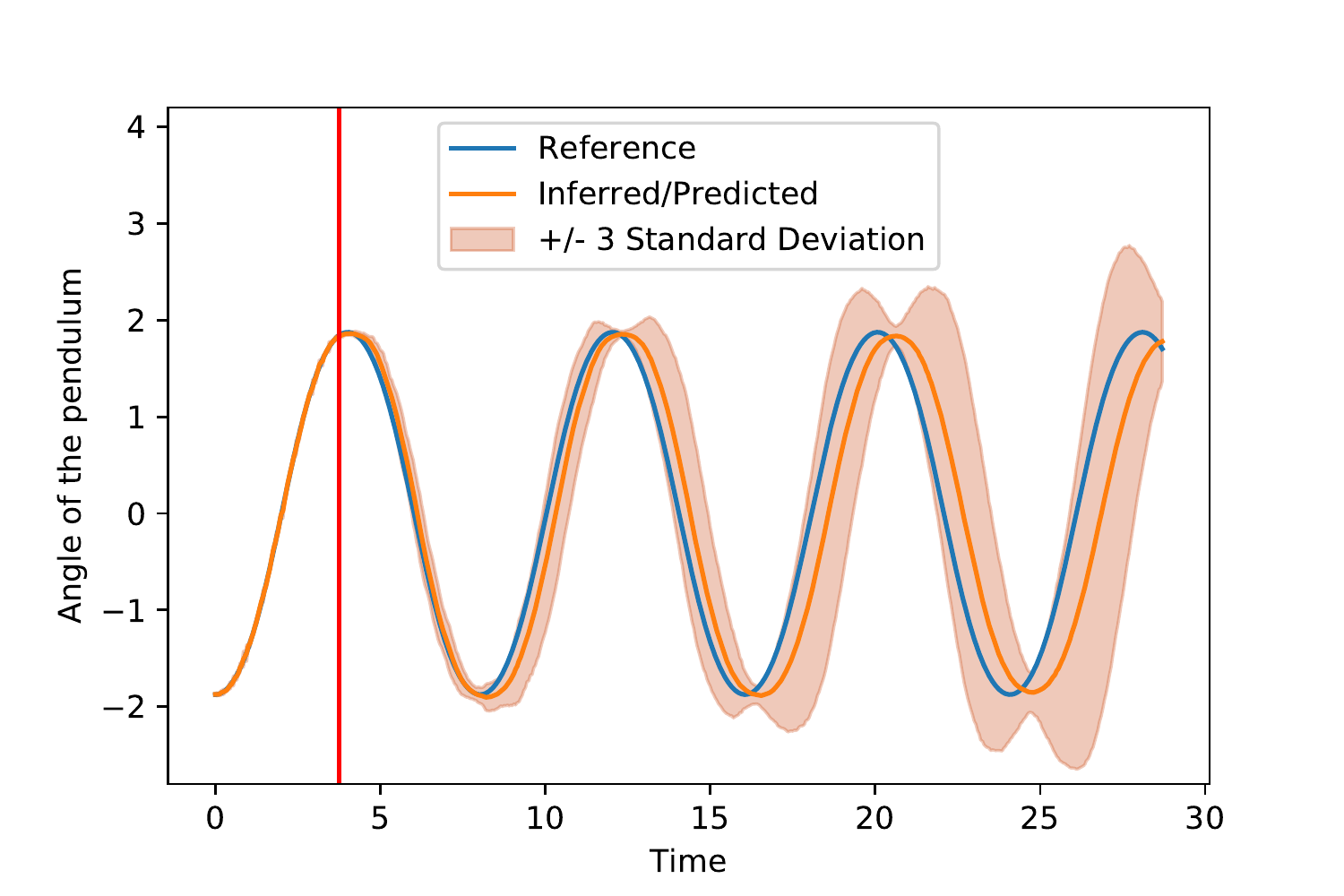}
    \caption{Inferred/Predicted evolution of the center of mass of the pendulum for the missing data case. The vertical line separates the inferred states from the predictions}
    \label{fig:md_uq}
\end{figure}

\review{
In particular, we investigated the performance of the model when every second FG state $\bx_t$ in the training sequences was {\em not} observed i.e. the FG observables consisted of $\{\bx^{(i)}_{0}, \bx^{(i)}_{2\Delta t}, \bx^{(i)}_{4\Delta t}, \ldots, \bx^{(i)}_{T \Delta t}\}$ for each data sequence $i$ (where $T=74$ as before). As one would expect, fewer observations lead to higher inferential uncertainties as seen when comparing Figure \ref{fig:pendulum_map} (fully observed case) with Figure \ref{fig:md_cg} (partially observed case). More importantly, fewer observations lead to higher predictive uncertainty as seen when comparing the predictions for the center of pendulum in Figure \ref{fig:pendulum_center} (fully observed case) with Figure \ref{fig:md_uq} (partially observed case).
}
%

%% file: conclusions.tex
We proposed a probabilistic generative model for the automated discovery of coarse-grained   variables and  dynamics based on fine-grained simulation data. The FG simulation data are   augmented in a fully Bayesian fashion by virtual observables that enable the incorporation of physical constraints at the CG level that appear in the form of equalities. These could be residuals of the CG evolution law or more importantly conservation laws that are available when CG variables have physical meaning. This is particularly important in the context of physical modeling as in many cases such domain knowledge is a priori available and its inclusion can, not only reduce the amount of training data, but endow the CG model learned with the necessary features that would allow it to provide accurate predictions in out-of-distribution settings. 
Our approach learns simultaneously a coarse-to-fine mapping and   an evolution law for the coarse-grained dynamics by employing probabilistic inference tools for the latent variables and model parameters. The use of deep neural nets for the former component can endow great expressiveness and flexibility. The concept of sparsity, which is invoked in learning  CG dynamics from a large vocabulary of right-hand-side terms, is readily incorporated using sparsity-inducing Bayesian priors without any hyperparameter tuning. \review{Furthermore, appropriate priors can promote the discovery of slow-varying CG variables which better capture the macroscopic features of the system.}
As  a result of the aforementioned characteristics, the framework can learn from {\em Small Data} \review{(i.e. shorter and fewer FG time-sequences)} which is a crucial advantage in multiscale models where the simulation of the FG dynamics is expensive and slow in exploring the state-space.  
The model proposed  was successfully tested on  coarse-graining tasks from different areas. In all three examples, the method performed well  under interpolative, and more importantly under extrapolative settings i.e. in cases where initial conditions different from the ones seen during training, are prescribed. \review{Partial or incomplete FG observations can readily be  handled due to its generative nature.} Moreover, as it is able to reconstruct the entire FG state vector at any future time instant, it is capable of producing predictions of any FG observable of interest as well as quantify the associated predictive uncertainty.

There exists various possibilities to extend the proposed framework, both methodologically as well as in terms of applications. In the latter case and apart from using it for predictive purposes, the CG model learned  could also be employed in optimization and control applications. 
On the methodological front an obvious extension would be to account for the virtual observables at future time-instants as well. This would ensure their enforcement by future CG states but would unavoidably complicate their simulation as a probabilistic inference scheme would need to be employed in order to draw samples.

\review{ Another important question pertains to the stability of the CG dynamics identified \citep{pan2020physics}. This is not currently guaranteed in the discretized nor in the continuous version. This could potentially be achieved   by  an a-priori parametrization of the CG dynamics in a way that guarantees stability which could  in turn reduce the expressivity of the model.
}
\review{Finally, we note that, in our opinion, the most difficult question in coarse-graining multiscale systems, is finding the number of CG state variables that are needed. In physics problems, very often one has an idea of which variables would be suitable either based on the analysis-objectives and/or  physical insight. Almost never though does one have a guarantee that these variables are sufficient. Assuming they are, the problem then reduces to finding the appropriate closures (i.e. right-hand sides in the CG dynamics) which is the problem we try to address in this paper.
The discovery of additional, potentially non-physical CG state variables, would require additional advances for which we believe the ELBO, i.e. the (approximate)  model evidence, could serve as the guiding objective.
}

%% file: main.bbl
\begin{thebibliography}{68}
\providecommand{\natexlab}[1]{#1}
\providecommand{\url}[1]{\texttt{#1}}
\expandafter\ifx\csname urlstyle\endcsname\relax
  \providecommand{\doi}[1]{doi: #1}\else
  \providecommand{\doi}{doi: \begingroup \urlstyle{rm}\Url}\fi

\bibitem[Givon et~al.(2004)Givon, Kupferman, and Stuart]{givon_extracting_2004}
D.~Givon, R.~Kupferman, and A.~Stuart.
\newblock Extracting {Macroscopic} {Dynamics}: {Model} {Problems} and
  {Algorithms}.
\newblock \emph{Nonlinearity}, 2004.

\bibitem[Bialek(2012)]{bialek_biophysics:_2012}
William Bialek.
\newblock \emph{Biophysics: {Searching} for {Principles}}.
\newblock Princeton University Press, October 2012.
\newblock ISBN 978-0-691-13891-6.

\bibitem[Alber et~al.(2019)Alber, Buganza~Tepole, Cannon, De, Dura-Bernal,
  Garikipati, Karniadakis, Lytton, Perdikaris, Petzold, and
  Kuhl]{alber_integrating_2019}
Mark Alber, Adrian Buganza~Tepole, William~R. Cannon, Suvranu De, Salvador
  Dura-Bernal, Krishna Garikipati, George Karniadakis, William~W. Lytton, Paris
  Perdikaris, Linda Petzold, and Ellen Kuhl.
\newblock Integrating machine learning and multiscale modeling-perspectives,
  challenges, and opportunities in the biological, biomedical, and behavioral
  sciences.
\newblock \emph{NPJ digital medicine}, 2:\penalty0 115, 2019.
\newblock ISSN 2398-6352.
\newblock \doi{10.1038/s41746-019-0193-y}.

\bibitem[Ghahramani(2015)]{ghahramani_probabilistic_2015}
Zoubin Ghahramani.
\newblock Probabilistic machine learning and artificial intelligence.
\newblock \emph{Nature}, 521\penalty0 (7553):\penalty0 452--459, May 2015.
\newblock ISSN 0028-0836.
\newblock \doi{10.1038/nature14541}.
\newblock URL
  \url{http://www.nature.com/nature/journal/v521/n7553/full/nature14541.html}.

\bibitem[LeCun et~al.(2015)LeCun, Bengio, and Hinton]{lecun2015deep}
Yann LeCun, Yoshua Bengio, and Geoffrey Hinton.
\newblock Deep learning.
\newblock \emph{nature}, 521\penalty0 (7553):\penalty0 436--444, 2015.

\bibitem[Koutsourelakis et~al.(2016)Koutsourelakis, Zabaras, and
  Girolami]{koutsourelakis2016big}
Phaedon-Stelios Koutsourelakis, Nicholas Zabaras, and Michele Girolami.
\newblock Big data and predictive computational modeling.
\newblock \emph{Journal of Computational Physics}, 321:\penalty0 1252--1254,
  2016.

\bibitem[Marcus and Davis(2019)]{marcus_rebooting_2019}
Gary Marcus and Ernest Davis.
\newblock \emph{Rebooting {AI}: {Building} {Artificial} {Intelligence} {We}
  {Can} {Trust}}.
\newblock Pantheon, September 2019.

\bibitem[Stinis et~al.(2019)Stinis, Hagge, Tartakovsky, and
  Yeung]{stinis2019enforcing}
Panos Stinis, Tobias Hagge, Alexandre~M Tartakovsky, and Enoch Yeung.
\newblock Enforcing constraints for interpolation and extrapolation in
  generative adversarial networks.
\newblock \emph{Journal of Computational Physics}, 397:\penalty0 108844, 2019.

\bibitem[Koutsourelakis and Bilionis(2011)]{koutsourelakis_scalable_2011}
Phaedon-Stelios Koutsourelakis and Elias Bilionis.
\newblock Scalable {Bayesian} {Reduced}-{Order} {Models} for {Simulating}
  {High}-{Dimensional} {Multiscale} {Dynamical} {Systems}.
\newblock \emph{Multiscale Modeling \& Simulation}, 9\penalty0 (1):\penalty0
  449--485, 2011.
\newblock \doi{10.1137/100783790}.

\bibitem[Kevrekidis et~al.(2003)Kevrekidis, Gear, Hyman, Kevrekidis, Runborg,
  and Theodoropoulos]{kevrekidis_equation-free_2003}
IG~Kevrekidis, CW~Gear, JM~Hyman, PG~Kevrekidis, O~Runborg, and
  K~Theodoropoulos.
\newblock Equation-free multiscale computation: enabling microscopic simulators
  to perform system-level tasks.
\newblock \emph{Communications in Mathematical Sciences}, 1\penalty0
  (4):\penalty0 715--762, 2003.

\bibitem[Schmid(2010)]{schmid_dynamic_2010}
Peter~J. Schmid.
\newblock Dynamic mode decomposition of numerical and experimental data.
\newblock \emph{Journal of Fluid Mechanics}, 656:\penalty0 5--28, August 2010.
\newblock ISSN 1469-7645, 0022-1120.
\newblock \doi{10.1017/S0022112010001217}.
\newblock URL \url{https://www.cambridge.org/core/journals/journal-of-fluid-
  mechanics/article/dynamic-mode-decomposition-of-numerical-
  and-experimental-data/AA4C763B525515AD4521A6CC5E10DBD4}.

\bibitem[Williams et~al.(2015)Williams, Kevrekidis, and
  Rowley]{williams_datadriven_2015}
Matthew~O. Williams, Ioannis~G. Kevrekidis, and Clarence~W. Rowley.
\newblock A {Data}–{Driven} {Approximation} of the {Koopman} {Operator}:
  {Extending} {Dynamic} {Mode} {Decomposition}.
\newblock \emph{Journal of Nonlinear Science}, 25:\penalty0 1307--1346, 2015.
\newblock \doi{10.1007/s00332-015-9258-5}.

\bibitem[Wu and No\'e(2017)]{wu_variational_2017}
Hao Wu and Frank No\'e.
\newblock Variational approach for learning {Markov} processes from time series
  data.
\newblock \emph{arXiv:1707.04659 [math, stat]}, July 2017.
\newblock URL \url{http://arxiv.org/abs/1707.04659}.

\bibitem[Froyland et~al.(2014)Froyland, Gottwald, and
  Hammerlindl]{froyland_computational_2014}
Gary. Froyland, Georg~A. Gottwald, and Andy. Hammerlindl.
\newblock A {Computational} {Method} to {Extract} {Macroscopic} {Variables} and
  {Their} {Dynamics} in {Multiscale} {Systems}.
\newblock \emph{SIAM Journal on Applied Dynamical Systems}, 13\penalty0
  (4):\penalty0 1816--1846, January 2014.
\newblock \doi{10.1137/130943637}.
\newblock URL \url{https://epubs.siam.org/doi/abs/10.1137/130943637}.

\bibitem[Felsberger and
  Koutsourelakis(2019)]{felsberger_physics-constrained_2019}
L~Felsberger and PS~Koutsourelakis.
\newblock Physics-constrained, data-driven discovery of coarse-grained
  dynamics.
\newblock \emph{Communications in Computational Physics}, 25\penalty0
  (5):\penalty0 1259--1301, 2019.
\newblock \doi{10.4208/cicp.OA-2018-0174}.

\bibitem[Sch{\"o}berl et~al.(2017)Sch{\"o}berl, Zabaras, and
  Koutsourelakis]{schoberl2017predictive}
Markus Sch{\"o}berl, Nicholas Zabaras, and Phaedon-Stelios Koutsourelakis.
\newblock Predictive coarse-graining.
\newblock \emph{Journal of Computational Physics}, 333:\penalty0 49--77, 2017.

\bibitem[Raissi et~al.(2017)Raissi, Perdikaris, and
  Karniadakis]{raissi2017physics}
Maziar Raissi, Paris Perdikaris, and George~Em Karniadakis.
\newblock Physics informed deep learning (part i): Data-driven solutions of
  nonlinear partial differential equations.
\newblock \emph{arXiv preprint arXiv:1711.10561}, 2017.

\bibitem[Raissi et~al.(2019)Raissi, Perdikaris, and
  Karniadakis]{raissi2019physics}
Maziar Raissi, Paris Perdikaris, and George~E Karniadakis.
\newblock Physics-informed neural networks: A deep learning framework for
  solving forward and inverse problems involving nonlinear partial differential
  equations.
\newblock \emph{Journal of Computational Physics}, 378:\penalty0 686--707,
  2019.

\bibitem[Yang and Perdikaris(2019)]{yang2019conditional}
Yibo Yang and Paris Perdikaris.
\newblock Conditional deep surrogate models for stochastic, high-dimensional,
  and multi-fidelity systems.
\newblock \emph{Computational Mechanics}, pages 1--18, 2019.

\bibitem[Mardt et~al.(2018)Mardt, Pasquali, Wu, and No\'e]{mardt_vampnets_2018}
Andreas Mardt, Luca Pasquali, Hao Wu, and Frank No\'e.
\newblock {VAMPnets} for deep learning of molecular kinetics.
\newblock \emph{Nature Communications}, 9, January 2018.
\newblock ISSN 2041-1723.
\newblock \doi{10.1038/s41467-017-02388-1}.
\newblock URL \url{https://www.ncbi.nlm.nih.gov/pmc/articles/PMC5750224/}.

\bibitem[Wu et~al.(2018)Wu, Mardt, Pasquali, and Noe]{wu2018deep}
Hao Wu, Andreas Mardt, Luca Pasquali, and Frank Noe.
\newblock Deep generative markov state models.
\newblock In \emph{Advances in Neural Information Processing Systems}, pages
  3975--3984, 2018.

\bibitem[Duncker et~al.(2019)Duncker, Bohner, Boussard, and
  Sahani]{duncker_learning_2019}
Lea Duncker, Gergo Bohner, Julien Boussard, and Maneesh Sahani.
\newblock Learning interpretable continuous-time models of latent stochastic
  dynamical systems.
\newblock \emph{arXiv preprint arXiv:1902.04420}, 2019.

\bibitem[Grigo and Koutsourelakis(2019{\natexlab{a}})]{grigo2019physics}
Constantin Grigo and Phaedon-Stelios Koutsourelakis.
\newblock A physics-aware, probabilistic machine learning framework for
  coarse-graining high-dimensional systems in the small data regime.
\newblock \emph{arXiv preprint arXiv:1902.03968}, 2019{\natexlab{a}}.

\bibitem[Pantazis and Tsamardinos(2019)]{pantazis_unified_2019}
Yannis Pantazis and Ioannis Tsamardinos.
\newblock A unified approach for sparse dynamical system inference from
  temporal measurements.
\newblock \emph{Bioinformatics}, 35\penalty0 (18):\penalty0 3387--3396,
  September 2019.
\newblock ISSN 1367-4803.
\newblock \doi{10.1093/bioinformatics/btz065}.
\newblock URL
  \url{https://academic.oup.com/bioinformatics/article/35/18/3387/5305020}.

\bibitem[Brunton et~al.(2016{\natexlab{a}})Brunton, Proctor, and
  Kutz]{brunton2016discovering}
Steven~L Brunton, Joshua~L Proctor, and J~Nathan Kutz.
\newblock Discovering governing equations from data by sparse identification of
  nonlinear dynamical systems.
\newblock \emph{Proceedings of the National Academy of Sciences}, 113\penalty0
  (15):\penalty0 3932--3937, 2016{\natexlab{a}}.

\bibitem[Kaiser et~al.(2018)Kaiser, Kutz, and Brunton]{kaiser2018sparse}
Eurika Kaiser, J~Nathan Kutz, and Steven~L Brunton.
\newblock Sparse identification of nonlinear dynamics for model predictive
  control in the low-data limit.
\newblock \emph{Proceedings of the Royal Society A}, 474\penalty0
  (2219):\penalty0 20180335, 2018.

\bibitem[Champion et~al.(2019)Champion, Lusch, Kutz, and
  Brunton]{champion2019data}
Kathleen Champion, Bethany Lusch, J~Nathan Kutz, and Steven~L Brunton.
\newblock Data-driven discovery of coordinates and governing equations.
\newblock \emph{arXiv preprint arXiv:1904.02107}, 2019.

\bibitem[Ohkubo(2011)]{ohkubo_nonparametric_2011}
Jun Ohkubo.
\newblock Nonparametric model reconstruction for stochastic differential
  equations from discretely observed time-series data.
\newblock \emph{Physical Review E}, 84\penalty0 (6):\penalty0 066702, December
  2011.
\newblock \doi{10.1103/PhysRevE.84.066702}.
\newblock URL \url{https://link.aps.org/doi/10.1103/PhysRevE.84.066702}.

\bibitem[Klus et~al.(2018)Klus, Nüske, Koltai, Wu, Kevrekidis, Schütte, and
  Noé]{klus_data-driven_2018}
Stefan Klus, Feliks Nüske, Péter Koltai, Hao Wu, Ioannis Kevrekidis, Christof
  Schütte, and Frank Noé.
\newblock Data-{Driven} {Model} {Reduction} and {Transfer} {Operator}
  {Approximation}.
\newblock \emph{Journal of Nonlinear Science}, 28\penalty0 (3):\penalty0
  985--1010, June 2018.
\newblock ISSN 1432-1467.
\newblock \doi{10.1007/s00332-017-9437-7}.
\newblock URL \url{https://doi.org/10.1007/s00332-017-9437-7}.

\bibitem[Koopman(1931)]{koopman_hamiltonian_1931}
B.~O. Koopman.
\newblock Hamiltonian {Systems} and {Transformations} in {Hilbert} {Space}.
\newblock \emph{Proceedings of the National Academy of Sciences of the United
  States of America}, 17\penalty0 (5):\penalty0 315--318, 1931.
\newblock ISSN 0027-8424.
\newblock URL \url{https://www.jstor.org/stable/86114}.

\bibitem[Mezi{\'c}(2005)]{mezic2005spectral}
Igor Mezi{\'c}.
\newblock Spectral properties of dynamical systems, model reduction and
  decompositions.
\newblock \emph{Nonlinear Dynamics}, 41\penalty0 (1-3):\penalty0 309--325,
  2005.

\bibitem[Brunton et~al.(2016{\natexlab{b}})Brunton, Brunton, Proctor, and
  Kutz]{brunton_koopman_2016}
Steven~L Brunton, Bingni~W Brunton, Joshua~L Proctor, and J~Nathan Kutz.
\newblock Koopman invariant subspaces and finite linear representations of
  nonlinear dynamical systems for control.
\newblock \emph{PloS one}, 11\penalty0 (2):\penalty0 e0150171,
  2016{\natexlab{b}}.

\bibitem[Katsoulakis and Plech{\'a}{\v{c}}(2013)]{katsoulakis2013information}
Markos~A Katsoulakis and Petr Plech{\'a}{\v{c}}.
\newblock Information-theoretic tools for parametrized coarse-graining of
  non-equilibrium extended systems.
\newblock \emph{The Journal of chemical physics}, 139\penalty0 (7):\penalty0
  074115, 2013.

\bibitem[Harmandaris et~al.(2016)Harmandaris, Kalligiannaki, Katsoulakis, and
  Plecháč]{harmandaris_path-space_2016}
Vagelis Harmandaris, Evangelia Kalligiannaki, Markos Katsoulakis, and Petr
  Plecháč.
\newblock Path-space variational inference for non-equilibrium coarse-grained
  systems.
\newblock \emph{Journal of Computational Physics}, 314:\penalty0 355--383, June
  2016.
\newblock ISSN 0021-9991.
\newblock \doi{10.1016/j.jcp.2016.03.021}.
\newblock URL \url{http://www.sciencedirect.com/science/article/pii/
  S002199911600173X}.

\bibitem[Katsoulakis and Vilanova(2019)]{katsoulakis2019data}
Markos~A Katsoulakis and Pedro Vilanova.
\newblock Data-driven, variational model reduction of high-dimensional reaction
  networks.
\newblock \emph{Journal of Computational Physics}, page 108997, 2019.

\bibitem[Mori(1965)]{mori1965transport}
Hazime Mori.
\newblock Transport, collective motion, and brownian motion.
\newblock \emph{Progress of theoretical physics}, 33\penalty0 (3):\penalty0
  423--455, 1965.

\bibitem[Zwanzig(1973)]{zwanzig1973nonlinear}
Robert Zwanzig.
\newblock Nonlinear generalized langevin equations.
\newblock \emph{Journal of Statistical Physics}, 9\penalty0 (3):\penalty0
  215--220, 1973.

\bibitem[Chorin and Stinis(2007)]{chorin2007problem}
Alexandre Chorin and Panagiotis Stinis.
\newblock Problem reduction, renormalization, and memory.
\newblock \emph{Communications in Applied Mathematics and Computational
  Science}, 1\penalty0 (1):\penalty0 1--27, 2007.

\bibitem[Lei et~al.(2016)Lei, Baker, and Li]{lei2016data}
Huan Lei, Nathan~A Baker, and Xiantao Li.
\newblock Data-driven parameterization of the generalized langevin equation.
\newblock \emph{Proceedings of the National Academy of Sciences}, 113\penalty0
  (50):\penalty0 14183--14188, 2016.

\bibitem[Zhu et~al.(2018)Zhu, Dominy, and Venturi]{zhu_estimation_2018}
Yuanran Zhu, Jason~M. Dominy, and Daniele Venturi.
\newblock On the estimation of the {Mori}-{Zwanzig} memory integral.
\newblock \emph{Journal of Mathematical Physics}, 59\penalty0 (10):\penalty0
  103501, September 2018.
\newblock ISSN 0022-2488.
\newblock \doi{10.1063/1.5003467}.
\newblock URL \url{https://aip.scitation.org/doi/10.1063/1.5003467}.

\bibitem[Hoffman et~al.(2013)Hoffman, Blei, Wang, and
  Paisley]{hoffman2013stochastic}
Matthew~D Hoffman, David~M Blei, Chong Wang, and John Paisley.
\newblock Stochastic variational inference.
\newblock \emph{The Journal of Machine Learning Research}, 14\penalty0
  (1):\penalty0 1303--1347, 2013.

\bibitem[Archambeau and Opper(2011)]{archambeau_approximate_2011}
Cedric Archambeau and Manfred Opper.
\newblock Approximate inference for continuous-time {Markov} processes.
\newblock \emph{Bayesian Time Series Models}, pages 125--140, 2011.

\bibitem[Krishnan et~al.(2017)Krishnan, Shalit, and
  Sontag]{krishnan2017structured}
Rahul~G Krishnan, Uri Shalit, and David Sontag.
\newblock Structured inference networks for nonlinear state space models.
\newblock In \emph{Thirty-First AAAI Conference on Artificial Intelligence},
  2017.

\bibitem[Fortuin et~al.(2019)Fortuin, R{\"a}tsch, and
  Mandt]{fortuin2019multivariate}
Vincent Fortuin, Gunnar R{\"a}tsch, and Stephan Mandt.
\newblock Multivariate time series imputation with variational autoencoders.
\newblock \emph{arXiv preprint arXiv:1907.04155}, 2019.

\bibitem[Butcher(2016)]{butcher2016numerical}
John~Charles Butcher.
\newblock \emph{Numerical methods for ordinary differential equations}.
\newblock John Wiley \& Sons, 2016.

\bibitem[Coifman et~al.(2008)Coifman, Kevrekidis, Lafon, Maggioni, and
  Nadler]{coifman_diffusion_2008}
R.R. Coifman, I.G. Kevrekidis, S.~Lafon, M.~Maggioni, and B.~Nadler.
\newblock Diffusion maps, reduction coordinates and low dimensional
  representation of stochastic systems.
\newblock \emph{Multiscale Modeling \& Simulation}, 7\penalty0 (2):\penalty0
  842 -- 864, 2008.
\newblock ISSN 1540-3459.

\bibitem[Trashorras and Tsagkarogiannis(2010)]{trashorras_mesoscale_2010}
J.~Trashorras and D.~Tsagkarogiannis.
\newblock From {Mesoscale} {Back} to {Microscale}: {Reconstruction} {Schemes}
  for {Coarse}-{Grained} {Stochastic} {Lattice} {Systems}.
\newblock \emph{SIAM Journal on Numerical Analysis}, 48\penalty0 (5):\penalty0
  1647--1677, January 2010.
\newblock ISSN 0036-1429.
\newblock \doi{10.1137/080722382}.
\newblock URL \url{http://epubs.siam.org/doi/abs/10.1137/080722382}.

\bibitem[Katsoulakis and Trashorras(2006)]{katsoulakis_information_2006}
Markos~A. Katsoulakis and José Trashorras.
\newblock Information loss in coarse-graining of stochastic particle dynamics.
\newblock \emph{Journal of statistical physics}, 122\penalty0 (1):\penalty0
  115--135, 2006.
\newblock URL \url{http://link.springer.com/article/10.1007/s10955-005-8063-1}.

\bibitem[Kondrashov et~al.(2015)Kondrashov, Chekroun, and
  Ghil]{kondrashov_data-driven_2015}
Dmitri Kondrashov, Mickaël~D. Chekroun, and Michael Ghil.
\newblock Data-driven non-{Markovian} closure models.
\newblock \emph{Physica D: Nonlinear Phenomena}, 297:\penalty0 33--55, March
  2015.
\newblock ISSN 0167-2789.
\newblock \doi{10.1016/j.physd.2014.12.005}.
\newblock URL \url{http://www.sciencedirect.com/science/article/pii/
  S0167278914002413}.

\bibitem[Coleman and Gurtin(1967)]{coleman_thermodynamics_1967}
Bernard~D. Coleman and Morton~E. Gurtin.
\newblock Thermodynamics with {Internal} {State} {Variables}.
\newblock \emph{The Journal of Chemical Physics}, 47\penalty0 (2):\penalty0
  597--613, July 1967.
\newblock ISSN 0021-9606, 1089-7690.
\newblock \doi{10.1063/1.1711937}.
\newblock URL \url{http://scitation.aip.org/content/aip/journal/jcp/47/2/
  10.1063/1.1711937}.

\bibitem[Cappe et~al.(2005)Cappe, Moulines, and Ryden]{cappe_inference_2005}
O.~Cappe, E.~Moulines, and T.~Ryden.
\newblock \emph{Inference in {Hidden} {Markov} {Models}}.
\newblock Springer-Verlag, 2005.

\bibitem[Ghahramani(2004)]{ghahramani_unsupervised_2004}
Z.~Ghahramani.
\newblock Unsupervised {Learning}.
\newblock In O.~Bousquet, G.~Raetsch, and U.~von Luxburg, editors,
  \emph{Advanced {Lectures} on {Machine} {Learning} {LNAI} 3176}.
  Springer-Verlag, 2004.

\bibitem[Durstewitz(2017)]{durstewitz_state_2017}
Daniel Durstewitz.
\newblock A state space approach for piecewise-linear recurrent neural networks
  for identifying computational dynamics from neural measurements.
\newblock \emph{PLOS Computational Biology}, 13\penalty0 (6):\penalty0
  e1005542, June 2017.
\newblock ISSN 1553-7358.
\newblock \doi{10.1371/journal.pcbi.1005542}.
\newblock URL \url{http://journals.plos.org/ploscompbiol/article?id=10.1371/
  journal.pcbi.1005542}.

\bibitem[Wiskott and Sejnowski(2002)]{wiskott_slow_2002}
L.~Wiskott and T.~J. Sejnowski.
\newblock Slow feature analysis: {Unsupervised} learning of invariances.
\newblock \emph{Neural Computation}, 14\penalty0 (4):\penalty0 715--770, April
  2002.
\newblock \doi{10.1162/089976602317318938}.

\bibitem[Bishop(2006)]{bishop_pattern_2006}
Christopher~M. Bishop.
\newblock \emph{Pattern {Recognition} and {Machine} {Learning}}.
\newblock Springer, 2006.

\bibitem[Kingma and Welling(2014)]{kingma_auto-encoding_2014}
Diederik~P. Kingma and Max Welling.
\newblock Auto-{Encoding} {Variational} {Bayes}.
\newblock In \emph{The {International} {Conference} on {Learning}
  {Representations} ({ICLR})}, volume abs/1312.6114, Banff, Alberta, Canada,
  2014.
\newblock URL \url{http://arxiv.org/abs/1312.6114}.

\bibitem[Kim et~al.(2018)Kim, Wiseman, Miller, Sontag, and
  Rush]{kim_semi-amortized_2018}
Yoon Kim, Sam Wiseman, Andrew~C Miller, David Sontag, and Alexander~M Rush.
\newblock Semi-amortized variational autoencoders.
\newblock \emph{arXiv preprint arXiv:1802.02550}, 2018.

\bibitem[Grigo and Koutsourelakis(2019{\natexlab{b}})]{grigo2019bayesian}
Constantin Grigo and Phaedon-Stelios Koutsourelakis.
\newblock Bayesian model and dimension reduction for uncertainty propagation:
  applications in random media.
\newblock \emph{SIAM/ASA Journal on Uncertainty Quantification}, 7\penalty0
  (1):\penalty0 292--323, 2019{\natexlab{b}}.

\bibitem[Li et~al.(2007)Li, Kevrekidis, Gear, and Kevrekidis]{li_deciding_2007}
Ju~Li, Panayotis~G. Kevrekidis, C.~William Gear, and Ioannis~G. Kevrekidis.
\newblock Deciding the {Nature} of the {Coarse} {Equation} {Through}
  {Microscopic} {Simulations}: {The} {Baby}-{Bathwater} {Scheme}.
\newblock \emph{SIAM Rev.}, 49\penalty0 (3):\penalty0 469--487, July 2007.
\newblock ISSN 0036-1445.
\newblock \doi{10.1137/070692303}.
\newblock URL \url{http://dx.doi.org/10.1137/070692303}.

\bibitem[Noid(2013)]{noid_perspective:_2013}
W.~G. Noid.
\newblock Perspective: {Coarse}-grained models for biomolecular systems.
\newblock \emph{The Journal of Chemical Physics}, 139\penalty0 (9), 2013.
\newblock \doi{http://dx.doi.org/10.1063/1.4818908}.
\newblock URL
  \url{http://scitation.aip.org/content/aip/journal/jcp/139/9/10.1063/1.4818908}.

\bibitem[Mackay(1995)]{mackay_probable_1995}
DJC Mackay.
\newblock Probable {Networks} and {Plausible} {Predictions} - a {Review} of
  {Practical} {Bayesian} {Methods} for {Supervised} {Neural} {Networks}.
\newblock \emph{Network-Computation in Neural Systems}, 6\penalty0
  (3):\penalty0 469--505, August 1995.
\newblock \doi{10.1088/0954-898X/6/3/011}.

\bibitem[Bishop and Tipping(2000)]{bishop2000variational}
Christopher~M Bishop and Michael~E Tipping.
\newblock Variational relevance vector machines.
\newblock In \emph{Proceedings of the Sixteenth conference on Uncertainty in
  artificial intelligence}, pages 46--53. Morgan Kaufmann Publishers Inc.,
  2000.

\bibitem[Kingma and Ba(2014)]{kingma2014adam}
Diederik~P Kingma and Jimmy Ba.
\newblock Adam: A method for stochastic optimization.
\newblock \emph{arXiv preprint arXiv:1412.6980}, 2014.

\bibitem[Cottet and Koumoutsakos(2000)]{cottet_vortex_2000}
Georges-Henri Cottet and Petros~D. Koumoutsakos.
\newblock \emph{Vortex {Methods}: {Theory} and {Practice}}.
\newblock Cambridge University Press, Cambridge ; New York, 2 edition edition,
  March 2000.
\newblock ISBN 978-0-521-62186-1.

\bibitem[Roberts(1989)]{roberts_convergence_1989}
Stephen Roberts.
\newblock Convergence of a {Random} {Walk} {Method} for the {Burgers}
  {Equation}.
\newblock \emph{Mathematics of Computation}, 52\penalty0 (186):\penalty0
  647--673, 1989.
\newblock ISSN 0025-5718.
\newblock \doi{10.2307/2008486}.
\newblock URL \url{http://www.jstor.org/stable/2008486}.

\bibitem[Chertock and Levy(2001)]{chertock_particle_2001}
Alina Chertock and Doron Levy.
\newblock Particle {Methods} for {Dispersive} {Equations}.
\newblock \emph{Journal of Computational Physics}, 171\penalty0 (2):\penalty0
  708--730, August 2001.
\newblock ISSN 0021-9991.
\newblock \doi{10.1006/jcph.2001.6803}.
\newblock URL \url{http://www.sciencedirect.com/science/article/pii/
  S0021999101968032}.

\bibitem[Abadi et~al.(2016)Abadi, Agarwal, Barham, Brevdo, Chen, Citro,
  Corrado, Davis, Dean, Devin, et~al.]{abadi2016tensorflow}
Mart{\'\i}n Abadi, Ashish Agarwal, Paul Barham, Eugene Brevdo, Zhifeng Chen,
  Craig Citro, Greg~S Corrado, Andy Davis, Jeffrey Dean, Matthieu Devin, et~al.
\newblock Tensorflow: Large-scale machine learning on heterogeneous distributed
  systems.
\newblock \emph{arXiv preprint arXiv:1603.04467}, 2016.

\bibitem[Pan and Duraisamy(2020)]{pan2020physics}
Shaowu Pan and Karthik Duraisamy.
\newblock Physics-informed probabilistic learning of linear embeddings of
  nonlinear dynamics with guaranteed stability.
\newblock \emph{SIAM Journal on Applied Dynamical Systems}, 19\penalty0
  (1):\penalty0 480--509, 2020.

\end{thebibliography}
